\documentclass[twocolumn,amsmath,showkeys,amssymb,superscriptaddress,longbibliography,floatfix, prl]{revtex4-2}

\pdfoutput=1
\usepackage{graphicx, subfigure}
\usepackage{amssymb, amsmath,amssymb,amsfonts}
\usepackage{amsthm,mathrsfs,amsopn}
\usepackage{dcolumn}
\usepackage{bm}
\usepackage{color}
\usepackage[utf8]{inputenc}
\usepackage{mathtools}
\usepackage[normalem]{ulem}

\theoremstyle{plain} 

\definecolor{brickred}{rgb}{0.7, 0.25, 0.33}
\definecolor{applegreen}{rgb}{0.55, 0.71, 0.0}

\def\be{\begin{equation}}
\def\ee{\end{equation}}
	
\def\bc{\begin{center}}
\def\ec{\end{center}}
\def\bea{\begin{eqnarray}}
\def\eea{\end{eqnarray}}
\newcommand{\avg}[1]{\langle{#1}\rangle}
\newcommand{\Avg}[1]{\left\langle{#1}\right\rangle}

\begin{document}

\title{Network science Ising states of matter}

\author{Hanlin Sun}
\affiliation{School of Mathematical Sciences, Queen Mary University of London, London, E1 4NS, United Kingdom}
\affiliation{Nordita, KTH Royal Institute of Technology and Stockholm University, Hannes Alfvéns väg 12, SE-106 91 Stockholm, Sweden}

\author{Rajat Kumar Panda}
\affiliation{The Abdus Salam International Centre for Theoretical Physics (ICTP), Strada Costiera 11, 34151 Trieste, Italy}
\affiliation{SISSA — International School of Advanced Studies, via Bonomea 265, 34136 Trieste, Italy}
\affiliation{INFN Sezione di Trieste, Via Valerio 2, 34127 Trieste, Italy}

\author{Roberto Verdel}
\affiliation{The Abdus Salam International Centre for Theoretical Physics (ICTP), Strada Costiera 11, 34151 Trieste, Italy}

\author{Alex Rodriguez}
\affiliation{The Abdus Salam International Centre for Theoretical Physics (ICTP), Strada Costiera 11, 34151 Trieste, Italy}
\affiliation{Dipartimento di Matematica e Geoscienze, Universitá degli Studi di Trieste, via Alfonso Valerio 12/1, 34127, Trieste, Italy}

\author{Marcello Dalmonte}
\affiliation{The Abdus Salam International Centre for Theoretical Physics (ICTP), Strada Costiera 11, 34151 Trieste, Italy}
\affiliation{SISSA — International School of Advanced Studies, via Bonomea 265, 34136 Trieste, Italy}
\author{Ginestra Bianconi}
\affiliation{School of Mathematical Sciences, Queen Mary University of London, London, E1 4NS, United Kingdom}
\affiliation{The  Alan  Turing  Institute,  96  Euston  Road,  London,  NW1  2DB,  United  Kingdom}

\begin{abstract}
Network science provides very powerful tools for extracting information from interacting data.
Although recently the unsupervised detection of phases of matter using machine learning has raised significant interest, the full prediction power of network science has not yet been systematically explored in this context.
Here we fill this gap by providing an in-depth statistical, combinatorial, geometrical and topological characterization of 2D Ising snapshot networks (IsingNets) extracted from Monte Carlo simulations of the $2$D  Ising model at different temperatures, going across the phase transition.
Our analysis reveals the complex organization properties of IsingNets in both the ferromagnetic and paramagnetic phases and demonstrates the significant deviations of the  IsingNets with respect to randomized null models. In particular percolation properties of the IsingNets reflect the existence of the symmetry between configurations with opposite magnetization below the critical temperature and the very compact nature of the two emerging giant clusters revealed by our persistent homology analysis of the IsingNets. Moreover, the IsingNets display a very broad degree distribution and significant degree-degree correlations and weight-degree correlations demonstrating that they encode relevant information present in the configuration space of the $2$D Ising model. The geometrical organization of the critical IsingNets is reflected in their spectral properties deviating from the one of the null model. This work reveals the important insights that network science can bring to the characterization of phases of matter. The set of tools described hereby can be applied as well to numerical and experimental data. 
\end{abstract}

\maketitle
\section{Introduction}

Networks \cite{barabasi2016network,newman2018networks,barrat2008dynamical,estrada2012structure,dorogovtsev2022nature} encode the information present in a large variety of natural and artificial interacting systems by representing them as graphs, i.e. a set of nodes (representing the element of the system) connected by links or edges (representing typically the interactions). In particular, the underlying architecture of complex systems is encoded in networks that strongly deviate from random graphs whose information content can be mined by exploiting their statistical, combinatorial as well as the geometrical and topological nature \cite{bianconi2021higher}.
Networks are hence a  simple yet very powerful framework that has been able in the last twenty years to transform our understanding of complex systems. These complex networks obey relevant organization principles while retaining a stochastic nature.

Recently great attention has been addressed to formulating unsupervised machine learning algorithms to detect different phases of matter \cite{MEHTA20191, PhysRevB.94.195105, PhysRevE.95.062122, PhysRevE.96.022140, Rodriguez-Nieva2019, PhysRevX.11.011040}. {\color{black}In this research line, important progress in characterizing phase transitions has recently been made by using interpretable machine learning tools such as graphical models and restricted Boltzmann machines \cite{morningstar2018deep,aoki2016restricted,carrasquilla2019reconstructing,benedetti2017quantum,melko2019restricted}.} The core of such approaches is that learning methods - in particular, unsupervised - shall be capable of revealing hidden structures in data sets, that correspond to physical information (such as, e.g., the existence of order parameters). However the very advanced, complementary tools of network science to extract information from complex data of interacting systems have not yet been systematically employed for this task.

Here we want to show how network science can enrich and enhance our unsupervised characterization of phases of matter. Indeed we will show evidence that network science provides a very transparent set of methods to investigate the characteristics of different phases across critical phase transitions and allows the determination of unsupervised indicators of their critical points.

Historically networks have been used in condensed matter for describing physical interactions existing among the elements of a system, as well as structured in configuration and Hilbert space. In principle, they can be used as well to represent abstract data structures coming from numerical simulations or directly from experiments, giving access to a whole new toolbox to define and interpret many-body correlations (of arbitrary order, if expressed in terms of local observables). 

Reflecting this two-fold possible application of network science, recently several works have explored the use of networks to model or represent interactions in condensed matter systems.
For instance, networks can be considered to define Hamiltonians whose interaction terms are determined by a complex network rather than by a lattice. In particular networks can be used to define different critical phenomena including the Ising model \cite{dorogovtsev2002ising,bianconi2002mean,leone2002ferromagnetic} and the inverse Ising model \cite{nguyen2017inverse},   as well as quantum critical phenomena including the transverse Ising model \cite{bianconi2012superconductor,bianconi2013superconductor,chepuri2022complex}, the Bose-Hubbard model~\cite{halu2012phase}, the Jaynes-Cummings-Hubbard model~\cite{halu2013phase}, in addition to others classical collective phenomena and inverse problems \cite{vicsek1995novel,ballerini2008interaction,morcos2011direct,mora2011biological}.
Alternatively, networks can define quantum environments \cite{nokkala2016complex,nokkala2018reconfigurable} or even multilayer couplings between interdependent superconductor networks \cite{bonamassa2023interdependent}.  Networks also can be used to represent correlations in complex and financial networks \cite{tumminello2005tool,bonanno2003topology} as well as in quantum systems. 
In particular recently network structures have been shown to encode the quantum long-range mutual information {(and other measures of quantum correlations)} existing among the nodes of quantum {lattice models in one~\cite{valdez2017quantifying, PhysRevA.97.052320, doi:10.1098/rsta.2020.0421} and two dimensions~\cite{Bagrov2020}}, providing {in some cases} indicators for quantum critical points.  

{As we will see in this work, weighted networks \cite{petri2013topological} are amenable to be analysed and treated with Topological Data Analysis (TDA) \cite{edelsbrunner2022computational,otter2017roadmap,vaccarino2022persistent,ghrist2008barcodes} which provides a very efficient way to probe topological, large scale and global network properties. TDA, although until now only applied to point clouds, is raising significant interest to do unsupervised inference of phase transitions~\cite{PhysRevE.93.052138, PhysRevResearch.2.043308, PhysRevB.104.104426, PhysRevE.103.052127, PhysRevB.104.235146, PhysRevE.105.024121, 10.1140/epjb/s10051-022-00453-3}, and has wide applications, including the characterization of universal dynamics in quantum gases~\cite{10.21468/SciPostPhys.11.3.060} and of confinement in lattice field theory~\cite{PhysRevB.106.085111, PhysRevD.107.034501, PhysRevD.107.034506}}.

Finally and most relevantly for our work, networks have been proposed to capture the underlying structure of quantum spin systems as revealed by wave function snapshots that can be probed experimentally as well as sampled from Monte Carlo simulations \cite{mendes2023wave}.
However, despite these very pioneering works \cite{valdez2017quantifying,mendes2023wave},  little attention has been so far addressed to study phases of matter using network science.

Here we launch a large-scale systematic study of the phase of matter based on network science. We leverage a multiplicity of tools developed in network science and we reveal the combinatorial, statistical, geometrical and topological network representation of different phases of matter. We provide an in-depth characterization of the networks generated from single snapshots of spin system configurations. 

In Ref. \cite{mendes2023wave} it was shown that wave function networks constructed starting from quantum wave function snapshots are strongly deviating from random graphs and for a wide range of values of the threshold distances they give rise to networks with very broad degree distribution. An open question is whether the complex properties of these networks are inherently quantum effects or they can be observed in classical systems as well.

{\color{black}To characterize phase transitions and critical behaviors, the Ising model is arguably the most studied model in statistical physics. Apart from the Ising model defined on lattices, the model has also been generalized on networked structures such as small-world networks \cite{herrero2002ising,pekalski2001ising}, random scale-free networks \cite{bianconi2002mean, leone2002ferromagnetic,dorogovtsev2002ising,herrero2004ising, herrero2015ising} and spatially embedded scale-free networks \cite{bradde2010critical}. The phase diagram of the model is shown to be highly sensible on the value of the branching ratio of the network. These results have been also extended to the Transverse Ising models \cite{bianconi2012superconductor,bianconi2013superconductor}. Moreover some of the machine learning tools to study the Ising model such as restricted Boltzmann machine and graphical models \cite{morningstar2018deep,aoki2016restricted,carrasquilla2019reconstructing,benedetti2017quantum,melko2019restricted} strongly leverage on their underlying network structure.
However, to the best of our knowledge, the tools from network science have not been used to the analysis the simulation snapshots of the Ising model.}

In this work we consider  2D Ising snapshot networks (IsingNets) following a construction proposed in Ref. \cite{mendes2023wave} applied to classical 2D Ising model snapshots and we characterize their structure using advanced statistical, combinatorial, geometrical and topological tools of network theory. 
In order to provide an in-depth analysis of the IsingNet, across different phases we focus our attention on IsingNets obtained starting from  Monte Carlo simulations of the   $2$D Ising model performed across the phase transition.
IsingNets are obtained from a sample of state configurations of the $2$D Ising model which constitute the set of nodes of the IsingNets. Each pair of nodes of an IsingNet is associated with a distance, here taken to be the Euclidean distance between the configuration snapshots.
IsingNets are constructed starting from the fully connected distance matrix between the nodes by connecting only the nodes whose distance is smaller than a threshold value of the distance.

Our in-depth network analysis of the IsingNets will allow us to go well beyond the characterization of these networks 
based solely on the degree distribution.  Possibly in the future, this in-depth analysis can be conducted also on networks built from quantum wave function snapshots in order to assess which are the properties inherently quantum in the latter networks.

Our analysis is conducted following two main directions whose goal is different but complementary. First, we will perform an analysis of the IsingNets that is agnostic about the choice of the distance threshold. In particular, we will study network properties as a function of the distance threshold. These include percolation properties, persistence homology, network embedding and statistical characterization of the distance matrices. Secondly, we will consider specific choices of the distance threshold and we will characterize the statistical, combinatorial and geometrical/spectral properties of the IsingNets, showing the important roles of degree-degree and weight-degree correlations in these systems.

Anticipating our main results we have found that IsingNets reflect the symmetry of the configuration space of the $2$D Ising model in a prominent way. In particular, the percolation properties of the IsingNets strongly deviate from the percolation properties of networks in which the same distances among the nodes are distributed randomly. In fact, below the critical temperature of the $2$D Ising model, the IsingNets are characterized by two giant components whereas their randomized counterpart displays only a single giant component. When the threshold distance defining the IsingNets is raised significantly these two giant clusters merge, but interestingly they keep a rather compact structure as revealed by our persistent homology results highlighting that the Betti number of their clique complex are strongly suppressed with respect to their random counterpart. {\color{black}The Weisserstein distance between the persistence diagrams of the IsingNets and their randomized couterpart is here shown to be an unsupervised indicator  of criticality as it displays a maximum in correspondence of the critical temperature.}
Our statistical and combinatorial analysis of the IsingNets strongly demonstrates the complex organization of these networks which display strong heterogeneity in both their topological (degree, clustering coefficient, degree correlations) and their weighted network properties.  In particular nodes of higher-degree are characterized by having neighbours connected by stronger affinity weights (smaller distances). Finally, the IsingNets possess a significant geometrical organization at criticality as it is revealed by their interesting spectral properties.

Among the most important benefits of our approach is that the proposed analysis is fully interpretable.
Moreover, we emphasize the applicability of the approach. Indeed while our analysis is here performed on data coming from Monte Carlo simulations the approach can be readily applied also to experimental data.
However, a possible limitation of the approach is due to the significant computational cost of our proposed unsupervised TDA analysis.

\section{The 2D Ising model Monte Carlo simulations}
We consider a square $2$D lattice of dimension $L\times L$ where on each site $n$ is located the spin $S_n\in \{-1,1\}$.
The nearest neighbour spins are interacting through the Hamiltonian 
\bea
H=-\sum_{\langle n,m \rangle}S_nS_m.
\eea
The $2D$ Ising model is characterized by $\mathbb{Z}_2$ spontaneous symmetry breaking and undergoes a  second-order phase transition
 at $T_c=2/\ln(1+\sqrt{2})\approx 2.269$ \cite{onsager1944crystal}.
Starting from Markov Chain Monte Carlo simulations of this model, for each temperature single snapshots $\vec{x}_i=\{S_1, S_2,\ldots, S_{L^2}\}$ of the spin system are sampled at equilibrium \cite{panda23}. More specifically, we use the Wolff cluster algorithm~\cite{wolff1989collective,wolff1989comparison}, starting from the configuration with either all up spins or all down spins, chosen at random. Next, $30000$ to $50000$ `cluster flips' are performed for the system to equilibrate.
After this, we collect snapshots every $1000$ to $1500$ cluster flips to ensure that the collected state configurations are as uncorrelated as possible~\cite{panda23}. In total, we gather $10000$ snapshots during a Monte Carlo run.

For each temperature, five independent Monte Carlo simulations are performed as prescribed. By combining the sampled configuration snapshots of these runs, we thus obtain a data set with $N_r=50000$ independent thermal configurations $\{\vec{x}_i\}_{i=1}^{N_r}$. 
The starting point to construct the IsingNets is a set of $N$ configuration snapshots $i\in \{1,2,\ldots, N\}$  randomly selected from the data set described above, and the fully connected distance matrix ${\bf d}$ of elements $d_{ij}$  between these states. 
Here the distance $d_{ij}$ between two generic snapshots $\vec{x}_i$ and $\vec{x}_j$ is taken to be their Euclidean distance. As discussed in previous works, such manifolds are typically living in very high dimensional subspaces~\cite{PhysRevX.11.011040,panda23}, so that simple dimensional reductions are not applicable, and a full-fledged network analysis is needed.

\section{Network characterization across the distance filtration}
\subsection{Weight filtration}
As anticipated in the introduction, in this first Section, our analysis focuses on the properties of IsingNets observed as a function of the distance filtration. 
We consider IsingNets which are graphs $G=(V,E)$ formed by a set of $N$ nodes $V$  and a set of links  $E$ with $(i,j)\in E$ only if the nodes (state configurations) $i$ and $j$ have distance $d_{ij}<r$. Here $r$ determines the distance filtration and indicates a tunable parameter that ranges from the minimum of the distances $r_{min}=\mbox{min}_{i,j}d_{ij}$  between the $N$ nodes to their maximum distance  $r_{max}=\mbox{max}_{i,j}d_{ij}$. We are thus here completely agnostic about the best choice of $r$ as we study the properties of IsingNets across all possible choices of the distance threshold $r$.

In particular as a function of $r$ we will explore the percolation properties of the IsingNets which define an agglomeration from $N$ disconnected nodes to a single connected component as $r$ is raised from $r_{min}$  to $r_{max}$ and we will compare this process with the corresponding null model obtained from the same process applied to a randomized distance ${\bf d}^{rand}$ matrix. The randomized distance matrix ${\bf d}^{rand}$ is constructed by randomly reshuffling the upper triangular elements of $d$ and subsequently symmetrizing the matrix. Therefore the null model networks display the same distribution of ``distances" as the true IsingNet while being completely randomized. Note however that one of the main differences between the distance matrix ${\bf d}$ and the randomized distance matrix ${\bf d}^{rand}$ is that the entries of ${\bf d}^{rand}$  are not proper distances as they do not obey the triangular inequality.

In this section, we will use a combination of tools coming from network science to analyse the IsingNets described above. In particular, anticipating the detailed description of the methods used to perform this analysis in the following paragraphs,  we will use persistent homology  \cite{ghrist2008barcodes,otter2017roadmap,vaccarino2022persistent} to further characterize topologically the mentioned aggregation process. In this way, we will show that persistent homology is able to detect the position of the critical temperature of the $2$D Ising model under study. This analysis will be accompanied by the visualization of the network using Minimum Spanning Trees \cite{graham1985history,bonanno2003topology},  the results of the network embedding conducted using the UMAP (Uniform manifold approximation and projection for dimension reduction) algorithm  \cite{mcinnes2018umap},  and the statistical characterization of the distance matrix as a function of the temperature conducted using the closeness centrality \cite{bavelas1950communication} distribution.

\subsection{Percolation process}

We start our investigation exploring the percolation process \cite{dorogovtsev2008critical,newman2018networks,li2021percolation,cohen2000resilience,barrat2008dynamical} monitoring the connected component of the network as a function of the filtration parameter $r$. To contrast the behavior of the percolation process of IsingNets with a null-hypothesis percolation process, we consider the process defined on the actual IsingNet distance matrix ${\bf d}$ with the same process defined starting from a matrix ${\bf d}^{rand}$ obtained by randomly permuting distances among pair of nodes. 

\begin{figure}[!htb]
  \includegraphics[width=\columnwidth]{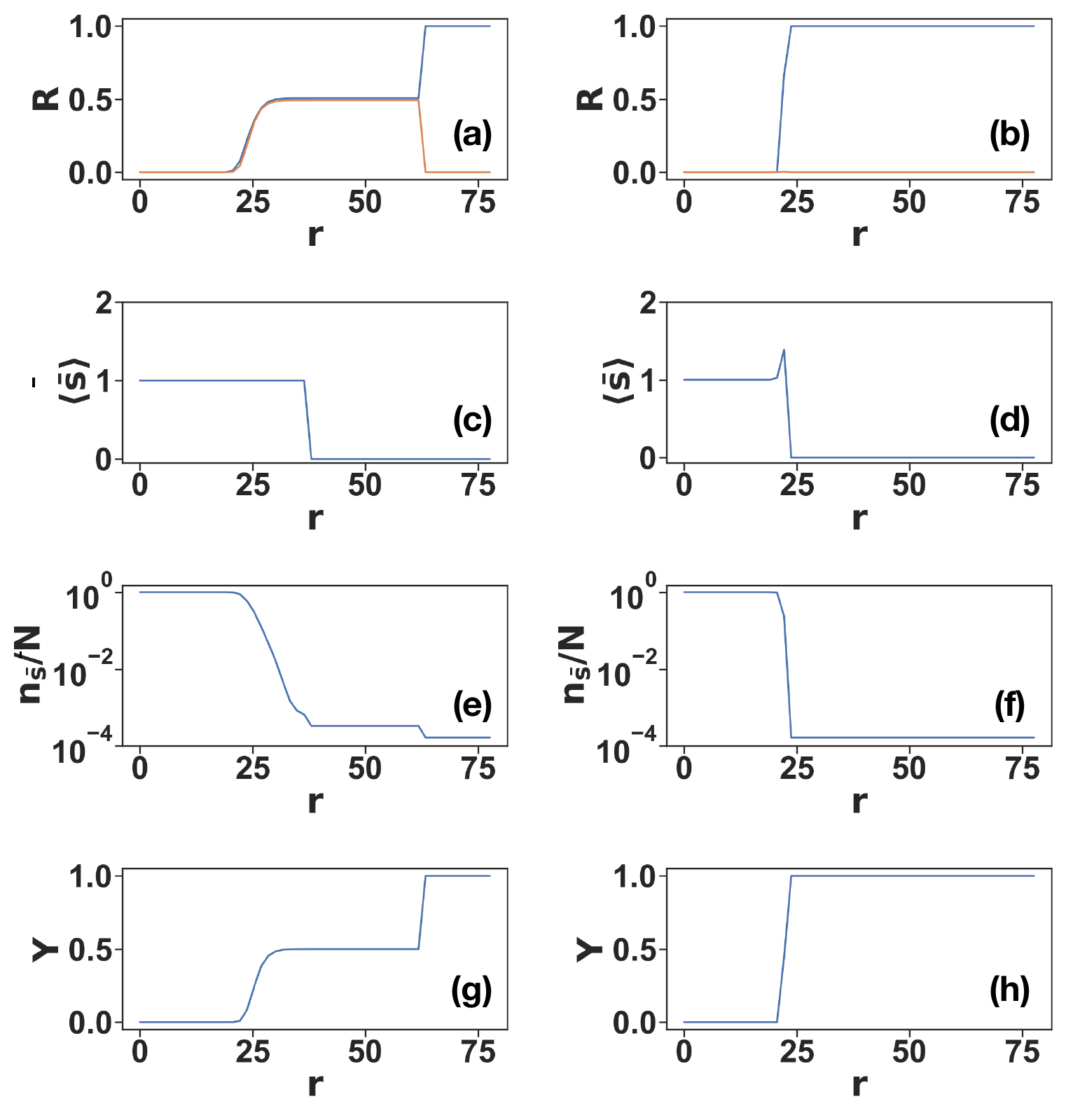}
  \caption{The percolation properties of the IsingNet (left panels) generated from  $2$D Ising model Monte Carlo simulations on spin systems of linear size $L=40$ at temperature $T=2.12<T_c$  are shown as a function of filtration parameter $r$. Nodes are connected if their distance is less than $r$. Five quantities are measured: the fraction of nodes in the largest connected component (the first row, blue line) and the fraction of nodes in the second largest connected component (the first row, orange line), the average size of components that are smaller than the second largest component $\langle s \rangle$ (the second row), the number of components $n_s$ (the third row) and the inverse participation ratio $Y$ (the fourth row). The results are compared with these quantities obtained from corresponding percolation properties obtained from a randomly permuted distance matrix (right panels). The number of nodes of the IsingNets is $N=6000.$ }
  \label{fig:1}
 \end{figure}

\begin{figure}[!htb]
 \includegraphics[width=\columnwidth]{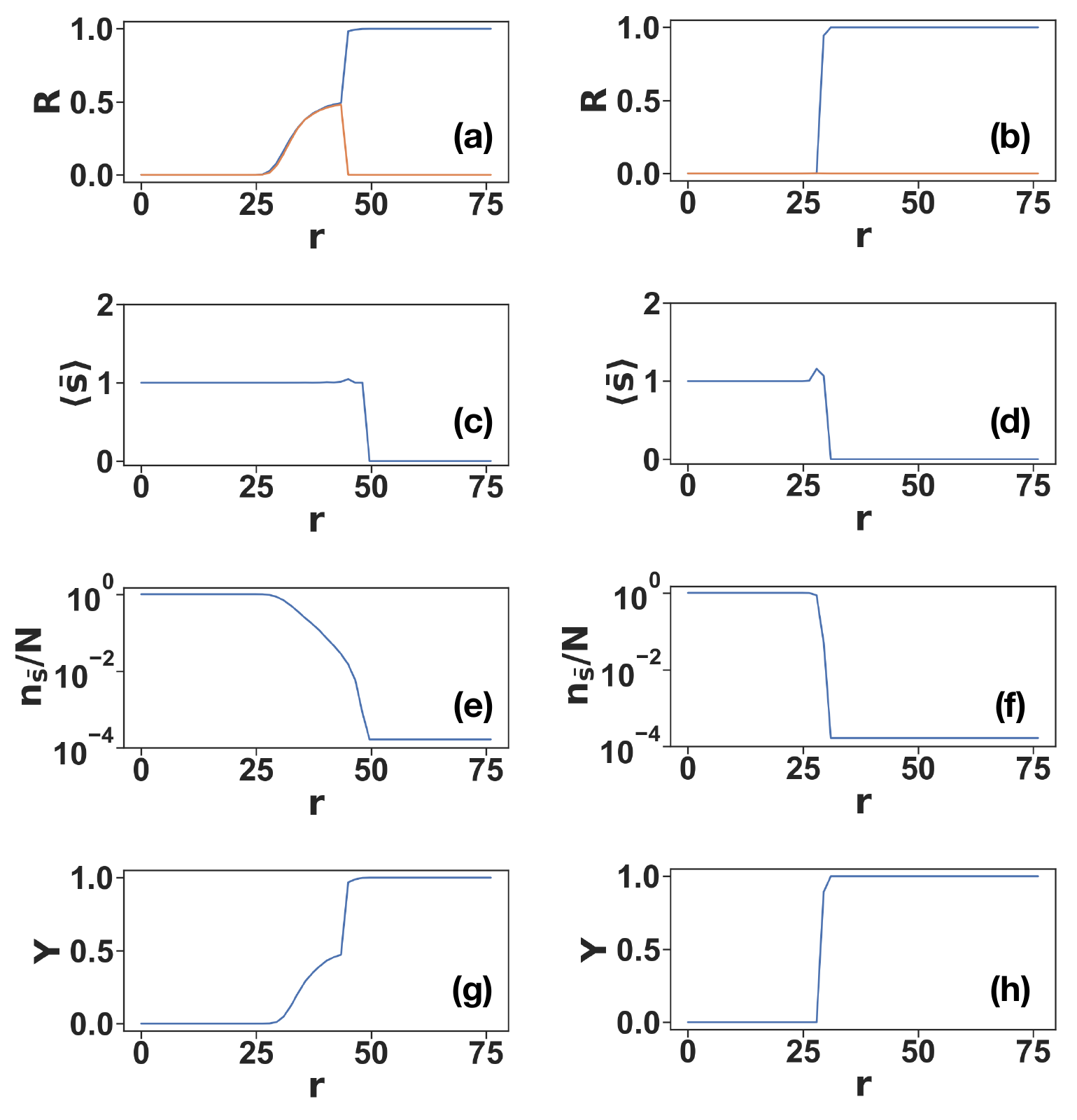}
 \caption{Same as Figure \ref{fig:1} but with  IsingNets  obtained from $2$D Ising model simulations at $T=2.25$. }
 \label{fig:2}
\end{figure}

\begin{figure}[!htb]
  \includegraphics[width=\columnwidth]{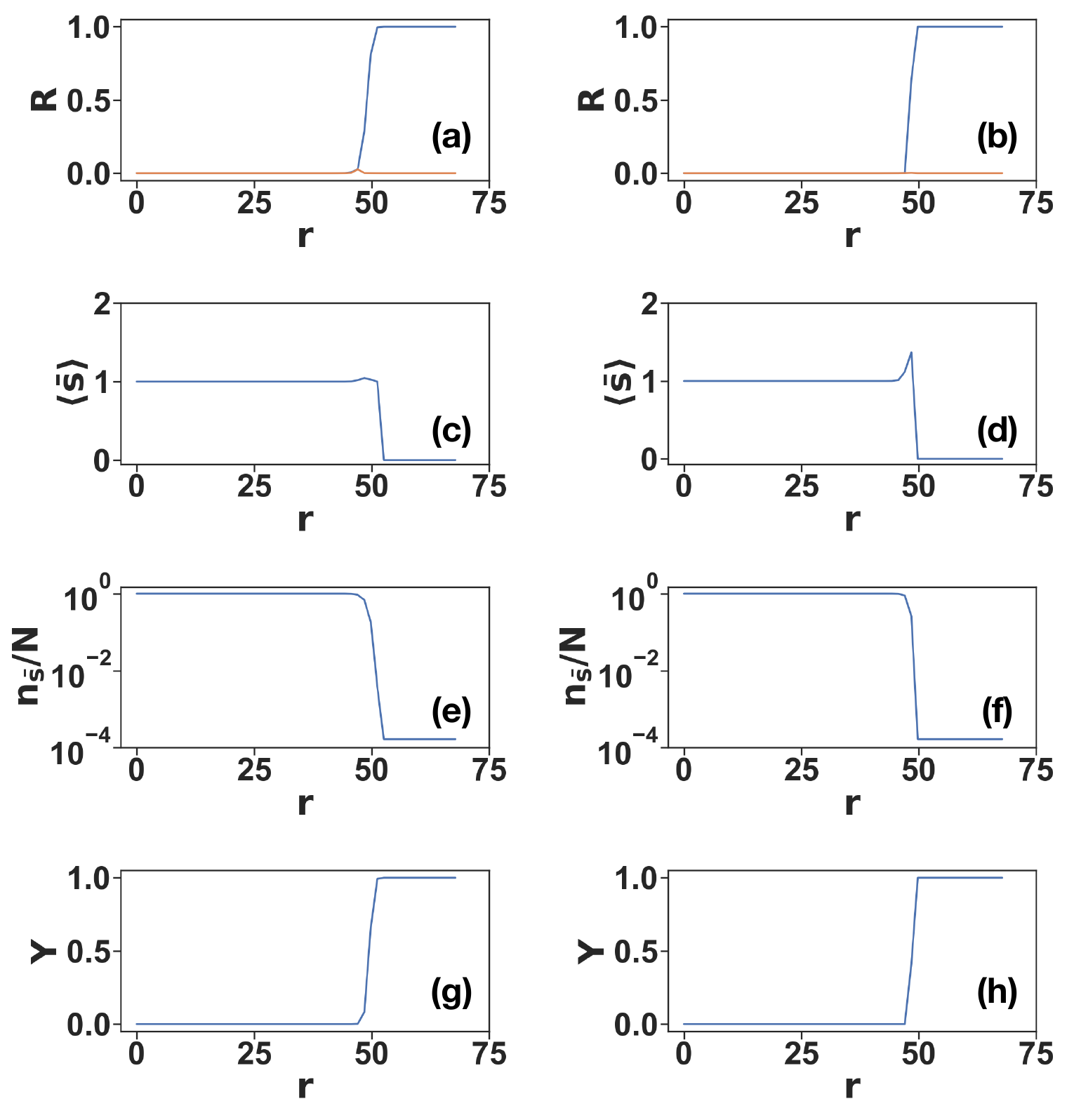}
  \caption{Same as Figure \ref{fig:1} but with  IsingNets  obtained from $2$D Ising model simulations at $T=2.50$. }
  \label{fig:3}
 \end{figure}

The percolation process of IsingNets reveals a major difference with the percolation process on the randomized distance matrix: mainly the IsingNets obtained for the $2$D Ising model below the phase transition, i.e. $T<T_c$ display for a very significant range of values of the filtration parameter $r$, two giant components while the randomized process only displays one giant component.
This phenomenon is evident from the plot in Figures \ref{fig:1} and \ref{fig:2} showing the relative size of the largest component $R$ and the second largest component $R_2$, which are both giant, i.e. extensive for a wide range of $r$ values.
Indeed below the critical temperature $T_c$, the IsingNets display two transitions as the value of the filtration parameter is raised (see Figures \ref{fig:1} and \ref{fig:2}). The first transition is characterized by the emergence of the two equal size giant components corresponding to the symmetry of the configuration snapshots of the $2$D Ising model for $T<T_c$ and the second one is characterized by the merging of these two giant components for very large values of $r$, characterized by the disappearance of a significant second largest connected component (orange line in Figure \ref{fig:1} (a) and Figure \ref{fig:2} (a).  This phenomenology is dramatically different from the percolation obtained in the randomized null model where the giant component is unique for every value of $r$ (see Figures \ref{fig:1} and \ref{fig:2}). 
For temperatures above the critical one (see Figure $\ref{fig:3}$), instead, only one giant component is observed corresponding to the paramagnetic state of the $2$D Ising model.
In order to further characterize the percolation process we also monitor as a function of $r$
the average size of finite components $\langle \bar{s} \rangle$, the number of components $n_{\bar{s}}$ and the inverse participation ratio $Y$ whose inverse determines the number of typical clusters. 
Specifically the inverse participation ratio $Y$ is defined as
\bea
Y = \sum_p\left(\frac{\bar{s}_p}{\sum_q \bar{s}_q}\right)^2 = \frac{1}{N^2}{\sum_p \bar{s}_p^2}.
\eea
where $\bar{s}_p$ indicates the size of the $p$-th largest component. The abrupt increase of $Y$ at large $r$ further indicates the merging of two giant components (Figure \ref{fig:1} (g) and Figure \ref{fig:2} (g)).

On an Erd\"os-Renyi random graph the average size of finite component $\Avg{\bar{s}}$ plays the role of the percolation susceptibility~\cite{dorogovtsev2008critical}, diverging in correspondence with the emergence of the (single) giant component. In the IsingNets (see Figures \ref{fig:1}-\ref{fig:3}) the average size of finite component $\Avg{\bar{s}}$ has a very suppressed maximum with respect to the same quantity measured on the considered null model, indicating that the agglomeration of the two giant clusters proceeds by subsequent agglomeration of very small components and isolated nodes rather than by the agglomeration of finite clusters of diverging average size as in a random graph.
This is also confirmed by the behavior of the number of clusters $n_{\bar{s}}$ as a function of the filtration parameter $r$ which for the IsingNets decays less steeply than in the randomized null model.
Finally, the inverse participation ratio $Y$ for low temperatures reveals a significant plateau at $Y=1/2$ indicating the existence of two giant components of approximately equal size (see Figure \ref{fig:1}).
Above the critical temperature, for $T>T_c$ as the filtration parameter is raised only one giant component emerges and the difference with respect to the randomized null model is reduced (see Fig. \ref{fig:3}).

 \subsection{Persistent homology}

Topology is the study of shapes and their invariant properties under continuous deformations (see for an introduction \cite{bianconi2021higher,edelsbrunner2022computational}). Major examples of topological invariants are the Betti numbers. 
The Betti number $\beta_0$ indicates the number of connected components, the Betti number $\beta_1$ indicates the number of one-dimensional holes, the number of $\beta_2$ indicates the number of two-dimensional holes, etc.
For instance, a point has Betti numbers $\beta_0=1$ and $\beta_n=0$ for any other value of $n$, a circle has non-zero Betti numbers $\beta_0=\beta_1=1$ and a sphere has non-zero Betti numbers $\beta_0=\beta_2=1$.
An important result of algebraic topology~\cite{bianconi2021higher,edelsbrunner2022computational,ghrist2008barcodes,otter2017roadmap,vaccarino2022persistent} is that the $\bar{n}$-dimensional Betti number $\beta_{\bar{n}}$ is the rank  of the $\bar{n}$-dimensional homology group of the considered topological space.

In the discrete setting, the Betti numbers are defined in general for simplicial complexes. Simplicial complexes are a type of higher-order network formed by a set of simplices such as nodes, links, triangles, tetrahedra, etc. They have the additional property of being closed under the inclusion of faces of each simplex. This last property implies that if a triangle belongs to the simplicial complexes also all its links and nodes belong to the simplicial complex.

\begin{figure*}[!htb]
  \includegraphics[width=0.7\textwidth]{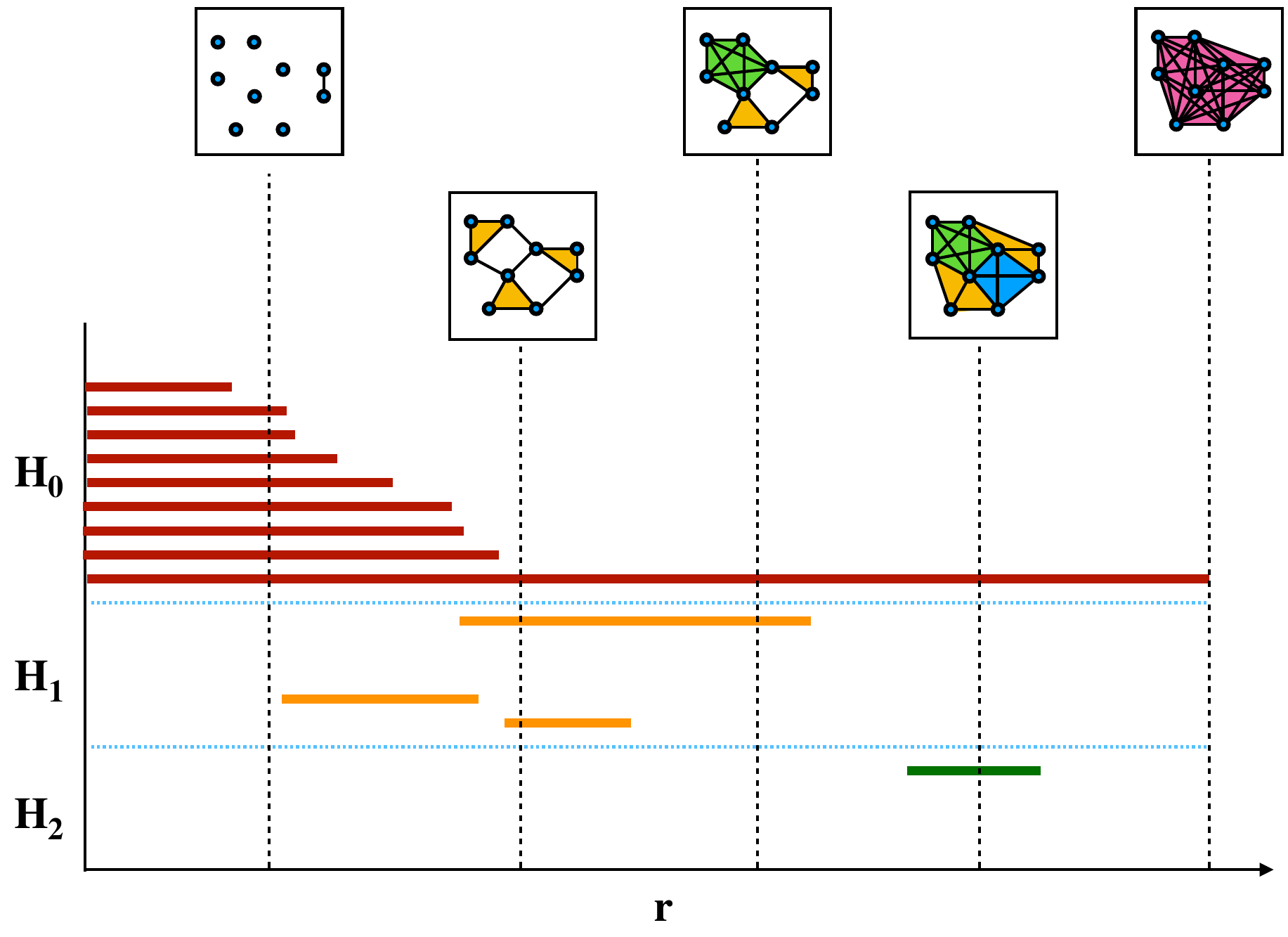}
  \caption{A schematic illustration of the filtration process. The barcodes are used to show the appearance and disappearance of topological features corresponding to different homology classes as the filtration parameter $r$ is increased. The filtration process ends at $r=r_{\max}$ when all $N$ nodes are fully connected and forms a $N$-simplex. Simplices of different dimensions are indicated by different colors.}
  \label{fig:barcode}
 \end{figure*}
 
\begin{figure*}[!htb]
  \includegraphics[width=\textwidth]{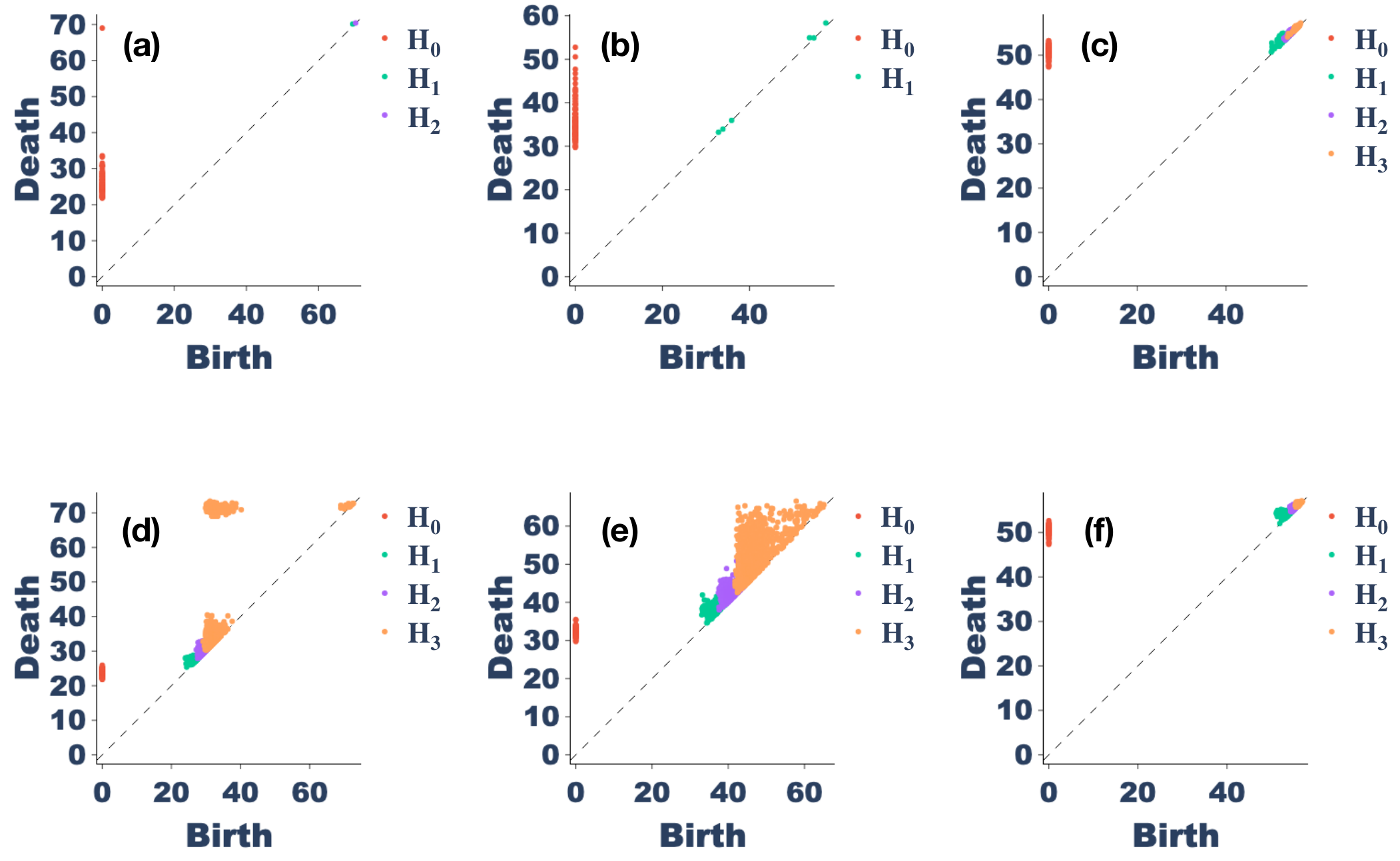}
  \caption{The persistent diagram corresponding to homology classes in $H_0$, $H_1$, $H_2$, and $H_3$ of the IsingNet clique complexes are plotted as a function of the filtration parameter $r$. Panels (a), (b), and (c) show the persistent diagrams of IsingNets {obtained from the spin system of linear size $L=40$} at $T=2.12$ (a), $T=2.25$ (b), and $T=2.50$ (c). Panels (d), (e), and (f) show the persistent diagram of corresponding randomized null models obtained at $T=2.12$ (d), $T=2.25$ (e), and $T=2.50$ (f). The networks are formed by $N=100$ nodes.}
  \label{fig:homology}
 \end{figure*}
 
 \begin{figure*}[!htb]
  \includegraphics[width=\textwidth]{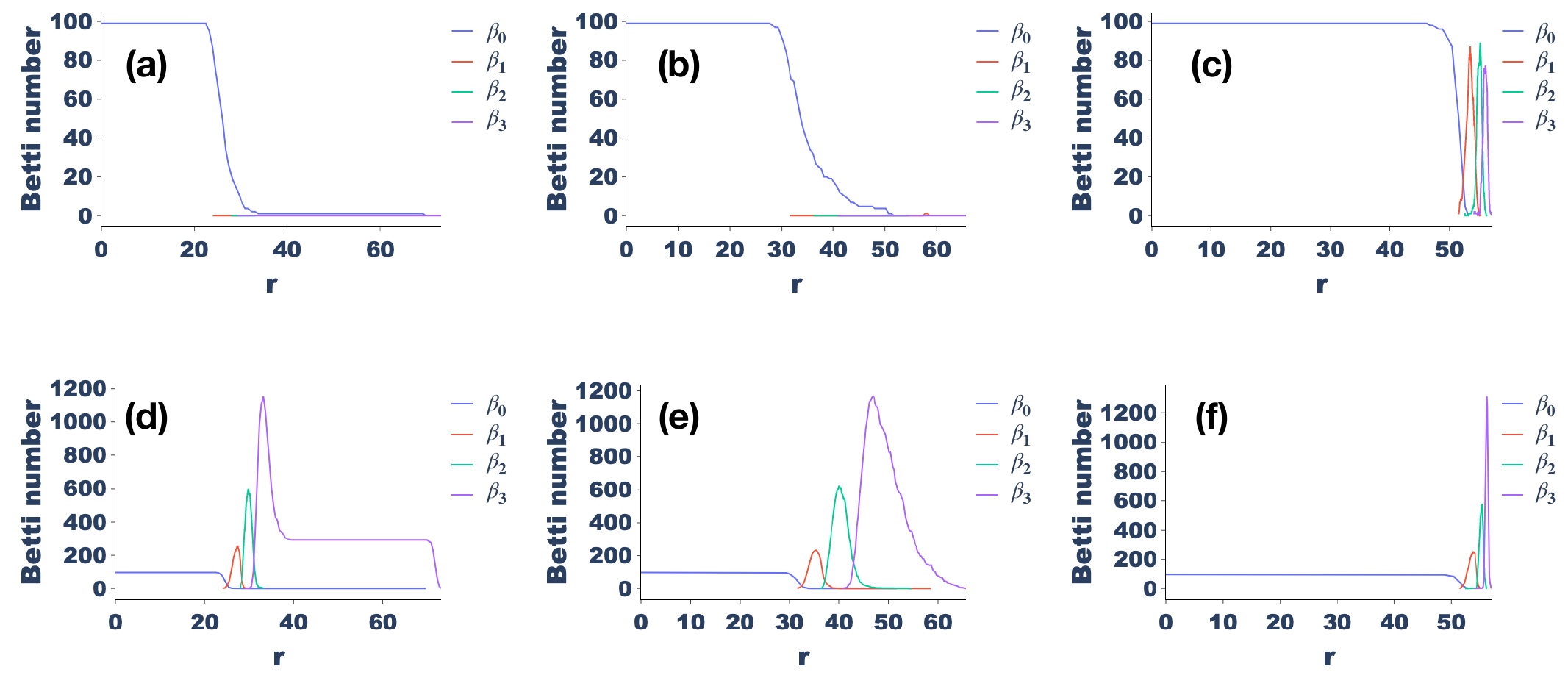}
  \caption{The Betti numbers  $\beta_0$, $\beta_1$, $\beta_2$, and $\beta_3$ of the IsingNet clique complexes are plotted as a function of the filtration parameter $r$. Panels (a), (b), and (c) show the persistent diagrams of IsingNets {obtained from the spin system of linear size $L=40$} at $T=2.12$ (a), $T=2.25$ (b), and $T=2.50$ (c). Panels (d), (e), and (f) show the persistent diagram of corresponding randomized null models obtained at $T=2.12$ (d), $T=2.25$ (e), and $T=2.50$ (f). The networks are formed by $N=100$ nodes.
  }
  \label{fig:Betti}
 \end{figure*}

\begin{figure}[!htb]
  \includegraphics[width=\columnwidth]{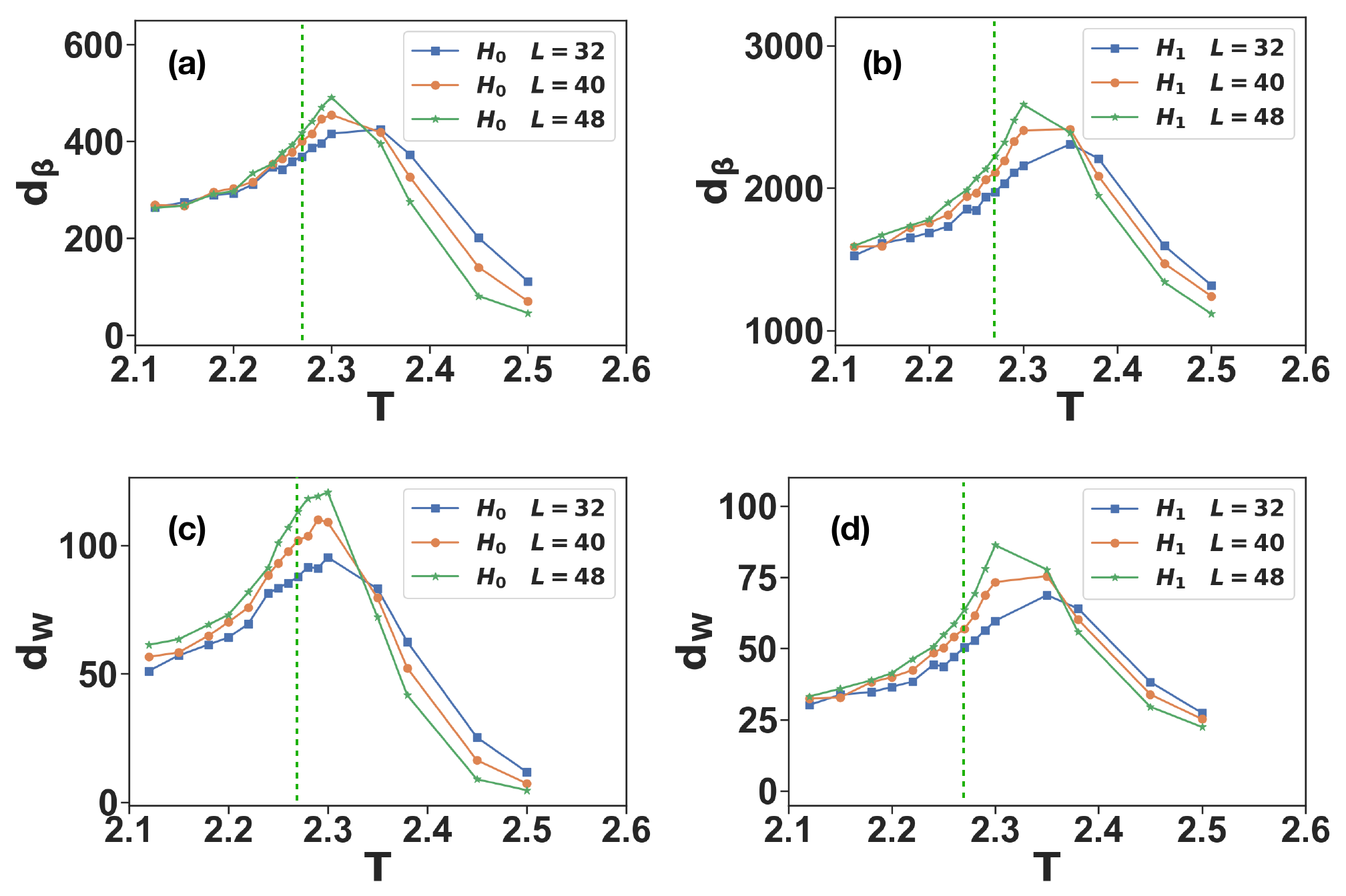}
  \caption{The Betti distance $d_\beta$ (panel (a) and (b)) and the Wasserstein distance $d_W$ (panel (c) and (d)) between the persistent diagrams of the IsingNets and its randomized null models are plotted as a function of the temperature $T$ for homology $H_0$ and $H_1$. The plot shows the finite size scaling of the distances on systems of size $L=32$, $L=40$, and $L=48$. The dashed line indicates the critical temperature $T_c$. The IsingNets on which the persistent diagrams have been calculated have $N=300$ nodes.}
  \label{fig:Persistent_distance}
 \end{figure}

Topological data analysis \cite{edelsbrunner2022computational,otter2017roadmap,vaccarino2022persistent,ghrist2008barcodes,petri2013topological,bobrowski2023universal} and in particular persistent homology allows to characterize the topological properties of data as a function of a filtration parameter and is becoming increasingly important in network and data science with applications ranging from the study of gene-expression to the investigation of brain networks. As mentioned in the introduction TDA is recently becoming a very popular computational tool to study phase transitions as well~\cite{10.21468/SciPostPhys.11.3.060,PhysRevB.106.085111, PhysRevD.107.034501, PhysRevD.107.034506,PhysRevE.93.052138, PhysRevResearch.2.043308, PhysRevB.104.104426, PhysRevE.103.052127, PhysRevB.104.235146, PhysRevE.105.024121, 10.1140/epjb/s10051-022-00453-3}. However, in this context, most of the TDA so far are performed on point clouds rather than on networks.  In our setting, when each pair of nodes $(i,j)$ is assigned a distance $d_{ij}$, persistent homology characterizes the topology of data by forming a simplicial complex representation of the data and characterizing its homology, i.e. the connected components ($H_0$ homology classes); the independent cycles -one-dimensional holes- ($H_1$ homology classes); the two-dimensional holes ($H_2$ homology classes) etc. 
The simplicial complexes  \cite{bianconi2021higher,petri2013topological,otter2017roadmap,vaccarino2022persistent} that we use to perform the topological data analysis are the so-called Vietoris-Rips complexes of the network generated by filling all the simplices having all links at distance $d_{ij}<r$. Thus as a function of $r$ the set of simplicial complexes forms a filtration.

As the filtration parameter  $r$ is raised,  first each node belongs to a different connected component, then connected components merge progressively as described also by the previous paragraph (see Figure $\ref{fig:barcode}$ for a schematic description of the filtration). However, as $r$ increases there is the potential also for one-dimensional holes (or network cycles) to emerge with each independent cycle represented by a different $H_1$ homological class. Eventually, as $r$ increases these cycles will become filled and thus disappear. Moreover, also higher-dimensional holes represented by higher dimensional homological classes $H_{\bar{n}}$ might arise and eventually coalesce as $r$ is further increased giving rise to barcodes representing the topology of the data.

By monitoring the homology classes as a function of $r$, the results of persistent homology are typically summarized by barcodes where each homology class is represented as a bar that extends through the corresponding range of values of $r$ for which the homology class is observed (see Figure $\ref{fig:barcode}$). The barcodes are then represented by persistent diagrams where for each homology class the value of $r$ where the homology class is disappearing  (death) is plotted versus the value of $r$ corresponding to the first appearance of the homology class (birth), see for instance Figure $\ref{fig:homology}$. Significant homology classes are the ones represented by points far from the diagonal which last for a large interval of values of the filtration parameter $r$.

The investigation of the persistent diagrams for the IsingNets can further characterize these discrete structures by revealing important properties about how the clusters merge as a function of $r$. In particular, the persistent diagrams of the IsingNets allow us to show that clusters remain compact with a suppression of the Betti numbers with respect to the corresponding persistent diagram of the randomized null models.
This result is evident from  Figure $\ref{fig:homology}$ and Figure $\ref{fig:Betti}$ where we plot the persistent diagram (for homology $H_{\bar{n}}$ with $\bar{n}\in \{0,1,2,3\}$  and the Betti numbers $\beta_{\bar{n}}$ with $\bar{n}\in \{0,1,2,3\}$ as a function of the filtration parameter $r$ for IsingNets and their randomized null models as a function of the temperature $T$.
Indeed the persistent diagrams of IsingNets corresponding to homology $H_{\bar{n}}$ with $\bar{n}\in \{0,1,2,3\}$ (see Figure $\ref{fig:homology}$) reveal that homology classes for the IsingNets are less persistent than the homology classes of the null model as they are represented in the diagram by points closer to the diagonal than in the null model. Moreover also the Betti numbers $\beta_{\bar{n}}$ with $\bar{n}\in \{0,1,2,3\}$ are strongly suppressed with respect to their null model counterpart.
\begin{figure*}[!htb]
  \includegraphics[width=2\columnwidth]{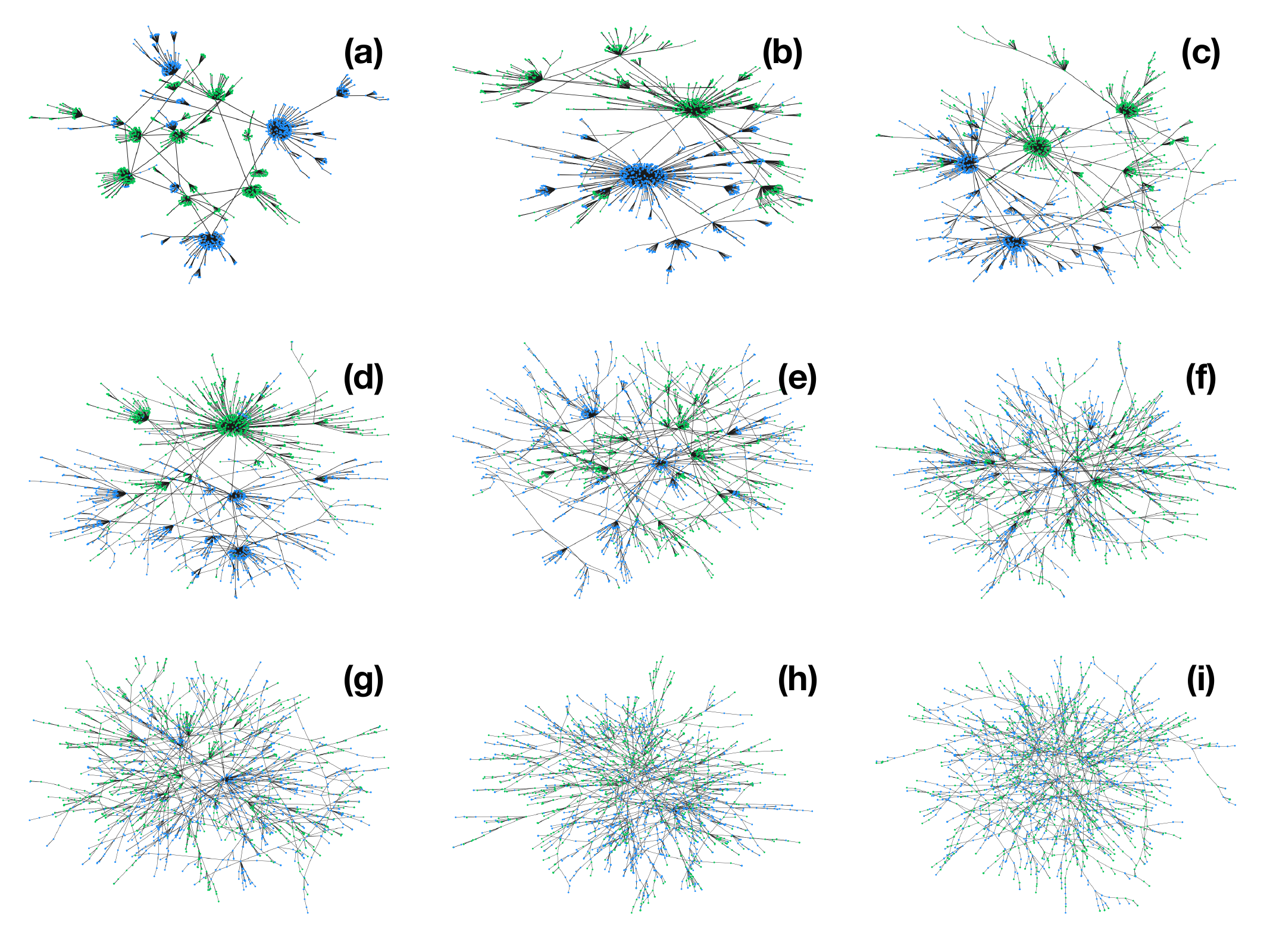}
  \caption{Minimum spanning trees (MST) of the IsingNets {obtained from the spin system of linear size $L=40$ are} plotted for different temperatures $T$ above and below the phase transition at $T_c=2.269\ldots$. The nodes are colored using a $K$-means clustering algorithm with $K=2$. The number of nodes of the MSTs is $N=2000$. The MSTs are generated at {$T=2.12$ (a), $T=2.20$ (b)}, $T=2.26$ (c), $T=2.28$ (d), $T=2.30$ (e), $T=2.35$ (f), $T=2.38$ (g), $T=2.50$ (h), $T=3.50$ (i).}
  \label{fig:mst}
 \end{figure*}
The persistent diagrams of IsingNets can be compared to the persistent diagrams of their randomized null model as a function of the temperature. This comparison can be performed by considering different measures of distances between persistent diagrams.

Here we show both the Weisserstein distance and the Betti distance among the persistent diagrams of the IsingNets and their corresponding randomized null model. The Weissestein distance captures the minimum distance over all perfect matchings between points in two persistent diagrams, i.e.,
\bea
d_W = \left[\inf_{\eta:I \to N}\sum_{x \in I}||x-\eta(x)||_{\infty}^2\right]^{1/2}
\eea
where $x\in I$  indicates a point $x=(b,d)$ with birth $b$ and dead $d$ in the persistent diagram of the IsingNet, while $\eta(x)\in N$ indicates the matched point in the persistent diagram of the corresponding null model. The map $\eta$ denotes any bijection between $I$ and $N$.
The Betti distance computes the $L_2$ distance between Betti curves of two persistent diagrams.
\bea
d_\beta = \left[\sum_r \left(\beta^{[I]}(r)-\beta^{[N]}(r)\right)^2\right]^{1/2}
\eea
where $\beta^{[I]}(r)$ and $\beta^{[N]}(r)$ indicate the Betti number of the IsingNet and the corresponding null model with filtration parameter $r$. 

These distance measures provide good indicators of the critical points as they display a maximum approaching the critical temperature of the $2$D Ising model as the size $L$ of the $2$D lattice increases (see Figure $\ref{fig:Persistent_distance}$). Note that here,  due to the computational cost of calculating the persistent diagrams corresponding to higher-order homological classes, we focus here only on homology classes $H_0$ and $H_1$. 


This is rather clear evidence that the IsingNets have a topology that encodes fundamental properties of the underlying spin system.

\subsection{Minimum Spanning Tree Visualization}
Our unsupervised analysis of the IsingNets includes their visualization which reveals their highly heterogeneous structure below the critical temperature $T_c$ and at criticality.
A very efficient way to visualize the IsingNets is by plotting their Minimum Spanning Trees (MST) \cite{graham1985history,bonanno2003topology}. The MST is the subtree of the network whose sum of the distances between the connected nodes is minimal, and its topology reveals important properties of the fully connected weighted distance matrix of the IsingNets. In particular, as shown in Figure $\ref{fig:mst}$ for $T$ deep in the ferromagnetic phase the topology of the MST of the IsingNets is dominated by very relevant hub nodes, and the network displays a clear partition between the two clusters detected by the $K$-means algorithm with $K=2$ indicated in the figure by two different colors of the nodes.  As the temperature is raised to the critical region, $T\simeq T_c$  the hubs of MST become less dominant. Above the phase transition, the MST becomes clearly more random with a suppression of the hubs in the MST.

\subsection{Network embedding}
\begin{figure}[!htb]
  \includegraphics[width=\columnwidth]{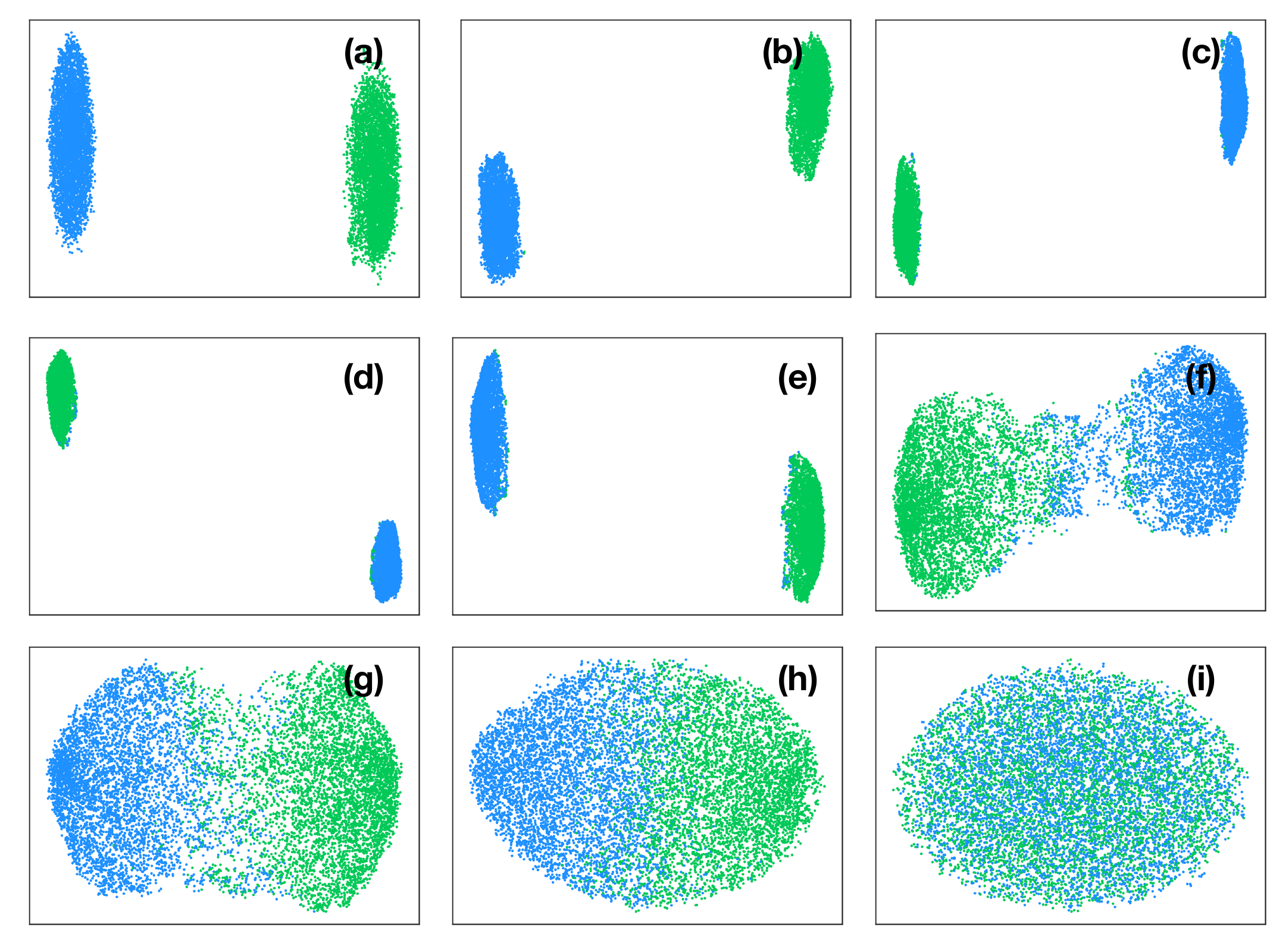}
  \caption{IsingNets {obtained from the spin system of linear size $L=40$} at different temperatures $T$ are embedded into a two-dimensional space using the Uniform Manifold Approximation and Projection (UMAP) embedding algorithm. The nodes are colored using a $K$-means clustering algorithm. The number of nodes of the IsingNets is $N=10^4$. The IsingNets are generated at $T=2.12$ (a), $T=2.20$ (b), $T=2.26$ (c), $T=2.28$ (d), $T=2.30$ (e), $T=2.35$ (f), $T=2.38$ (g), $T=2.50$ (h), $T=3.50$ (i).}
  \label{fig:embedding}
 \end{figure}

Our analysis of the IsingNets is here enriched by considering the UMAP $2$-dimensional embedding of the fully connected network endowed with the distance matrix ${\bf d}$. UMAP is a widely used embedding algorithm exploiting dimension reduction (for more detail see for instance Ref.\cite{baptista2023zoo}). The embedding is here performed as a function of the temperature $T$ and the nodes are colored according to their clustering in two groups performed using $K$-means (see Figure \ref{fig:embedding}).

The UMAP embedding provides a clear visualization of the two main clusters of the nodes of the IsingNets present for $T<T_c$ and corresponding to the symmetry among configuration snapshots and of their merging as the temperature $T$ is raised above the critical temperature.  

 \subsection{Metric-based centrality measures}
 To conclude our characterization of the IsingNets without imposing a fixed value of the filtration parameter, we present here the statistical properties of some important geometrical aspects of the IsingNet captured by the closeness centrality of the nodes.
 
\begin{figure}[!htb]
    \includegraphics[width=\columnwidth]{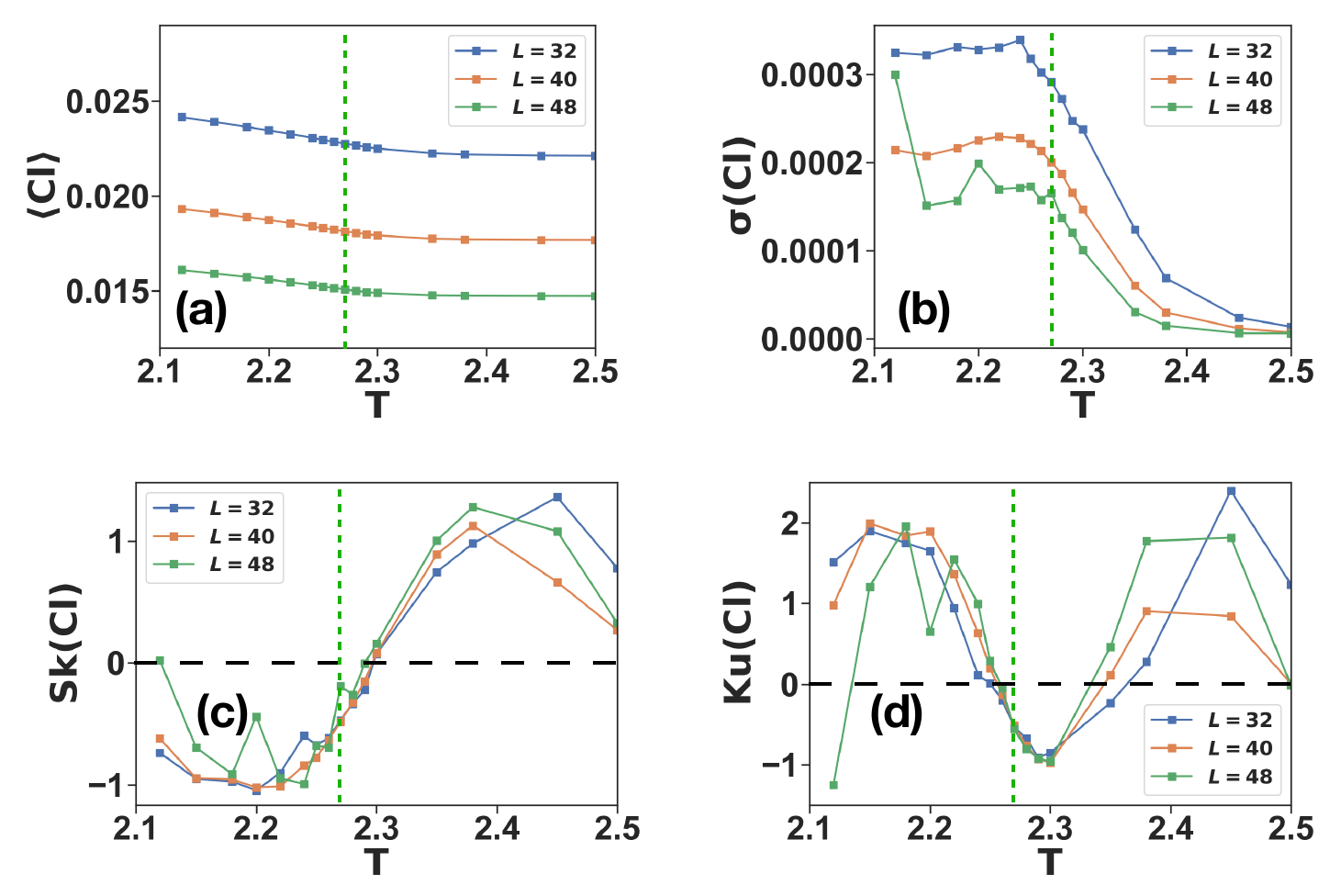}
  \caption{The mean $\Avg{Cl}$, standard deviation $\sigma(Cl)$, skewness $Sk(Cl)$, and kurtosis $Ku(Cl)$ of closeness distribution at different temperatures $T$ and sizes $L$ are plotted on the IsingNets where all the distance between each pair of nodes is retained. The closeness is calculated via Eq. \ref{eq:clo_eff}. The corresponding random networks are formed by randomly permuting the distances between node pairs. The average closeness centrality of the IsingNets coincides by construction with the average of the null model (not shown) however the higher-order statistics strongly depart from the null model behavior.  The {vertical green} dashed lines indicate the critical temperature $T_c$ the horizontal black dashed lines indicate $Sk(Cl)=0$ in panel (c) and $Ku(Cl)=0$ in panel (d).} 
  \label{fig:closeness}
 \end{figure}

 The closeness centrality $Cl_i$  of a node $i$ measures how close is the node to the other nodes of the network, and is defined as the inverse of the average distance  of node $i$ to the other nodes of the network, i.e.
 \bea
 CL_i = \frac{N-1}{\sum_{j \neq i} d_{ij}}
 \label{eq:clo_eff}
 \eea
 We investigate the statistical properties (the mean $\Avg{Cl}$, the standard deviation $\sigma(Cl)$, the skewness $Sk(Cl)$ and the kurtosis $Ku(Cl)$) of the distribution of the closeness centrality in IsingNets as a function of the temperature $T$.
 In Figure $\ref{fig:closeness}$ we show that the average closeness centrality decreases as a function of the temperature, demonstrating that on average the distances between the nodes are higher at higher temperatures.
 The higher-order statistics of the closeness centrality distribution are even more revealing of the IsingNets organization and the skewness and kurtosis of the closeness distribution provide a good unsupervised indicator of the critical point (see Figure $\ref{fig:closeness}$).
 Indeed the standard deviation of the closeness centrality displays a maximum for temperatures close to the critical temperature; the skewness of the closeness centrality is negative below the critical temperature and positive above the critical temperature and the kurtosis becomes negative above the critical temperature, while at very high temperature is strongly affected by noise.

\section{Network characterization of IsingNets at a given value of the threshold distance $r$}
\subsection{IsingNets at given  threshold distance $r$}
In this section, we study the properties of IsingNets where we fix a given choice of filtration parameter $r$.
In particular we will consider the IsingNets, whose $N\times N$ adjacency matrix ${\bf A}$ has elements
\bea
A_{ij}=\theta(r-d_{ij}),
\eea
with $\theta(x)=1$ for $x>0$ and $\theta(x)=0$ otherwise. In the following, we will indicate with $i\sim j$ two neighbour nodes for which $A_{ij}=1$.

We adopt the same choice of the parameter $r$ proposed in Ref. \cite{mendes2023wave} where it was shown that for a wide range of choices of $r$ the IsingNets are scale-free presenting often power-law exponents less than two which are known to occur in a variety of context \cite{seyed2006scale,courtney2018dense,caron2017sparse,timar2016scale}.
In particular, here we take $r$ equal to the average distance of the $5^{\mbox{th}}$ nearest neighbour.

Similarly, the randomized network forming our null model in this section is performed by thresholding the randomized distance matrix ${\bf d}^{rand}$ with the same threshold used for the corresponding IsingNet.

We consider the statistical and combinatorial properties of these networks where we assign to each link $(i,j)$  a weight $w_{ij}$ equal to the inverse of the distance $d_{ij}$, provided this distance is smaller than $r$, i.e. 
\bea
w_{ij}=\frac{1}{d_{ij}}A_{ij}\, 
\eea

We provide an in-depth network analysis of these networks investigating their degree and strength distribution, and going beyond these statistical properties providing evidence of the presence of degree correlations, of a nontrivial $k$-core structure and of weight degree correlations. Moreover, the investigation of the spectral properties of the IsingNets will demonstrate non-trivial signatures of criticality.
This analysis provides clear evidence that not only the degree distribution of IsingNets is strongly different from the one of an Erd\"os-Renyi random graph, but the network also obeys important higher-order correlations reflecting the correlations existing in the spin system configuration snapshots.

\subsection{Degree and strength distribution}
One of the most simple yet fundamental property of a network is its degree distribution $P(k)$ which characterize globally the network starting from the knowledge of the node's degrees where the degree $k_i$ of the node $i$ indicates the sum of the links incident to it, i.e.
\bea
k_i=\sum_{j=1}^NA_{ij}.
\eea
Thus while the degree is a local property of the nodes the degree distribution $P(k)$ is able to characterize the global properties of the networks.
In particular scale-free \cite{barabasi1999emergence,albert2002statistical,voitalov2019scale} and in general degree distribution with second $\Avg{k^2}$ (and eventually first $\Avg{k}$) moment diverging with the network size have been shown to have a significant role in determining the response of the network to random damage and the critical behavior of the dynamics defined on top of these networks, such as epidemic spreading and the Ising model \cite{albert2002statistical,barabasi2016network}.

 For weighted networks it is also possible to define weighted degree also called {\em strength} $s_i$ of the generic node $i$ \cite{barthelemy2005characterization} given by the sum of the weights of its incident links, 
 i.e.
 \bea
 s_i=\sum_{j=1}^Nw_{ij}.
 \eea
 From the sequence of the node's strengths, it is possible to extract also the strength distribution $P(s/s_0)$  where $s_0$ is the minimal strength of the links. This distribution has been shown to display broad distributions in a number of real weighted networks, such as collaboration networks and airport networks \cite{barrat2004architecture}.

\begin{figure}[!htb]
  \includegraphics[width=\columnwidth]{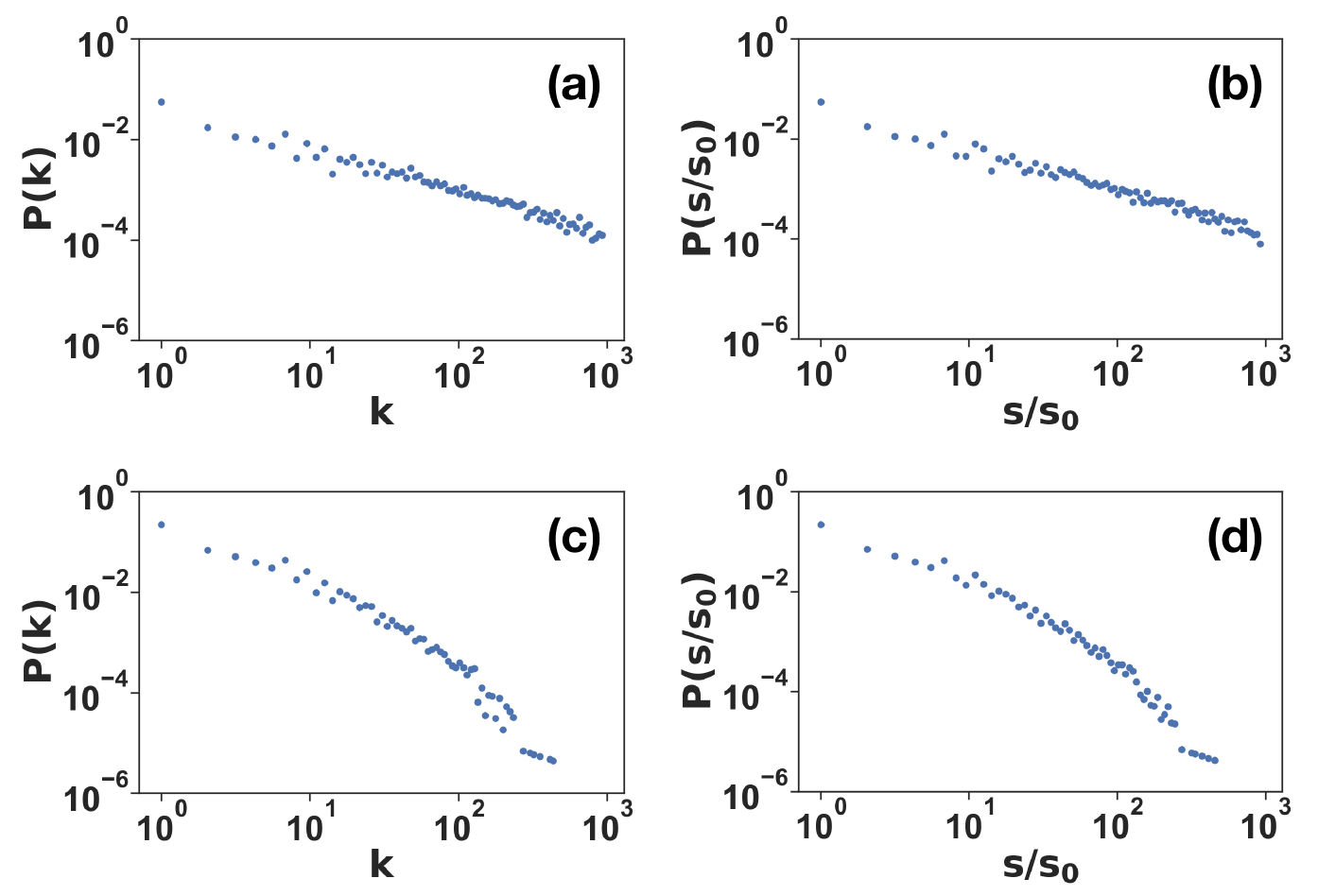}
  \caption{Degree distribution $P(k)$ and strength distribution $P(s/s_0)$ of IsingNets {obtained from the spin system of linear size $L=40$} formed by $N=10^4$ nodes at temperature  $T=2.12$ and $T=2.50$. Panel (a) shows the degree distribution of IsingNet at $T=2.12$. Panel (b) shows the strength distribution of the same network as panel (a). Panel (c) is the same as panel (a) but obtained at $T=2.50$. Panel (d) is the same as panel (b) but obtained at $T=2.50$.
  }
  \label{fig:degree_strength}
 \end{figure}

In Figure \ref{fig:degree_strength} we plot the degree $P(k)$ and strength $P(s/s_0)$ distribution for temperatures below and above the phase transition demonstrating that these distributions are broad.  We note that below the critical temperature $T_c$, the networks are not only broad but also dense, i.e. having an average degree growing with the network size (for models of these networks see \cite{seyed2006scale,courtney2018dense,caron2017sparse,timar2016scale}). 

\begin{figure}[!htb]
    \includegraphics[width=\columnwidth]{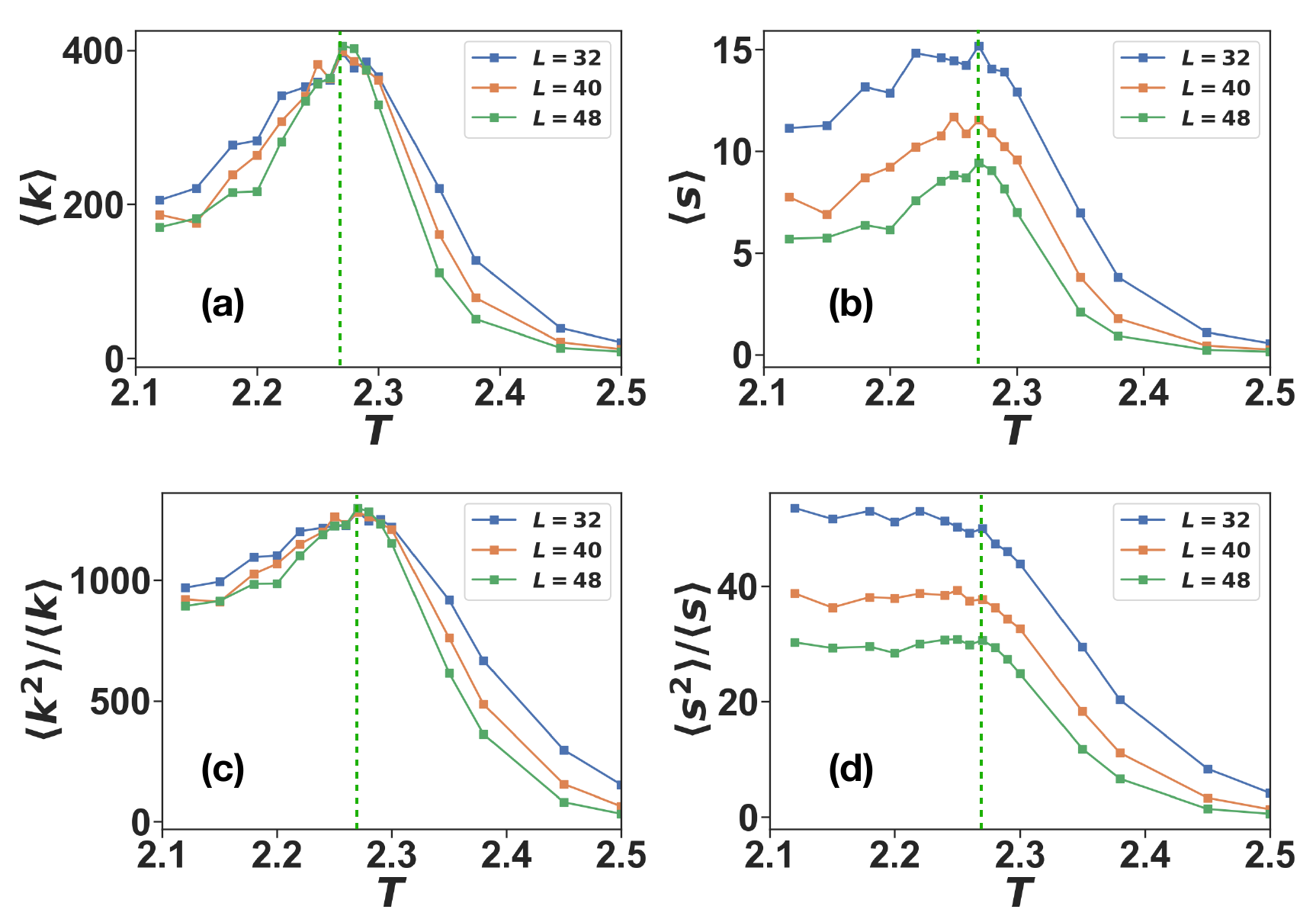}
  \caption{The average degree $\avg{k}$ (panel (a)) and the average strength $\avg{s}$ (panel (b))  are plotted together with the ratio $\avg{k^2}/\avg{k}$ (panel (c)) and $\avg{s^2}/\avg{s}$ (panel (d)) as a function of temperature $T$. The dashed line indicates the critical temperature $T_c$. The IsingNets are formed by $N=10^4$ nodes.}
  \label{fig:first_second_moment}
 \end{figure}

To quantify the scale-free nature of these distributions we plot the ratio between the second moment, the average degree $\avg{k}$ and the average strength $\avg{s}$ as a function of the temperature, showing that that $\avg{k}$, and $\avg{k^2}/\avg{k}$  are good indicators of the critical point displaying a maximum for $T=T_c$ without noticeable finite size effects for the sizes investigated in this work (see Figure \ref{fig:first_second_moment}).{\color{black} We note that origin of such power laws is not necessarily related to the presence of a critical point, as the data sets in Figure  \ref{fig:degree_strength} are representative of regimes where the correlation length is not much larger than $L$ (and, for $T=2.5$, is smaller). We attribute these features of $P(k)$ to the strong correlations present in the system (so: not critical behavior, but rather, just correlation length much larger than the lattice spacing).}

\subsection{Degree correlations}
\begin{figure}[!htb]
  \includegraphics[width=\columnwidth]{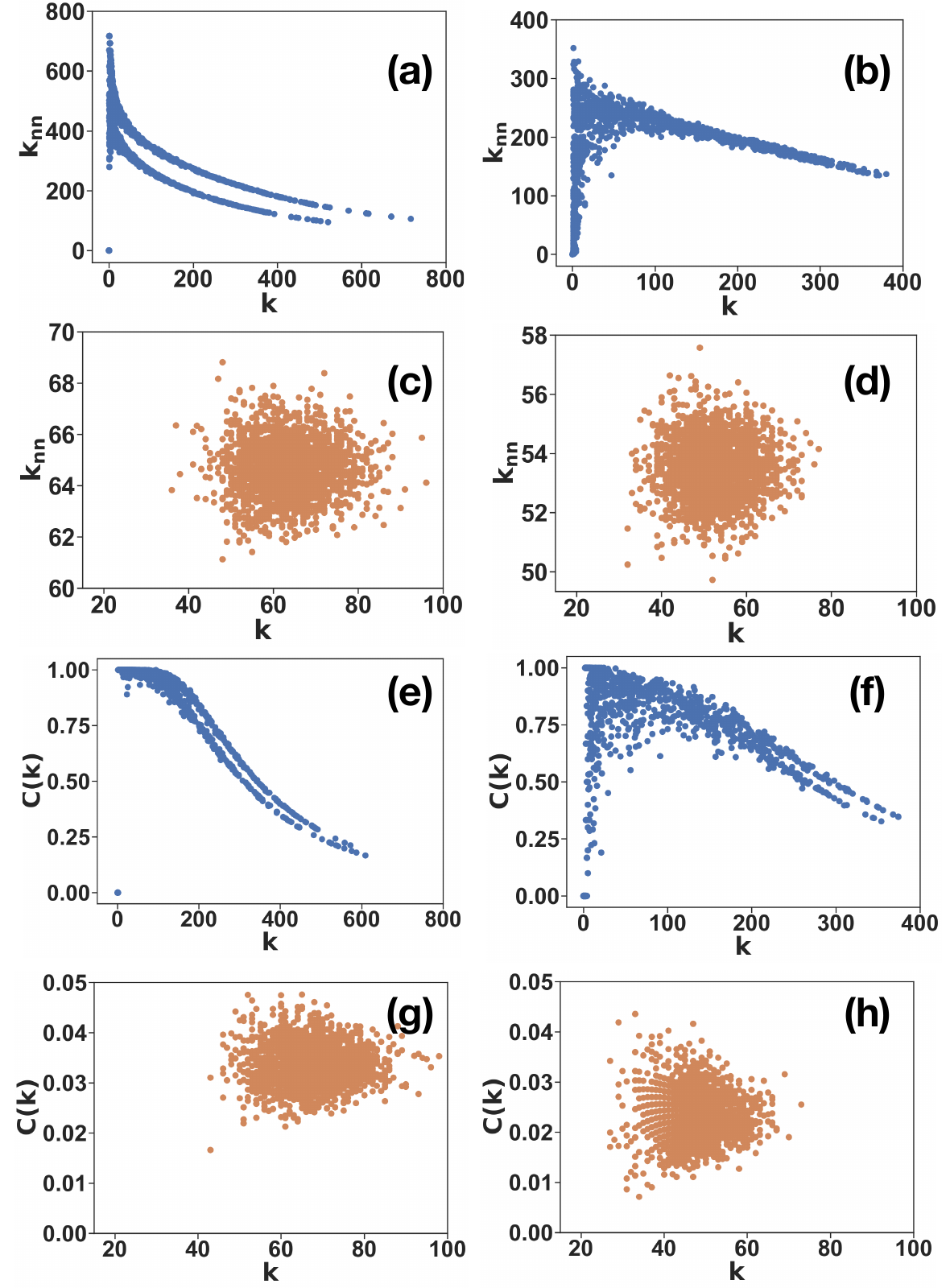}
  \caption{{Average nearest neighbour degree $k_{nn}$  and clustering coefficient $C(k)$ on IsingNets obtained from Monte Carlo simulations of the spin system of linear size $L=40$ and corresponding null models with $N=2000$ nodes. The threshold distance of connecting two nodes is the 5th nearest neighbour average distance. The random network is obtained by randomly permuting distances between node pairs and nodes are connected with the same threshold. The average neighbour degree $knn$ is shown as a function of degree $k$ on IsingNets and corresponding null models. Panel (a) and (b) show $knn$ obtained at $T=2.12$ (a) and $T=2.35$ (b). Panel (c) and (d) show $knn$ of the corresponding null model at $T=2.12$ (c) and $T=2.35$ (d). Panel (e) and (f) show $C(k)$ obtained at $T=2.12$ (e) and $T=2.35$ (f). Panel (g) and (h) show $C(k)$ of the corresponding null model at $T=2.12$ (g) and $T=2.35$ (h)}. }
  \label{fig:knn}
 \end{figure}

In order to go beyond single node statistics and to explore how far IsingNets are from random networks with a given degree distribution, in this section we characterize the degree correlations \cite{newman2018networks,barrat2008dynamical} of the IsingNets. Degree-degree correlations measure to what extent the degree of two end nodes of the same link are correlated. Degree correlations are typically classified as either assortative or disassortative. Assortative degree correlations imply that highly connected nodes are more likely to be connected to highly connected nodes while low degree nodes are more likely to be connected to low degree nodes than in a maximally random network with the same degree distribution. Conversely, disassortative networks are networks in which highly connected nodes are more likely to be connected to low degree nodes than in the maximally random network with the same degree distribution. Examples of assortative networks are social networks while examples of disassortative networks include the Internet and the protein interaction networks.
The degree-degree correlations can be quantified by considering the average degree of the neighbour of a node $k_{nn}(i)$ defined as 
\bea
k_{nn}(i)=\frac{1}{k_i}\sum_{j\sim i}k_j
\eea
When  $k_{nn}(i)$ tends to be higher for nodes of higher degree $k_i$ the network is classified as assortative. Instead when $k_{nn}(i)$ is typically lower for nodes of higher degree $k_i$ then the network is classified as disassortative.
The IsingNets are clearly displaying a disassortative behavior for $T<T_c$ that deviates strongly from the behavior of the null model in which the distance matrix ${\bf d}$ is reshuffled (see Figure \ref{fig:knn}). On the contrary, for $T>T_c$ the trend of $k_{nn}$ versus $k$ does not appear to be fully monotonic, while nodes of larger degrees remain more likely to connect to low degree nodes. 

Interestingly the degree correlations are also affecting the average clustering coefficient $C(k)$ \cite{watts1998collective,barrat2008dynamical,bianconi2006effect} of nodes of degree $k$ which display a decay as a function of $k$ typical of networks with disassortative degree correlations (see Figure \ref{fig:knn}).

\begin{figure}[!htb]
  \includegraphics[width=\columnwidth]{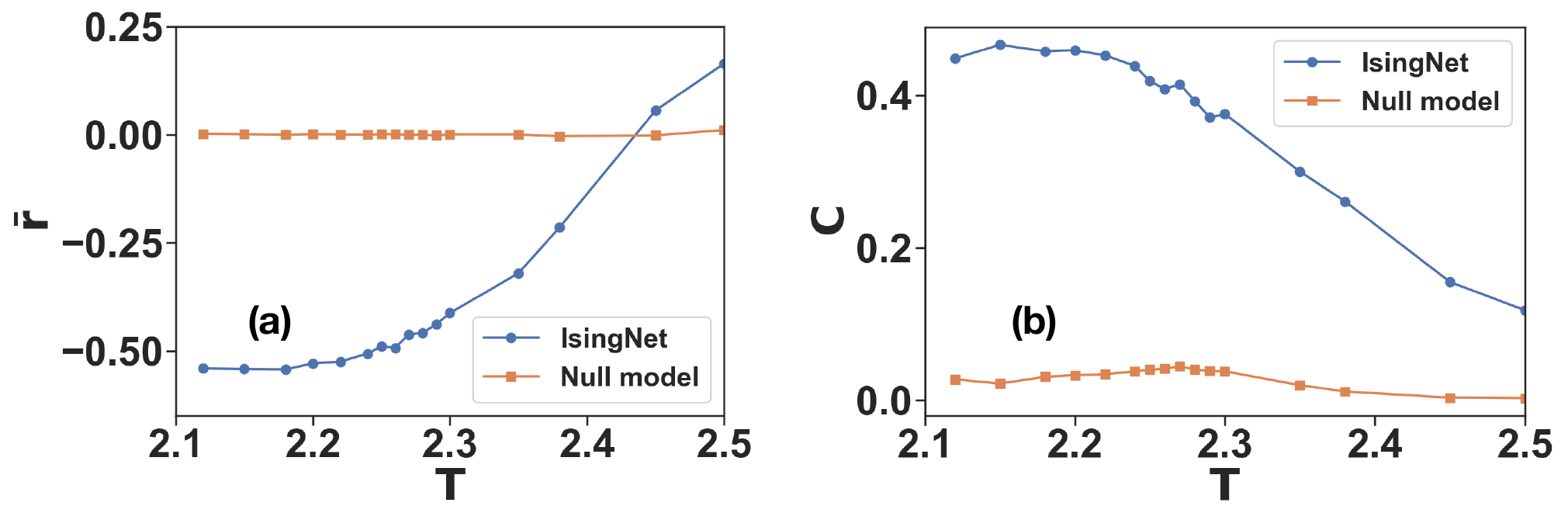}
  \caption{Pearson correlation coefficient $\bar{r}$ (a) and average clustering coefficient $C$ (b) on IsingNets and corresponding null models formed at different temperatures. The Pearson correlation coefficient $r$ is calculated on networks with $10^4$ nodes and the average clustering coefficient $C$ is calculated for IsingNet with $N=2000$ nodes coming from Monte Carlo simulation with linear size {$L=40$}. The threshold distance of connecting two nodes is the 5th nearest neighbour average distance. The random network is obtained by randomly permuting distances between node pairs and nodes are connected with the same threshold.}
  \label{fig:pearson_clustering}
 \end{figure}

Degree-degree correlations  can also be detected by the 
 Pearson correlation coefficient $\bar{r}$ \cite{newman2018networks} between degrees of linked nodes which is plotted as a function of the temperature in Fig. \ref{fig:pearson_clustering}(a) showing a clear negative (disassortative) correlations for low temperatures which strongly deviates from the null model.
 In Fig. \ref{fig:pearson_clustering}(b) we also report the average clustering coefficient $C$  as a function of the temperature showing that IsingNets display a much larger average clustering coefficient than the null model counterpart and that this average clustering coefficient is higher deep in the ferromagnetic phase (lower temperatures).
 
\subsection{$K$-core structure}

\begin{figure}[!htb]
  \includegraphics[width=\columnwidth]{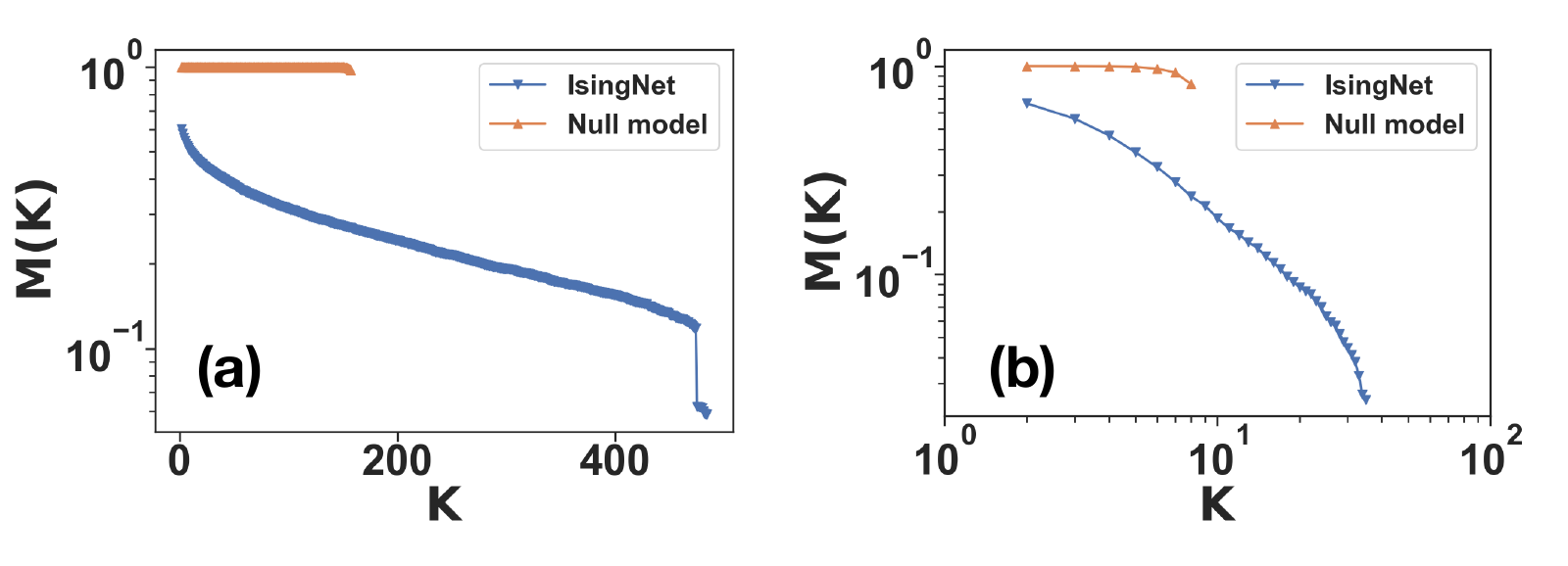}
  \caption{Fraction of nodes in the $K-$core $M(K)$ as a function of the core size $K$ on IsingNet and null model {with $N=10^4$ nodes} at $T=2.12$ (a) and $T=2.50$ (b).{The IsingNets are generated from 2D Ising model Monte Carlo simulations of the spin system of linear size $L=40$.} Panel (a) is shown with a linear-log scale and panel (b) is shown with a log-log scale. }
  \label{fig:kcore}
 \end{figure}

Networks can be decomposed in nested $K$-cores characterizing their core-periphery structure \cite{alvarez2005large,carmi2007model,dorogovtsev2006k}. A  $K$-core is a subgraph of the network formed by a set of $M(K)$  nodes each having at least $K$ connections with the other nodes of the set. Power-law networks with exponent $\gamma\in (2,3]$ display a significant $K$-core structure with the maximum $K$  diverging with the network size and a power-law decay of $M(K)$ as a function of $K$. On the contrary Erd\"os and Renyi networks have a finite number of $K$-cores also when the average degree diverges.
Here we show that IsingNets have a very rich $K$-core structure having statistical properties that change below and above the critical temperature (see Figure $\ref{fig:kcore}$). Indeed above the critical temperature, we observe a behavior similar to the expected behavior for sparse scale-free networks with power-law exponent between two and three presenting a broad (seemingly power-law straight line on a log-log plot) decay of $M(K)$ versus $K$. However, below the critical temperature where the average degree diverges, the $K$-cores include more nodes while the decay of $M(K)$ versus $K$ is better approximated by an exponential (straight line in a log-linear plot) rather than by a power-law.

\subsection{Weight-topology correlations}
\begin{figure}[!htb]
  \includegraphics[width=\columnwidth]{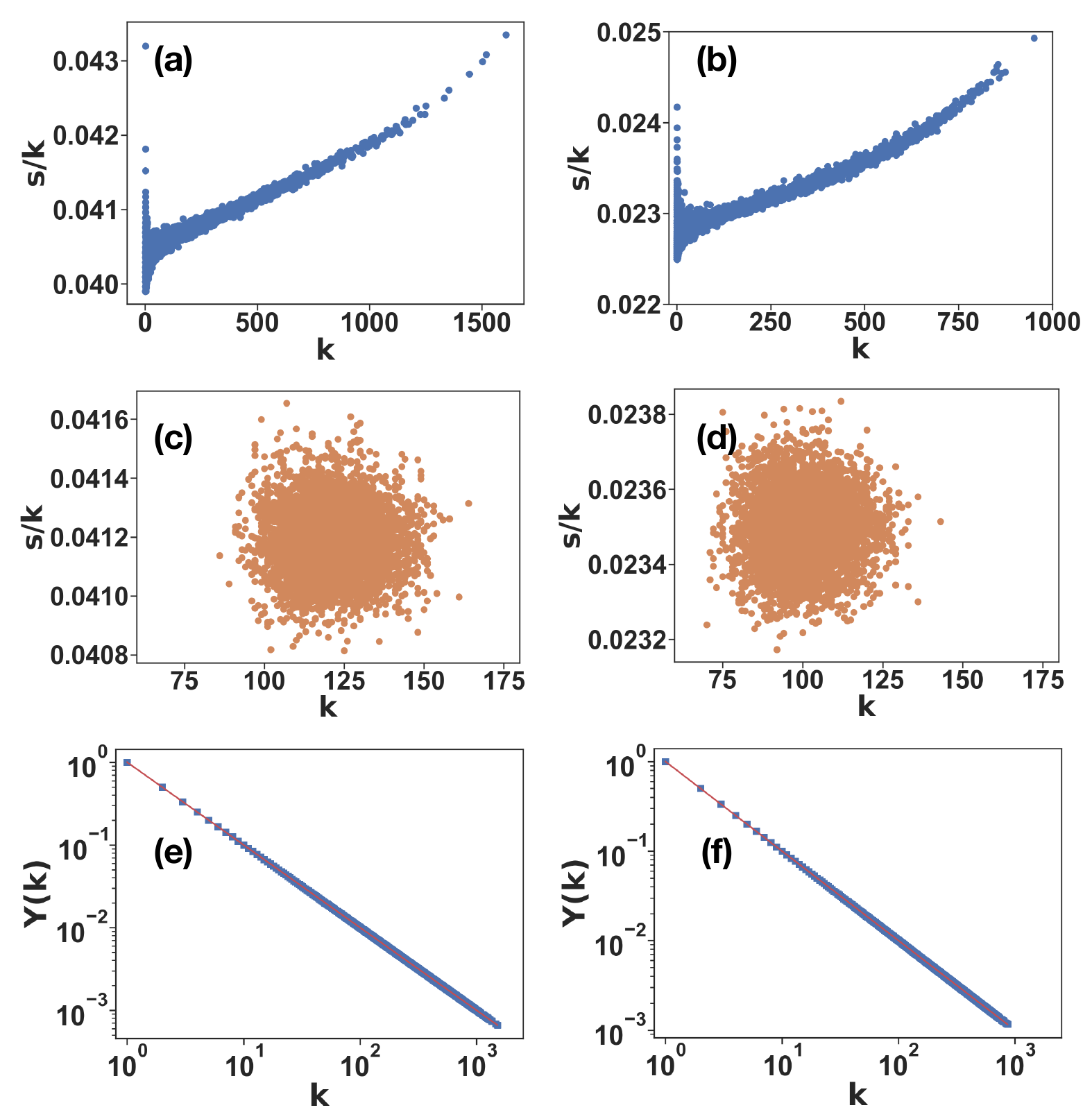}
  \caption{The ratio of strength and degree and the inverse participation ratio {on IsingNets obtained from Monte Carlo simulations of the spin system of linear size $L=40$} are shown versus degree on networks formed at different temperatures. The networks are formed by $N=5000$ samples. The threshold distance of connecting two nodes is the 5th nearest neighbour average distance. The random networks are obtained by randomly permuting distances between node pairs and nodes are connected with the same threshold. The first row shows the strength degree ratio $s/k$ versus degree $k$ on IsingNets and the second row shows which on corresponding random networks. The third row shows the inverse participation ratio $Y(k)$ versus degree $k$. The networks are formed by simulation obtained at temperature $T=2.12$ (left column), $T=2.35$ (right column). }
  \label{fig:strength_degree_Y}
 \end{figure}

Interestingly in weighted networks not only the network topology can reveal relevant degree correlations showing that the networks deviate from maximally random networks, but also the weights can be distributed in a non-random way.
In particular, there are two main network analyses that are able to detect weight-topology correlations.
The first analysis \cite{barthelemy2005characterization} involves studying the normalized strength $s/k$ versus the degree $k$ for each node of the network.  If the weights are distributed randomly and independently on the degree of the two end nodes there should not be any significant dependence of $s/k$ with $k$. Conversely if $s/k$ increases with $k$ it implies that nodes with higher degrees are incident in average to links with larger weights. 
The second analysis \cite{almaas2004global}  investigates the weight-topology correlations aiming at revealing the weight heterogeneity among links that connect to the same node. This heterogeneity if any can be quantified by calculating the inverse participation ratio $Y$ for the weights of the links ending to node $i$, defined as 
\bea
Y_i = \sum_{j\sim i} \left(\frac{w_{ij}}{s_i}\right)^2.
\eea
If the weights $w_{ij}$ of the links $(i,j)$ incident to node $i$ are homogeneous, $Y_i \sim 1/k_i$. If the weights are highly heterogeneous,  $Y_i^{-1}$ indicates the effective number of links with significant weight.
Interestingly when we measure $Y_i$ for the IsingNets we observe that this second type of weight heterogeneity is missing in the data and that $Y_i\sim 1/k_i$ indicating that for each node $i$ the weights of the links incident to it have all weights of comparable order or magnitude (see figure $\ref{fig:strength_degree_Y}$).

\subsection{Spectral properties of Ising networks}

The IsingNets do not only have very interesting combinatorial and statistical properties encoded in their highly correlated structure but display also relevant geometrical properties reflected in their spectrum.  
In particular, the critical IsingNets display non-trivial spectral properties characterized by a power-law scaling close to criticality and a highly degenerate spectral gap.
The spectral properties of networks are usually probed by considering the spectrum of the graph Laplacian describing diffusion processes. The graph Laplacian $\bm\Delta$ is defined as $\bm{\Delta}={\bf D}-{\bf A}$ where ${\bf D}$ is the diagonal matrix whose diagonal elements are the degrees of the nodes (i.e. $D_{ii}=k_i$), and ${\bf A}$ is the adjacency matrix of the network.
The graph Laplacian $\bm\Delta$ is semi-definite positive and the spectrum always includes a zero eigenvalue with degeneracy given by the number of connected components of the network, i.e. given by the Betti number $\beta_0$.
The smallest non-zero eigenvalue of the graph Laplacian of a  network is also called the Fiedler eigenvalue and is typically indicated as $\lambda_2$ (as it is the second smallest eigenvalue in a connected network).
In the literature,  often one distinguishes between network models displaying a finite Fiedler eigenvalue $\lambda_2\to \lambda_2^{\star}>0$ in the limit $N\to \infty$ and network models in which $\lambda_2\to 0$ as $N\to \infty$. In the first case, we say that the networks display a spectral gap whereas in the latter case, we say that the ``spectral gap closes".
Examples of networks with finite spectral dimension are random graphs above the percolation threshold and examples of networks in which the spectral gap closes are finite dimensional lattices. 

\begin{figure}[!htb]
 \includegraphics[width=\columnwidth]{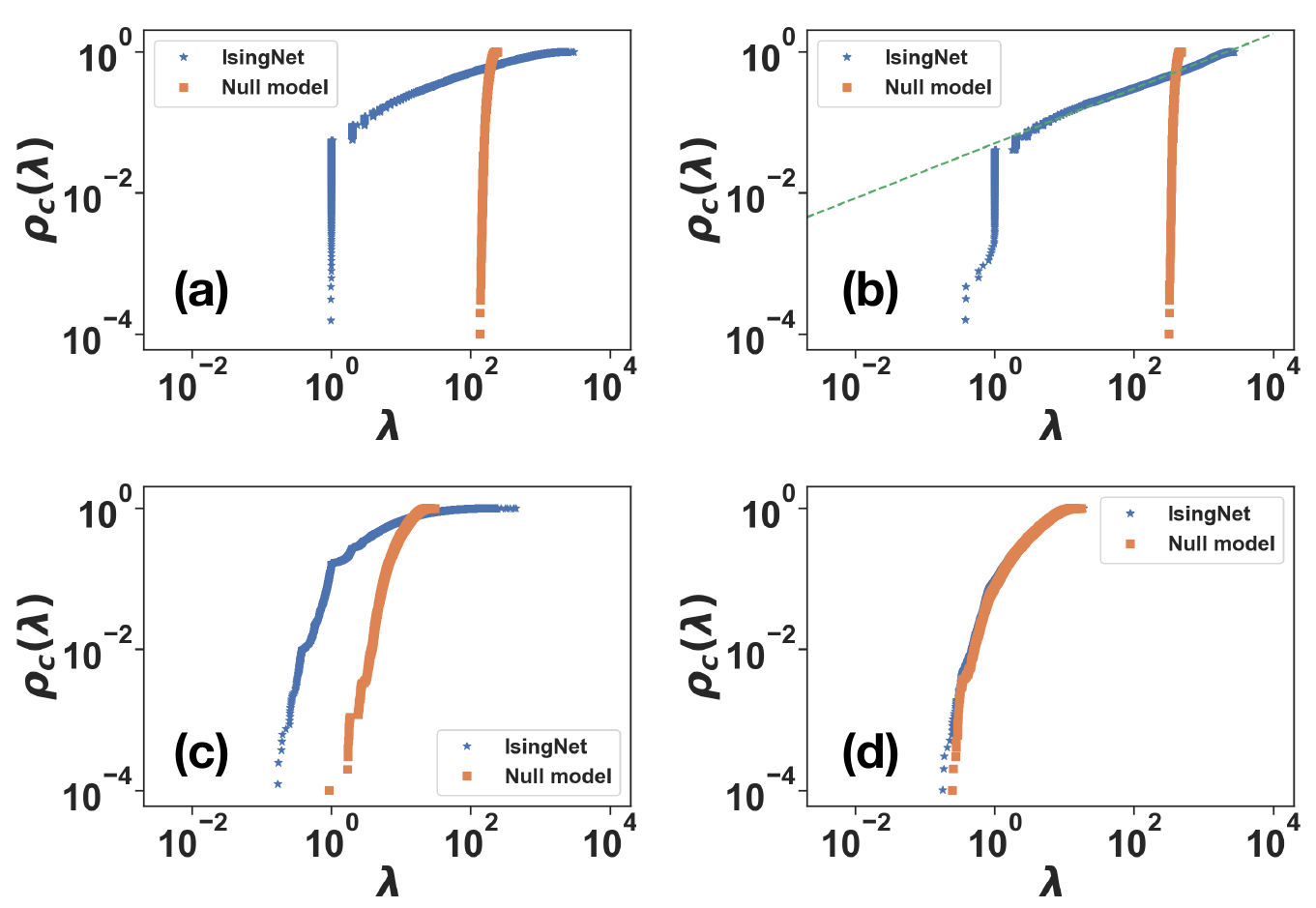}
  \caption{The cumulative distribution $\rho_c(\lambda)$ of the eigenvalues $\lambda$ of graph Laplacian $\bm\Delta$ of IsingNets and corresponding null models at different temperatures $T$. The distribution is shown at $T=2.12$ (a), $T=2.27$ (b), $T=2.50$ (c), and $T=3.50$ (d). Data are shown for Isingnets generated from Monte Carlo simulations of {spin system of linear size $L=40$}. In panel (b), the dashed lines indicate a power-law growth shown in Eq. \ref{eq:rhoc_critical} with exponent $\hat{d}=0.78\pm 0.04$.}
  \label{fig:spectrum}
 \end{figure}

In several networks in which the  spectral  gap closes, it is possible to observe the spectral dimension $d_S$
\cite{bianconi2021higher,burioni2005random}. The spectral dimension $d_S$ is the dimension perceived by diffusion processes on the networks encoded in the graph Laplacian.  On a lattice, the spectral dimension coincides with the Euclidean dimension $d$ of the lattice while on general network topology, the spectral dimension can be distinct from the Hausdorff dimension of the network. Interestingly also small world networks with infinite Hausdorff dimension can have a finite spectral dimension $d_S\geq 2$ \cite{bianconi2021higher,correia1998spectral}.
The spectral dimension determines the scaling of the cumulative density of the eigenvalues $\rho_c(\lambda)$ of the graph Laplacian for $\lambda\ll 1$ in networks where the spectral gap closes. In particular we have that networks with a spectral dimension $d_S$ have a cumulative distribution $\rho_c(\lambda)$ that obeys for $\lambda\ll 1$
\bea
\rho_c(\lambda)\simeq C \lambda^{d_S/2},
\label{eq:rhoc}
\eea
where $C$ is a constant.
In Figure $\ref{fig:spectrum}$ we show that the IsingNets display nontrivial spectral properties that have very peculiar characteristics strongly deviating from their corresponding null model.
Particularly noticeable are the spectral properties of IsingNets close to the critical point where one observes the coexistence  of a highly degenerate finite Fiedler eigenvalue $\lambda_2$  with a power-law scaling of the cumulative distribution 
\bea
\rho_c(\lambda)\simeq C \lambda^{\hat{d}/2},
\label{eq:rhoc_critical}
\eea
for $\lambda>\lambda_2$ with a exponent given by $\hat{d}\simeq 0.78\pm 0.04$ for $L=40$.
Above the critical temperature, the degeneracy of the Fiedler eigenvalue is reduced and one observes 
a nontrivial spectrum reminiscent of the scale-dependent spectral dimension discussed in Refs. \cite{ambjorn2005spectral} within the critical region 
that at higher temperatures converges with the spectrum of the null model.
Below the critical dimension, the spectral gap remains highly degenerate while the rest of the spectrum remains broadly distributed.

\begin{figure}[!htb]
\includegraphics[width=\columnwidth]{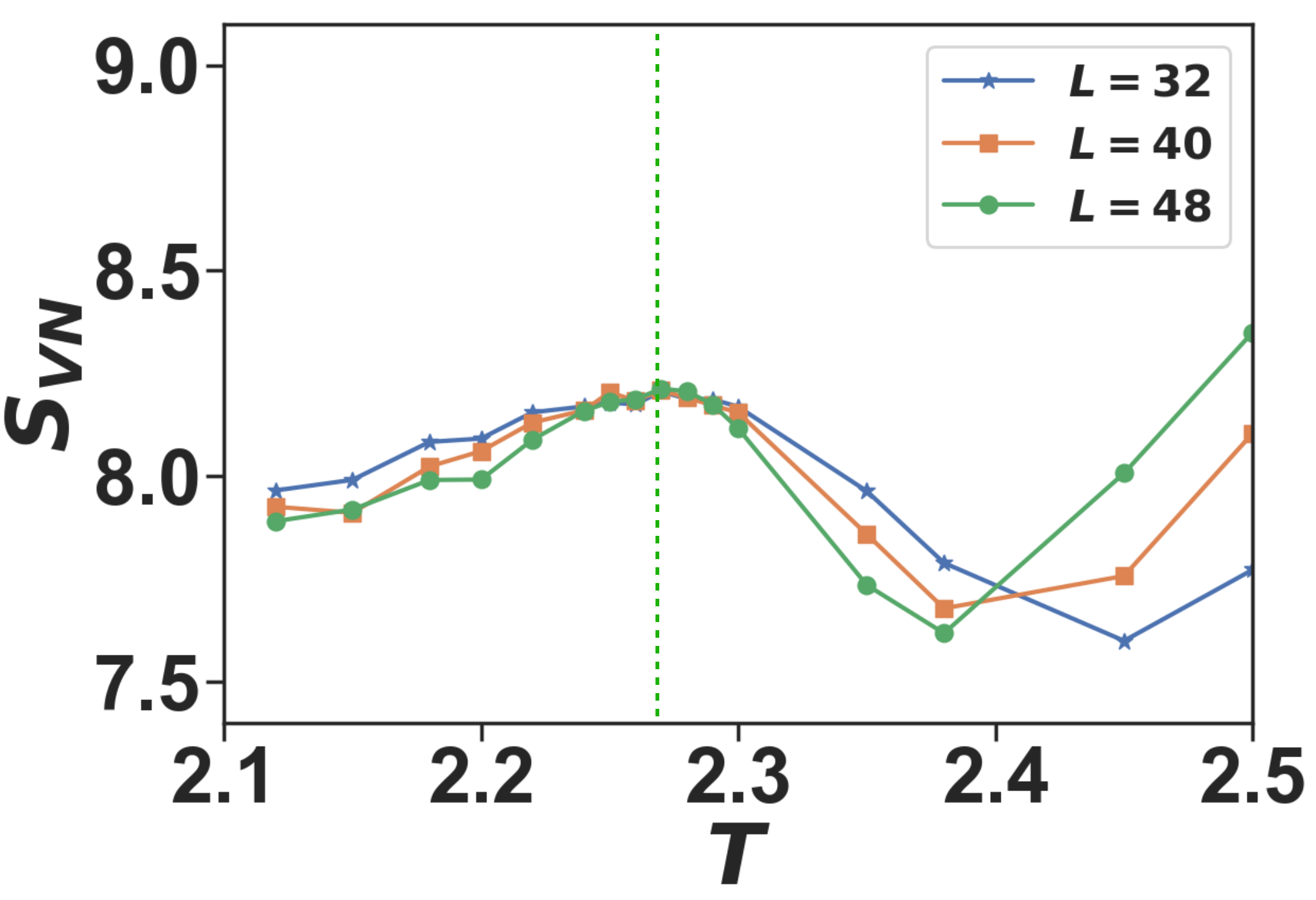}
  \caption{{The von Neumann entropy $S_{VN}$ of different system size $L$ is shown versus the temperature $T$. The IsingNets are formed by $N=10^4$ nodes. The dashed line indicates the critical temperature $T_c$.}}
  \label{fig:vonNeumann}
 \end{figure}

The spectrum of the graph Laplacian is also key to characterizing the network von Neumann entropy $S_{VN}$ \cite{anand2009entropy,anand2011shannon,de2015structural} defined as 
\bea
S_{VN}=-\sum_{\lambda}\frac{\lambda}{\avg{k}N}\ln \left(\frac{\lambda}{\avg{k}N}\right).
\eea
The von Neumann entropy strongly departs from the von Neumann entropy of the null model for low temperatures displaying a local maximum for $T=T_c$ (see  Figure \ref{fig:vonNeumann}).

{An interesting open question that will be addressed in the following works is the relation between  these spectral properties of the IsingNets graph Laplacians and the intrinsic dimension and the entropy measures that have been recently proposed starting from the unsupervised PCA analysis of spin systems \cite{PhysRevX.11.011040,panda23}.}

\section{Conclusions}
In this work, we have launched a systematic network analysis of unsupervised learning of different states of matter. The analyzed IsingNets obtained from Monte Carlo simulations of the $2$D Ising model are shown to reveal the statistical, combinatorial, geometrical and topological organization of these networks.
Through the paper, we have shown that true IsingNets are highly non-random by comparing their structural properties with the structural properties of the randomized counterparts.
Importantly, we have also identified several indicators of the phase transition.

We have addressed the characterization of IsingNets following two different approaches. In the first approach, we have studied the structure of IsingNets as the filtration parameter $r$ is increased, which enforces an effective percolation process in which nodes are subsequently aggregated by considering connected pairs of nodes at increasing distances. This percolation process reveals the presence of two giant components in IsingNets in the ferromagnetic state, each one corresponding to configurations with different magnetization. The same filtration scheme has also been used here to study the topology of the data by constructing the clique complexes of the IsingNets and calculating their persistent diagram. Interestingly, the persistent diagrams reveal that real IsingNets are formed by compact clusters as the Betti numbers of their clique complex are strongly suppressed with respect to the clique complex of the corresponding randomized null models. This network analysis conducted across the filtration is also enriched by effective visualization of the network embedding conducted using MST and the  UMAP embedding and by the statistical characterization of the distribution of closeness centralities. 

Secondly, our investigation of IsingNets has been conducted by considering only links at a distance less than a threshold value $r$ taken to be the average distance of the $5th$ nearest neighbours nodes according to the (fully connected) distance matrix. These networks display a broad degree and strength distribution but their complexity extends well beyond the degree and strength distribution because IsingNets have strong degree-degree correlations and weights-degree correlations, and a rich core structure that changes significantly across the phase transition reflecting the highly nontrivial structure of the spin system that they describe.

This work opens new perspectives for the unsupervised characterization of the study of phases of matter using the tools of network science. This work can be extended in different directions. The analysis performed here for the $2D$ Ising model can be extended to the study of other classical critical phenomena with the goal of characterizing the possible presence or the lack of universalities among the networks constructed from spin system configuration snapshots. Similarly, the same toolbox can be utilized to attack out-of-equilibrium critical behavior.{\color{black} Widening the class of models where our analysis can lead to physical insight is fundamental to establish ground for possible analytical treatments. In particular, before this is done, a key point to be understood is the relation between correlation length and sampling in models in different universality classes. The reason is that a naive connection (number of samples is proportional to the correlation length resolved, modulo dimensional factors) is likely not correct, based on earlier manifold learning characterization of path integrals~\cite{mendes2021intrinsic} (which did not observe such relation, at least in simple quantities such as the intrinsic dimension).}

Another natural extension is to consider path integrals of quantum systems~\cite{mendes2021intrinsic}. While the data structures of such objects might feature anisotropies due to the different roles of space and imaginary time correlations, they shall be equally amenable to the analysis discussed here. Moreover, single-sliced path integrals can also be represented as networks, as discussed in Ref.~\cite{mendes2023wave}: this last route provides a very promising venue for future investigation is the application of the proposed network science tools to study directly experimental data of many-body wave function snapshots.

We acknowledge useful discussions with M. Heyl, M. Schmitt and T. Mendes-Santos.
M.D. work was partly supported by
the MIUR Programme FARE (MEPH), by QUANTERA DYNAMITE PCI2022-132919, and by the
PNRR MUR project PE0000023-NQSTI. Nordita is supported in part by NordForsk.

\bibliography{bibliography}

\begin{thebibliography}{95}%
\makeatletter
\providecommand \@ifxundefined [1]{%
 \@ifx{#1\undefined}
}%
\providecommand \@ifnum [1]{%
 \ifnum #1\expandafter \@firstoftwo
 \else \expandafter \@secondoftwo
 \fi
}%
\providecommand \@ifx [1]{%
 \ifx #1\expandafter \@firstoftwo
 \else \expandafter \@secondoftwo
 \fi
}%
\providecommand \natexlab [1]{#1}%
\providecommand \enquote  [1]{``#1''}%
\providecommand \bibnamefont  [1]{#1}%
\providecommand \bibfnamefont [1]{#1}%
\providecommand \citenamefont [1]{#1}%
\providecommand \href@noop [0]{\@secondoftwo}%
\providecommand \href [0]{\begingroup \@sanitize@url \@href}%
\providecommand \@href[1]{\@@startlink{#1}\@@href}%
\providecommand \@@href[1]{\endgroup#1\@@endlink}%
\providecommand \@sanitize@url [0]{\catcode `\\12\catcode `\$12\catcode
  `\&12\catcode `\#12\catcode `\^12\catcode `\_12\catcode `\%12\relax}%
\providecommand \@@startlink[1]{}%
\providecommand \@@endlink[0]{}%
\providecommand \url  [0]{\begingroup\@sanitize@url \@url }%
\providecommand \@url [1]{\endgroup\@href {#1}{\urlprefix }}%
\providecommand \urlprefix  [0]{URL }%
\providecommand \Eprint [0]{\href }%
\providecommand \doibase [0]{https://doi.org/}%
\providecommand \selectlanguage [0]{\@gobble}%
\providecommand \bibinfo  [0]{\@secondoftwo}%
\providecommand \bibfield  [0]{\@secondoftwo}%
\providecommand \translation [1]{[#1]}%
\providecommand \BibitemOpen [0]{}%
\providecommand \bibitemStop [0]{}%
\providecommand \bibitemNoStop [0]{.\EOS\space}%
\providecommand \EOS [0]{\spacefactor3000\relax}%
\providecommand \BibitemShut  [1]{\csname bibitem#1\endcsname}%
\let\auto@bib@innerbib\@empty
\bibitem [{\citenamefont {Barabasi}(2016)}]{barabasi2016network}%
  \BibitemOpen
  \bibfield  {author} {\bibinfo {author} {\bibfnamefont {A.-L.}\ \bibnamefont
  {Barabasi}},\ }\href@noop {} {\emph {\bibinfo {title} {Network Science}}}\
  (\bibinfo  {publisher} {Cambridge University Press},\ \bibinfo {year}
  {2016})\BibitemShut {NoStop}%
\bibitem [{\citenamefont {Newman}(2018)}]{newman2018networks}%
  \BibitemOpen
  \bibfield  {author} {\bibinfo {author} {\bibfnamefont {M.}~\bibnamefont
  {Newman}},\ }\href@noop {} {\emph {\bibinfo {title} {Networks}}}\ (\bibinfo
  {publisher} {Oxford University Press},\ \bibinfo {year} {2018})\BibitemShut
  {NoStop}%
\bibitem [{\citenamefont {Barrat}\ \emph {et~al.}(2008)\citenamefont {Barrat},
  \citenamefont {Barthelemy},\ and\ \citenamefont
  {Vespignani}}]{barrat2008dynamical}%
  \BibitemOpen
  \bibfield  {author} {\bibinfo {author} {\bibfnamefont {A.}~\bibnamefont
  {Barrat}}, \bibinfo {author} {\bibfnamefont {M.}~\bibnamefont {Barthelemy}},\
  and\ \bibinfo {author} {\bibfnamefont {A.}~\bibnamefont {Vespignani}},\
  }\href@noop {} {\emph {\bibinfo {title} {Dynamical processes on complex
  networks}}}\ (\bibinfo  {publisher} {Cambridge University Press},\ \bibinfo
  {year} {2008})\BibitemShut {NoStop}%
\bibitem [{\citenamefont {Estrada}(2012)}]{estrada2012structure}%
  \BibitemOpen
  \bibfield  {author} {\bibinfo {author} {\bibfnamefont {E.}~\bibnamefont
  {Estrada}},\ }\href@noop {} {\emph {\bibinfo {title} {The structure of
  complex networks: theory and applications}}}\ (\bibinfo  {publisher} {Oxford
  University Press, USA},\ \bibinfo {year} {2012})\BibitemShut {NoStop}%
\bibitem [{\citenamefont {Dorogovtsev}\ and\ \citenamefont
  {Mendes}(2022)}]{dorogovtsev2022nature}%
  \BibitemOpen
  \bibfield  {author} {\bibinfo {author} {\bibfnamefont {S.~N.}\ \bibnamefont
  {Dorogovtsev}}\ and\ \bibinfo {author} {\bibfnamefont {J.~F.}\ \bibnamefont
  {Mendes}},\ }\href@noop {} {\emph {\bibinfo {title} {The nature of complex
  networks}}}\ (\bibinfo  {publisher} {Oxford University Press},\ \bibinfo
  {year} {2022})\BibitemShut {NoStop}%
\bibitem [{\citenamefont {Bianconi}(2021)}]{bianconi2021higher}%
  \BibitemOpen
  \bibfield  {author} {\bibinfo {author} {\bibfnamefont {G.}~\bibnamefont
  {Bianconi}},\ }\href@noop {} {\emph {\bibinfo {title} {Higher-order
  networks}}}\ (\bibinfo  {publisher} {Cambridge University Press},\ \bibinfo
  {year} {2021})\BibitemShut {NoStop}%
\bibitem [{\citenamefont {Mehta}\ \emph {et~al.}(2019)\citenamefont {Mehta},
  \citenamefont {Bukov}, \citenamefont {Wang}, \citenamefont {Day},
  \citenamefont {Richardson}, \citenamefont {Fisher},\ and\ \citenamefont
  {Schwab}}]{MEHTA20191}%
  \BibitemOpen
  \bibfield  {author} {\bibinfo {author} {\bibfnamefont {P.}~\bibnamefont
  {Mehta}}, \bibinfo {author} {\bibfnamefont {M.}~\bibnamefont {Bukov}},
  \bibinfo {author} {\bibfnamefont {C.-H.}\ \bibnamefont {Wang}}, \bibinfo
  {author} {\bibfnamefont {A.~G.}\ \bibnamefont {Day}}, \bibinfo {author}
  {\bibfnamefont {C.}~\bibnamefont {Richardson}}, \bibinfo {author}
  {\bibfnamefont {C.~K.}\ \bibnamefont {Fisher}},\ and\ \bibinfo {author}
  {\bibfnamefont {D.~J.}\ \bibnamefont {Schwab}},\ }\bibfield  {title}
  {\bibinfo {title} {A high-bias, low-variance introduction to machine learning
  for physicists},\ }\href
  {https://doi.org/https://doi.org/10.1016/j.physrep.2019.03.001} {\bibfield
  {journal} {\bibinfo  {journal} {Physics Reports}\ }\textbf {\bibinfo {volume}
  {810}},\ \bibinfo {pages} {1} (\bibinfo {year} {2019})}\BibitemShut {NoStop}%
\bibitem [{\citenamefont {Wang}(2016)}]{PhysRevB.94.195105}%
  \BibitemOpen
  \bibfield  {author} {\bibinfo {author} {\bibfnamefont {L.}~\bibnamefont
  {Wang}},\ }\bibfield  {title} {\bibinfo {title} {Discovering phase
  transitions with unsupervised learning},\ }\href
  {https://doi.org/10.1103/PhysRevB.94.195105} {\bibfield  {journal} {\bibinfo
  {journal} {Phys. Rev. B}\ }\textbf {\bibinfo {volume} {94}},\ \bibinfo
  {pages} {195105} (\bibinfo {year} {2016})}\BibitemShut {NoStop}%
\bibitem [{\citenamefont {Hu}\ \emph {et~al.}(2017)\citenamefont {Hu},
  \citenamefont {Singh},\ and\ \citenamefont {Scalettar}}]{PhysRevE.95.062122}%
  \BibitemOpen
  \bibfield  {author} {\bibinfo {author} {\bibfnamefont {W.}~\bibnamefont
  {Hu}}, \bibinfo {author} {\bibfnamefont {R.~R.~P.}\ \bibnamefont {Singh}},\
  and\ \bibinfo {author} {\bibfnamefont {R.~T.}\ \bibnamefont {Scalettar}},\
  }\bibfield  {title} {\bibinfo {title} {Discovering phases, phase transitions,
  and crossovers through unsupervised machine learning: A critical
  examination},\ }\href {https://doi.org/10.1103/PhysRevE.95.062122} {\bibfield
   {journal} {\bibinfo  {journal} {Phys. Rev. E}\ }\textbf {\bibinfo {volume}
  {95}},\ \bibinfo {pages} {062122} (\bibinfo {year} {2017})}\BibitemShut
  {NoStop}%
\bibitem [{\citenamefont {Wetzel}(2017)}]{PhysRevE.96.022140}%
  \BibitemOpen
  \bibfield  {author} {\bibinfo {author} {\bibfnamefont {S.~J.}\ \bibnamefont
  {Wetzel}},\ }\bibfield  {title} {\bibinfo {title} {Unsupervised learning of
  phase transitions: From principal component analysis to variational
  autoencoders},\ }\href {https://doi.org/10.1103/PhysRevE.96.022140}
  {\bibfield  {journal} {\bibinfo  {journal} {Phys. Rev. E}\ }\textbf {\bibinfo
  {volume} {96}},\ \bibinfo {pages} {022140} (\bibinfo {year}
  {2017})}\BibitemShut {NoStop}%
\bibitem [{\citenamefont {Rodriguez-Nieva}\ and\ \citenamefont
  {Scheurer}(2019)}]{Rodriguez-Nieva2019}%
  \BibitemOpen
  \bibfield  {author} {\bibinfo {author} {\bibfnamefont {J.~F.}\ \bibnamefont
  {Rodriguez-Nieva}}\ and\ \bibinfo {author} {\bibfnamefont {M.~S.}\
  \bibnamefont {Scheurer}},\ }\bibfield  {title} {\bibinfo {title} {Identifying
  topological order through unsupervised machine learning},\ }\href
  {https://doi.org/10.1038/s41567-019-0512-x} {\bibfield  {journal} {\bibinfo
  {journal} {Nature Physics}\ }\textbf {\bibinfo {volume} {15}},\ \bibinfo
  {pages} {790} (\bibinfo {year} {2019})}\BibitemShut {NoStop}%
\bibitem [{\citenamefont {Mendes-Santos}\ \emph
  {et~al.}(2021{\natexlab{a}})\citenamefont {Mendes-Santos}, \citenamefont
  {Turkeshi}, \citenamefont {Dalmonte},\ and\ \citenamefont
  {Rodriguez}}]{PhysRevX.11.011040}%
  \BibitemOpen
  \bibfield  {author} {\bibinfo {author} {\bibfnamefont {T.}~\bibnamefont
  {Mendes-Santos}}, \bibinfo {author} {\bibfnamefont {X.}~\bibnamefont
  {Turkeshi}}, \bibinfo {author} {\bibfnamefont {M.}~\bibnamefont {Dalmonte}},\
  and\ \bibinfo {author} {\bibfnamefont {A.}~\bibnamefont {Rodriguez}},\
  }\bibfield  {title} {\bibinfo {title} {Unsupervised learning universal
  critical behavior via the intrinsic dimension},\ }\href
  {https://doi.org/10.1103/PhysRevX.11.011040} {\bibfield  {journal} {\bibinfo
  {journal} {Phys. Rev. X}\ }\textbf {\bibinfo {volume} {11}},\ \bibinfo
  {pages} {011040} (\bibinfo {year} {2021}{\natexlab{a}})}\BibitemShut
  {NoStop}%
\bibitem [{\citenamefont {Morningstar}\ and\ \citenamefont
  {Melko}(2018)}]{morningstar2018deep}%
  \BibitemOpen
  \bibfield  {author} {\bibinfo {author} {\bibfnamefont {A.}~\bibnamefont
  {Morningstar}}\ and\ \bibinfo {author} {\bibfnamefont {R.~G.}\ \bibnamefont
  {Melko}},\ }\bibfield  {title} {\bibinfo {title} {Deep learning the ising
  model near criticality},\ }\href@noop {} {\bibfield  {journal} {\bibinfo
  {journal} {Journal of Machine Learning Research}\ }\textbf {\bibinfo {volume}
  {18}},\ \bibinfo {pages} {1} (\bibinfo {year} {2018})}\BibitemShut {NoStop}%
\bibitem [{\citenamefont {Aoki}\ and\ \citenamefont
  {Kobayashi}(2016)}]{aoki2016restricted}%
  \BibitemOpen
  \bibfield  {author} {\bibinfo {author} {\bibfnamefont {K.-I.}\ \bibnamefont
  {Aoki}}\ and\ \bibinfo {author} {\bibfnamefont {T.}~\bibnamefont
  {Kobayashi}},\ }\bibfield  {title} {\bibinfo {title} {Restricted {B}oltzmann
  machines for the long range {I}sing models},\ }\href@noop {} {\bibfield
  {journal} {\bibinfo  {journal} {Modern Physics Letters B}\ }\textbf {\bibinfo
  {volume} {30}},\ \bibinfo {pages} {1650401} (\bibinfo {year}
  {2016})}\BibitemShut {NoStop}%
\bibitem [{\citenamefont {Carrasquilla}\ \emph {et~al.}(2019)\citenamefont
  {Carrasquilla}, \citenamefont {Torlai}, \citenamefont {Melko},\ and\
  \citenamefont {Aolita}}]{carrasquilla2019reconstructing}%
  \BibitemOpen
  \bibfield  {author} {\bibinfo {author} {\bibfnamefont {J.}~\bibnamefont
  {Carrasquilla}}, \bibinfo {author} {\bibfnamefont {G.}~\bibnamefont
  {Torlai}}, \bibinfo {author} {\bibfnamefont {R.~G.}\ \bibnamefont {Melko}},\
  and\ \bibinfo {author} {\bibfnamefont {L.}~\bibnamefont {Aolita}},\
  }\bibfield  {title} {\bibinfo {title} {Reconstructing quantum states with
  generative models},\ }\href@noop {} {\bibfield  {journal} {\bibinfo
  {journal} {Nature Machine Intelligence}\ }\textbf {\bibinfo {volume} {1}},\
  \bibinfo {pages} {155} (\bibinfo {year} {2019})}\BibitemShut {NoStop}%
\bibitem [{\citenamefont {Benedetti}\ \emph {et~al.}(2017)\citenamefont
  {Benedetti}, \citenamefont {Realpe-G{\'o}mez}, \citenamefont {Biswas},\ and\
  \citenamefont {Perdomo-Ortiz}}]{benedetti2017quantum}%
  \BibitemOpen
  \bibfield  {author} {\bibinfo {author} {\bibfnamefont {M.}~\bibnamefont
  {Benedetti}}, \bibinfo {author} {\bibfnamefont {J.}~\bibnamefont
  {Realpe-G{\'o}mez}}, \bibinfo {author} {\bibfnamefont {R.}~\bibnamefont
  {Biswas}},\ and\ \bibinfo {author} {\bibfnamefont {A.}~\bibnamefont
  {Perdomo-Ortiz}},\ }\bibfield  {title} {\bibinfo {title} {Quantum-assisted
  learning of hardware-embedded probabilistic graphical models},\ }\href@noop
  {} {\bibfield  {journal} {\bibinfo  {journal} {Physical Review X}\ }\textbf
  {\bibinfo {volume} {7}},\ \bibinfo {pages} {041052} (\bibinfo {year}
  {2017})}\BibitemShut {NoStop}%
\bibitem [{\citenamefont {Melko}\ \emph {et~al.}(2019)\citenamefont {Melko},
  \citenamefont {Carleo}, \citenamefont {Carrasquilla},\ and\ \citenamefont
  {Cirac}}]{melko2019restricted}%
  \BibitemOpen
  \bibfield  {author} {\bibinfo {author} {\bibfnamefont {R.~G.}\ \bibnamefont
  {Melko}}, \bibinfo {author} {\bibfnamefont {G.}~\bibnamefont {Carleo}},
  \bibinfo {author} {\bibfnamefont {J.}~\bibnamefont {Carrasquilla}},\ and\
  \bibinfo {author} {\bibfnamefont {J.~I.}\ \bibnamefont {Cirac}},\ }\bibfield
  {title} {\bibinfo {title} {Restricted {B}oltzmann machines in quantum
  physics},\ }\href@noop {} {\bibfield  {journal} {\bibinfo  {journal} {Nature
  Physics}\ }\textbf {\bibinfo {volume} {15}},\ \bibinfo {pages} {887}
  (\bibinfo {year} {2019})}\BibitemShut {NoStop}%
\bibitem [{\citenamefont {Dorogovtsev}\ \emph {et~al.}(2002)\citenamefont
  {Dorogovtsev}, \citenamefont {Goltsev},\ and\ \citenamefont
  {Mendes}}]{dorogovtsev2002ising}%
  \BibitemOpen
  \bibfield  {author} {\bibinfo {author} {\bibfnamefont {S.~N.}\ \bibnamefont
  {Dorogovtsev}}, \bibinfo {author} {\bibfnamefont {A.~V.}\ \bibnamefont
  {Goltsev}},\ and\ \bibinfo {author} {\bibfnamefont {J.~F.~F.}\ \bibnamefont
  {Mendes}},\ }\bibfield  {title} {\bibinfo {title} {Ising model on networks
  with an arbitrary distribution of connections},\ }\href@noop {} {\bibfield
  {journal} {\bibinfo  {journal} {Physical Review E}\ }\textbf {\bibinfo
  {volume} {66}},\ \bibinfo {pages} {016104} (\bibinfo {year}
  {2002})}\BibitemShut {NoStop}%
\bibitem [{\citenamefont {Bianconi}(2002)}]{bianconi2002mean}%
  \BibitemOpen
  \bibfield  {author} {\bibinfo {author} {\bibfnamefont {G.}~\bibnamefont
  {Bianconi}},\ }\bibfield  {title} {\bibinfo {title} {Mean field solution of
  the {I}sing model on a barab{\'a}si--albert network},\ }\href@noop {}
  {\bibfield  {journal} {\bibinfo  {journal} {Physics Letters A}\ }\textbf
  {\bibinfo {volume} {303}},\ \bibinfo {pages} {166} (\bibinfo {year}
  {2002})}\BibitemShut {NoStop}%
\bibitem [{\citenamefont {Leone}\ \emph {et~al.}(2002)\citenamefont {Leone},
  \citenamefont {V{\'a}zquez}, \citenamefont {Vespignani},\ and\ \citenamefont
  {Zecchina}}]{leone2002ferromagnetic}%
  \BibitemOpen
  \bibfield  {author} {\bibinfo {author} {\bibfnamefont {M.}~\bibnamefont
  {Leone}}, \bibinfo {author} {\bibfnamefont {A.}~\bibnamefont {V{\'a}zquez}},
  \bibinfo {author} {\bibfnamefont {A.}~\bibnamefont {Vespignani}},\ and\
  \bibinfo {author} {\bibfnamefont {R.}~\bibnamefont {Zecchina}},\ }\bibfield
  {title} {\bibinfo {title} {Ferromagnetic ordering in graphs with arbitrary
  degree distribution},\ }\href@noop {} {\bibfield  {journal} {\bibinfo
  {journal} {The European Physical Journal B-Condensed Matter and Complex
  Systems}\ }\textbf {\bibinfo {volume} {28}},\ \bibinfo {pages} {191}
  (\bibinfo {year} {2002})}\BibitemShut {NoStop}%
\bibitem [{\citenamefont {Nguyen}\ \emph {et~al.}(2017)\citenamefont {Nguyen},
  \citenamefont {Zecchina},\ and\ \citenamefont {Berg}}]{nguyen2017inverse}%
  \BibitemOpen
  \bibfield  {author} {\bibinfo {author} {\bibfnamefont {H.~C.}\ \bibnamefont
  {Nguyen}}, \bibinfo {author} {\bibfnamefont {R.}~\bibnamefont {Zecchina}},\
  and\ \bibinfo {author} {\bibfnamefont {J.}~\bibnamefont {Berg}},\ }\bibfield
  {title} {\bibinfo {title} {Inverse statistical problems: from the inverse
  {I}sing problem to data science},\ }\href@noop {} {\bibfield  {journal}
  {\bibinfo  {journal} {Advances in Physics}\ }\textbf {\bibinfo {volume}
  {66}},\ \bibinfo {pages} {197} (\bibinfo {year} {2017})}\BibitemShut
  {NoStop}%
\bibitem [{\citenamefont {Bianconi}(2012)}]{bianconi2012superconductor}%
  \BibitemOpen
  \bibfield  {author} {\bibinfo {author} {\bibfnamefont {G.}~\bibnamefont
  {Bianconi}},\ }\bibfield  {title} {\bibinfo {title} {Superconductor-insulator
  transition on annealed complex networks},\ }\href@noop {} {\bibfield
  {journal} {\bibinfo  {journal} {Physical Review E}\ }\textbf {\bibinfo
  {volume} {85}},\ \bibinfo {pages} {061113} (\bibinfo {year}
  {2012})}\BibitemShut {NoStop}%
\bibitem [{\citenamefont {Bianconi}(2013)}]{bianconi2013superconductor}%
  \BibitemOpen
  \bibfield  {author} {\bibinfo {author} {\bibfnamefont {G.}~\bibnamefont
  {Bianconi}},\ }\bibfield  {title} {\bibinfo {title} {Superconductor-insulator
  transition in a network of 2d percolation clusters},\ }\href@noop {}
  {\bibfield  {journal} {\bibinfo  {journal} {Europhysics Letters}\ }\textbf
  {\bibinfo {volume} {101}},\ \bibinfo {pages} {26003} (\bibinfo {year}
  {2013})}\BibitemShut {NoStop}%
\bibitem [{\citenamefont {Chepuri}\ and\ \citenamefont
  {Kov{\'a}cs}(2022)}]{chepuri2022complex}%
  \BibitemOpen
  \bibfield  {author} {\bibinfo {author} {\bibfnamefont {R.}~\bibnamefont
  {Chepuri}}\ and\ \bibinfo {author} {\bibfnamefont {I.~A.}\ \bibnamefont
  {Kov{\'a}cs}},\ }\bibfield  {title} {\bibinfo {title} {Complex quantum
  network models from spin clusters},\ }\href@noop {} {\bibfield  {journal}
  {\bibinfo  {journal} {arXiv preprint arXiv:2210.15838}\ } (\bibinfo {year}
  {2022})}\BibitemShut {NoStop}%
\bibitem [{\citenamefont {Halu}\ \emph {et~al.}(2012)\citenamefont {Halu},
  \citenamefont {Ferretti}, \citenamefont {Vezzani},\ and\ \citenamefont
  {Bianconi}}]{halu2012phase}%
  \BibitemOpen
  \bibfield  {author} {\bibinfo {author} {\bibfnamefont {A.}~\bibnamefont
  {Halu}}, \bibinfo {author} {\bibfnamefont {L.}~\bibnamefont {Ferretti}},
  \bibinfo {author} {\bibfnamefont {A.}~\bibnamefont {Vezzani}},\ and\ \bibinfo
  {author} {\bibfnamefont {G.}~\bibnamefont {Bianconi}},\ }\bibfield  {title}
  {\bibinfo {title} {Phase diagram of the bose-hubbard model on complex
  networks},\ }\href@noop {} {\bibfield  {journal} {\bibinfo  {journal}
  {Europhysics Letters}\ }\textbf {\bibinfo {volume} {99}},\ \bibinfo {pages}
  {18001} (\bibinfo {year} {2012})}\BibitemShut {NoStop}%
\bibitem [{\citenamefont {Halu}\ \emph {et~al.}(2013)\citenamefont {Halu},
  \citenamefont {Garnerone}, \citenamefont {Vezzani},\ and\ \citenamefont
  {Bianconi}}]{halu2013phase}%
  \BibitemOpen
  \bibfield  {author} {\bibinfo {author} {\bibfnamefont {A.}~\bibnamefont
  {Halu}}, \bibinfo {author} {\bibfnamefont {S.}~\bibnamefont {Garnerone}},
  \bibinfo {author} {\bibfnamefont {A.}~\bibnamefont {Vezzani}},\ and\ \bibinfo
  {author} {\bibfnamefont {G.}~\bibnamefont {Bianconi}},\ }\bibfield  {title}
  {\bibinfo {title} {Phase transition of light on complex quantum networks},\
  }\href@noop {} {\bibfield  {journal} {\bibinfo  {journal} {Physical Review
  E}\ }\textbf {\bibinfo {volume} {87}},\ \bibinfo {pages} {022104} (\bibinfo
  {year} {2013})}\BibitemShut {NoStop}%
\bibitem [{\citenamefont {Vicsek}\ \emph {et~al.}(1995)\citenamefont {Vicsek},
  \citenamefont {Czir{\'o}k}, \citenamefont {Ben-Jacob}, \citenamefont
  {Cohen},\ and\ \citenamefont {Shochet}}]{vicsek1995novel}%
  \BibitemOpen
  \bibfield  {author} {\bibinfo {author} {\bibfnamefont {T.}~\bibnamefont
  {Vicsek}}, \bibinfo {author} {\bibfnamefont {A.}~\bibnamefont {Czir{\'o}k}},
  \bibinfo {author} {\bibfnamefont {E.}~\bibnamefont {Ben-Jacob}}, \bibinfo
  {author} {\bibfnamefont {I.}~\bibnamefont {Cohen}},\ and\ \bibinfo {author}
  {\bibfnamefont {O.}~\bibnamefont {Shochet}},\ }\bibfield  {title} {\bibinfo
  {title} {Novel type of phase transition in a system of self-driven
  particles},\ }\href@noop {} {\bibfield  {journal} {\bibinfo  {journal}
  {Physical Review Letters}\ }\textbf {\bibinfo {volume} {75}},\ \bibinfo
  {pages} {1226} (\bibinfo {year} {1995})}\BibitemShut {NoStop}%
\bibitem [{\citenamefont {Ballerini}\ \emph {et~al.}(2008)\citenamefont
  {Ballerini}, \citenamefont {Cabibbo}, \citenamefont {Candelier},
  \citenamefont {Cavagna}, \citenamefont {Cisbani}, \citenamefont {Giardina},
  \citenamefont {Lecomte}, \citenamefont {Orlandi}, \citenamefont {Parisi},
  \citenamefont {Procaccini} \emph {et~al.}}]{ballerini2008interaction}%
  \BibitemOpen
  \bibfield  {author} {\bibinfo {author} {\bibfnamefont {M.}~\bibnamefont
  {Ballerini}}, \bibinfo {author} {\bibfnamefont {N.}~\bibnamefont {Cabibbo}},
  \bibinfo {author} {\bibfnamefont {R.}~\bibnamefont {Candelier}}, \bibinfo
  {author} {\bibfnamefont {A.}~\bibnamefont {Cavagna}}, \bibinfo {author}
  {\bibfnamefont {E.}~\bibnamefont {Cisbani}}, \bibinfo {author} {\bibfnamefont
  {I.}~\bibnamefont {Giardina}}, \bibinfo {author} {\bibfnamefont
  {V.}~\bibnamefont {Lecomte}}, \bibinfo {author} {\bibfnamefont
  {A.}~\bibnamefont {Orlandi}}, \bibinfo {author} {\bibfnamefont
  {G.}~\bibnamefont {Parisi}}, \bibinfo {author} {\bibfnamefont
  {A.}~\bibnamefont {Procaccini}}, \emph {et~al.},\ }\bibfield  {title}
  {\bibinfo {title} {Interaction ruling animal collective behavior depends on
  topological rather than metric distance: Evidence from a field study},\
  }\href@noop {} {\bibfield  {journal} {\bibinfo  {journal} {Proceedings of the
  National Academy of Sciences}\ }\textbf {\bibinfo {volume} {105}},\ \bibinfo
  {pages} {1232} (\bibinfo {year} {2008})}\BibitemShut {NoStop}%
\bibitem [{\citenamefont {Morcos}\ \emph {et~al.}(2011)\citenamefont {Morcos},
  \citenamefont {Pagnani}, \citenamefont {Lunt}, \citenamefont {Bertolino},
  \citenamefont {Marks}, \citenamefont {Sander}, \citenamefont {Zecchina},
  \citenamefont {Onuchic}, \citenamefont {Hwa},\ and\ \citenamefont
  {Weigt}}]{morcos2011direct}%
  \BibitemOpen
  \bibfield  {author} {\bibinfo {author} {\bibfnamefont {F.}~\bibnamefont
  {Morcos}}, \bibinfo {author} {\bibfnamefont {A.}~\bibnamefont {Pagnani}},
  \bibinfo {author} {\bibfnamefont {B.}~\bibnamefont {Lunt}}, \bibinfo {author}
  {\bibfnamefont {A.}~\bibnamefont {Bertolino}}, \bibinfo {author}
  {\bibfnamefont {D.~S.}\ \bibnamefont {Marks}}, \bibinfo {author}
  {\bibfnamefont {C.}~\bibnamefont {Sander}}, \bibinfo {author} {\bibfnamefont
  {R.}~\bibnamefont {Zecchina}}, \bibinfo {author} {\bibfnamefont {J.~N.}\
  \bibnamefont {Onuchic}}, \bibinfo {author} {\bibfnamefont {T.}~\bibnamefont
  {Hwa}},\ and\ \bibinfo {author} {\bibfnamefont {M.}~\bibnamefont {Weigt}},\
  }\bibfield  {title} {\bibinfo {title} {Direct-coupling analysis of residue
  coevolution captures native contacts across many protein families},\
  }\href@noop {} {\bibfield  {journal} {\bibinfo  {journal} {Proceedings of the
  National Academy of Sciences}\ }\textbf {\bibinfo {volume} {108}},\ \bibinfo
  {pages} {E1293} (\bibinfo {year} {2011})}\BibitemShut {NoStop}%
\bibitem [{\citenamefont {Mora}\ and\ \citenamefont
  {Bialek}(2011)}]{mora2011biological}%
  \BibitemOpen
  \bibfield  {author} {\bibinfo {author} {\bibfnamefont {T.}~\bibnamefont
  {Mora}}\ and\ \bibinfo {author} {\bibfnamefont {W.}~\bibnamefont {Bialek}},\
  }\bibfield  {title} {\bibinfo {title} {Are biological systems poised at
  criticality?},\ }\href@noop {} {\bibfield  {journal} {\bibinfo  {journal}
  {Journal of Statistical Physics}\ }\textbf {\bibinfo {volume} {144}},\
  \bibinfo {pages} {268} (\bibinfo {year} {2011})}\BibitemShut {NoStop}%
\bibitem [{\citenamefont {Nokkala}\ \emph {et~al.}(2016)\citenamefont
  {Nokkala}, \citenamefont {Galve}, \citenamefont {Zambrini}, \citenamefont
  {Maniscalco},\ and\ \citenamefont {Piilo}}]{nokkala2016complex}%
  \BibitemOpen
  \bibfield  {author} {\bibinfo {author} {\bibfnamefont {J.}~\bibnamefont
  {Nokkala}}, \bibinfo {author} {\bibfnamefont {F.}~\bibnamefont {Galve}},
  \bibinfo {author} {\bibfnamefont {R.}~\bibnamefont {Zambrini}}, \bibinfo
  {author} {\bibfnamefont {S.}~\bibnamefont {Maniscalco}},\ and\ \bibinfo
  {author} {\bibfnamefont {J.}~\bibnamefont {Piilo}},\ }\bibfield  {title}
  {\bibinfo {title} {Complex quantum networks as structured environments:
  engineering and probing},\ }\href@noop {} {\bibfield  {journal} {\bibinfo
  {journal} {Scientific Reports}\ }\textbf {\bibinfo {volume} {6}},\ \bibinfo
  {pages} {26861} (\bibinfo {year} {2016})}\BibitemShut {NoStop}%
\bibitem [{\citenamefont {Nokkala}\ \emph {et~al.}(2018)\citenamefont
  {Nokkala}, \citenamefont {Arzani}, \citenamefont {Galve}, \citenamefont
  {Zambrini}, \citenamefont {Maniscalco}, \citenamefont {Piilo}, \citenamefont
  {Treps},\ and\ \citenamefont {Parigi}}]{nokkala2018reconfigurable}%
  \BibitemOpen
  \bibfield  {author} {\bibinfo {author} {\bibfnamefont {J.}~\bibnamefont
  {Nokkala}}, \bibinfo {author} {\bibfnamefont {F.}~\bibnamefont {Arzani}},
  \bibinfo {author} {\bibfnamefont {F.}~\bibnamefont {Galve}}, \bibinfo
  {author} {\bibfnamefont {R.}~\bibnamefont {Zambrini}}, \bibinfo {author}
  {\bibfnamefont {S.}~\bibnamefont {Maniscalco}}, \bibinfo {author}
  {\bibfnamefont {J.}~\bibnamefont {Piilo}}, \bibinfo {author} {\bibfnamefont
  {N.}~\bibnamefont {Treps}},\ and\ \bibinfo {author} {\bibfnamefont
  {V.}~\bibnamefont {Parigi}},\ }\bibfield  {title} {\bibinfo {title}
  {Reconfigurable optical implementation of quantum complex networks},\
  }\href@noop {} {\bibfield  {journal} {\bibinfo  {journal} {New Journal of
  Physics}\ }\textbf {\bibinfo {volume} {20}},\ \bibinfo {pages} {053024}
  (\bibinfo {year} {2018})}\BibitemShut {NoStop}%
\bibitem [{\citenamefont {Bonamassa}\ \emph {et~al.}(2023)\citenamefont
  {Bonamassa}, \citenamefont {Gross}, \citenamefont {Laav}, \citenamefont
  {Volotsenko}, \citenamefont {Frydman},\ and\ \citenamefont
  {Havlin}}]{bonamassa2023interdependent}%
  \BibitemOpen
  \bibfield  {author} {\bibinfo {author} {\bibfnamefont {I.}~\bibnamefont
  {Bonamassa}}, \bibinfo {author} {\bibfnamefont {B.}~\bibnamefont {Gross}},
  \bibinfo {author} {\bibfnamefont {M.}~\bibnamefont {Laav}}, \bibinfo {author}
  {\bibfnamefont {I.}~\bibnamefont {Volotsenko}}, \bibinfo {author}
  {\bibfnamefont {A.}~\bibnamefont {Frydman}},\ and\ \bibinfo {author}
  {\bibfnamefont {S.}~\bibnamefont {Havlin}},\ }\bibfield  {title} {\bibinfo
  {title} {Interdependent superconducting networks},\ }\href@noop {} {\bibfield
   {journal} {\bibinfo  {journal} {Nature Physics}\ ,\ \bibinfo {pages} {1}}
  (\bibinfo {year} {2023})}\BibitemShut {NoStop}%
\bibitem [{\citenamefont {Tumminello}\ \emph {et~al.}(2005)\citenamefont
  {Tumminello}, \citenamefont {Aste}, \citenamefont {Di~Matteo},\ and\
  \citenamefont {Mantegna}}]{tumminello2005tool}%
  \BibitemOpen
  \bibfield  {author} {\bibinfo {author} {\bibfnamefont {M.}~\bibnamefont
  {Tumminello}}, \bibinfo {author} {\bibfnamefont {T.}~\bibnamefont {Aste}},
  \bibinfo {author} {\bibfnamefont {T.}~\bibnamefont {Di~Matteo}},\ and\
  \bibinfo {author} {\bibfnamefont {R.~N.}\ \bibnamefont {Mantegna}},\
  }\bibfield  {title} {\bibinfo {title} {A tool for filtering information in
  complex systems},\ }\href@noop {} {\bibfield  {journal} {\bibinfo  {journal}
  {Proceedings of the National Academy of Sciences}\ }\textbf {\bibinfo
  {volume} {102}},\ \bibinfo {pages} {10421} (\bibinfo {year}
  {2005})}\BibitemShut {NoStop}%
\bibitem [{\citenamefont {Bonanno}\ \emph {et~al.}(2003)\citenamefont
  {Bonanno}, \citenamefont {Caldarelli}, \citenamefont {Lillo},\ and\
  \citenamefont {Mantegna}}]{bonanno2003topology}%
  \BibitemOpen
  \bibfield  {author} {\bibinfo {author} {\bibfnamefont {G.}~\bibnamefont
  {Bonanno}}, \bibinfo {author} {\bibfnamefont {G.}~\bibnamefont {Caldarelli}},
  \bibinfo {author} {\bibfnamefont {F.}~\bibnamefont {Lillo}},\ and\ \bibinfo
  {author} {\bibfnamefont {R.~N.}\ \bibnamefont {Mantegna}},\ }\bibfield
  {title} {\bibinfo {title} {Topology of correlation-based minimal spanning
  trees in real and model markets},\ }\href@noop {} {\bibfield  {journal}
  {\bibinfo  {journal} {Physical Review E}\ }\textbf {\bibinfo {volume} {68}},\
  \bibinfo {pages} {046130} (\bibinfo {year} {2003})}\BibitemShut {NoStop}%
\bibitem [{\citenamefont {Valdez}\ \emph {et~al.}(2017)\citenamefont {Valdez},
  \citenamefont {Jaschke}, \citenamefont {Vargas},\ and\ \citenamefont
  {Carr}}]{valdez2017quantifying}%
  \BibitemOpen
  \bibfield  {author} {\bibinfo {author} {\bibfnamefont {M.~A.}\ \bibnamefont
  {Valdez}}, \bibinfo {author} {\bibfnamefont {D.}~\bibnamefont {Jaschke}},
  \bibinfo {author} {\bibfnamefont {D.~L.}\ \bibnamefont {Vargas}},\ and\
  \bibinfo {author} {\bibfnamefont {L.~D.}\ \bibnamefont {Carr}},\ }\bibfield
  {title} {\bibinfo {title} {Quantifying complexity in quantum phase
  transitions via mutual information complex networks},\ }\href@noop {}
  {\bibfield  {journal} {\bibinfo  {journal} {Physical Review Letters}\
  }\textbf {\bibinfo {volume} {119}},\ \bibinfo {pages} {225301} (\bibinfo
  {year} {2017})}\BibitemShut {NoStop}%
\bibitem [{\citenamefont {Sundar}\ \emph {et~al.}(2018)\citenamefont {Sundar},
  \citenamefont {Valdez}, \citenamefont {Carr},\ and\ \citenamefont
  {Hazzard}}]{PhysRevA.97.052320}%
  \BibitemOpen
  \bibfield  {author} {\bibinfo {author} {\bibfnamefont {B.}~\bibnamefont
  {Sundar}}, \bibinfo {author} {\bibfnamefont {M.~A.}\ \bibnamefont {Valdez}},
  \bibinfo {author} {\bibfnamefont {L.~D.}\ \bibnamefont {Carr}},\ and\
  \bibinfo {author} {\bibfnamefont {K.~R.~A.}\ \bibnamefont {Hazzard}},\
  }\bibfield  {title} {\bibinfo {title} {Complex-network description of thermal
  quantum states in the {I}sing spin chain},\ }\href
  {https://doi.org/10.1103/PhysRevA.97.052320} {\bibfield  {journal} {\bibinfo
  {journal} {Physical Review A}\ }\textbf {\bibinfo {volume} {97}},\ \bibinfo
  {pages} {052320} (\bibinfo {year} {2018})}\BibitemShut {NoStop}%
\bibitem [{\citenamefont {Sokolov}\ \emph {et~al.}(2022)\citenamefont
  {Sokolov}, \citenamefont {Rossi}, \citenamefont {García-Pérez},\ and\
  \citenamefont {Maniscalco}}]{doi:10.1098/rsta.2020.0421}%
  \BibitemOpen
  \bibfield  {author} {\bibinfo {author} {\bibfnamefont {B.}~\bibnamefont
  {Sokolov}}, \bibinfo {author} {\bibfnamefont {M.~A.~C.}\ \bibnamefont
  {Rossi}}, \bibinfo {author} {\bibfnamefont {G.}~\bibnamefont
  {García-Pérez}},\ and\ \bibinfo {author} {\bibfnamefont {S.}~\bibnamefont
  {Maniscalco}},\ }\bibfield  {title} {\bibinfo {title} {Emergent entanglement
  structures and self-similarity in quantum spin chains},\ }\href
  {https://doi.org/10.1098/rsta.2020.0421} {\bibfield  {journal} {\bibinfo
  {journal} {Philosophical Transactions of the Royal Society A: Mathematical,
  Physical and Engineering Sciences}\ }\textbf {\bibinfo {volume} {380}},\
  \bibinfo {pages} {20200421} (\bibinfo {year} {2022})}\BibitemShut {NoStop}%
\bibitem [{\citenamefont {Bagrov}\ \emph {et~al.}(2020)\citenamefont {Bagrov},
  \citenamefont {Danilov}, \citenamefont {Brener}, \citenamefont {Harland},
  \citenamefont {Lichtenstein},\ and\ \citenamefont {Katsnelson}}]{Bagrov2020}%
  \BibitemOpen
  \bibfield  {author} {\bibinfo {author} {\bibfnamefont {A.~A.}\ \bibnamefont
  {Bagrov}}, \bibinfo {author} {\bibfnamefont {M.}~\bibnamefont {Danilov}},
  \bibinfo {author} {\bibfnamefont {S.}~\bibnamefont {Brener}}, \bibinfo
  {author} {\bibfnamefont {M.}~\bibnamefont {Harland}}, \bibinfo {author}
  {\bibfnamefont {A.~I.}\ \bibnamefont {Lichtenstein}},\ and\ \bibinfo {author}
  {\bibfnamefont {M.~I.}\ \bibnamefont {Katsnelson}},\ }\bibfield  {title}
  {\bibinfo {title} {Detecting quantum critical points in the $t-t'$
  {F}ermi-{H}ubbard model via complex network theory},\ }\href
  {https://doi.org/10.1038/s41598-020-77513-0} {\bibfield  {journal} {\bibinfo
  {journal} {Scientific Reports}\ }\textbf {\bibinfo {volume} {10}},\ \bibinfo
  {pages} {20470} (\bibinfo {year} {2020})}\BibitemShut {NoStop}%
\bibitem [{\citenamefont {Petri}\ \emph {et~al.}(2013)\citenamefont {Petri},
  \citenamefont {Scolamiero}, \citenamefont {Donato},\ and\ \citenamefont
  {Vaccarino}}]{petri2013topological}%
  \BibitemOpen
  \bibfield  {author} {\bibinfo {author} {\bibfnamefont {G.}~\bibnamefont
  {Petri}}, \bibinfo {author} {\bibfnamefont {M.}~\bibnamefont {Scolamiero}},
  \bibinfo {author} {\bibfnamefont {I.}~\bibnamefont {Donato}},\ and\ \bibinfo
  {author} {\bibfnamefont {F.}~\bibnamefont {Vaccarino}},\ }\bibfield  {title}
  {\bibinfo {title} {Topological strata of weighted complex networks},\
  }\href@noop {} {\bibfield  {journal} {\bibinfo  {journal} {PloS one}\
  }\textbf {\bibinfo {volume} {8}},\ \bibinfo {pages} {e66506} (\bibinfo {year}
  {2013})}\BibitemShut {NoStop}%
\bibitem [{\citenamefont {Edelsbrunner}\ and\ \citenamefont
  {Harer}(2022)}]{edelsbrunner2022computational}%
  \BibitemOpen
  \bibfield  {author} {\bibinfo {author} {\bibfnamefont {H.}~\bibnamefont
  {Edelsbrunner}}\ and\ \bibinfo {author} {\bibfnamefont {J.~L.}\ \bibnamefont
  {Harer}},\ }\href@noop {} {\emph {\bibinfo {title} {Computational topology:
  an introduction}}}\ (\bibinfo  {publisher} {American Mathematical Society},\
  \bibinfo {year} {2022})\BibitemShut {NoStop}%
\bibitem [{\citenamefont {Otter}\ \emph {et~al.}(2017)\citenamefont {Otter},
  \citenamefont {Porter}, \citenamefont {Tillmann}, \citenamefont {Grindrod},\
  and\ \citenamefont {Harrington}}]{otter2017roadmap}%
  \BibitemOpen
  \bibfield  {author} {\bibinfo {author} {\bibfnamefont {N.}~\bibnamefont
  {Otter}}, \bibinfo {author} {\bibfnamefont {M.~A.}\ \bibnamefont {Porter}},
  \bibinfo {author} {\bibfnamefont {U.}~\bibnamefont {Tillmann}}, \bibinfo
  {author} {\bibfnamefont {P.}~\bibnamefont {Grindrod}},\ and\ \bibinfo
  {author} {\bibfnamefont {H.~A.}\ \bibnamefont {Harrington}},\ }\bibfield
  {title} {\bibinfo {title} {A roadmap for the computation of persistent
  homology},\ }\href@noop {} {\bibfield  {journal} {\bibinfo  {journal} {EPJ
  Data Science}\ }\textbf {\bibinfo {volume} {6}},\ \bibinfo {pages} {1}
  (\bibinfo {year} {2017})}\BibitemShut {NoStop}%
\bibitem [{\citenamefont {Vaccarino}\ \emph {et~al.}(2022)\citenamefont
  {Vaccarino}, \citenamefont {Fugacci},\ and\ \citenamefont
  {Scaramuccia}}]{vaccarino2022persistent}%
  \BibitemOpen
  \bibfield  {author} {\bibinfo {author} {\bibfnamefont {F.}~\bibnamefont
  {Vaccarino}}, \bibinfo {author} {\bibfnamefont {U.}~\bibnamefont {Fugacci}},\
  and\ \bibinfo {author} {\bibfnamefont {S.}~\bibnamefont {Scaramuccia}},\
  }\bibfield  {title} {\bibinfo {title} {Persistent homology: A topological
  tool for higher-interaction systems},\ }in\ \href@noop {} {\emph {\bibinfo
  {booktitle} {Higher-Order Systems}}}\ (\bibinfo  {publisher} {Springer},\
  \bibinfo {year} {2022})\ pp.\ \bibinfo {pages} {97--139}\BibitemShut
  {NoStop}%
\bibitem [{\citenamefont {Ghrist}(2008)}]{ghrist2008barcodes}%
  \BibitemOpen
  \bibfield  {author} {\bibinfo {author} {\bibfnamefont {R.}~\bibnamefont
  {Ghrist}},\ }\bibfield  {title} {\bibinfo {title} {Barcodes: the persistent
  topology of data},\ }\href@noop {} {\bibfield  {journal} {\bibinfo  {journal}
  {Bulletin of the American Mathematical Society}\ }\textbf {\bibinfo {volume}
  {45}},\ \bibinfo {pages} {61} (\bibinfo {year} {2008})}\BibitemShut {NoStop}%
\bibitem [{\citenamefont {Donato}\ \emph {et~al.}(2016)\citenamefont {Donato},
  \citenamefont {Gori}, \citenamefont {Pettini}, \citenamefont {Petri},
  \citenamefont {De~Nigris}, \citenamefont {Franzosi},\ and\ \citenamefont
  {Vaccarino}}]{PhysRevE.93.052138}%
  \BibitemOpen
  \bibfield  {author} {\bibinfo {author} {\bibfnamefont {I.}~\bibnamefont
  {Donato}}, \bibinfo {author} {\bibfnamefont {M.}~\bibnamefont {Gori}},
  \bibinfo {author} {\bibfnamefont {M.}~\bibnamefont {Pettini}}, \bibinfo
  {author} {\bibfnamefont {G.}~\bibnamefont {Petri}}, \bibinfo {author}
  {\bibfnamefont {S.}~\bibnamefont {De~Nigris}}, \bibinfo {author}
  {\bibfnamefont {R.}~\bibnamefont {Franzosi}},\ and\ \bibinfo {author}
  {\bibfnamefont {F.}~\bibnamefont {Vaccarino}},\ }\bibfield  {title} {\bibinfo
  {title} {Persistent homology analysis of phase transitions},\ }\href
  {https://doi.org/10.1103/PhysRevE.93.052138} {\bibfield  {journal} {\bibinfo
  {journal} {Physical Review E}\ }\textbf {\bibinfo {volume} {93}},\ \bibinfo
  {pages} {052138} (\bibinfo {year} {2016})}\BibitemShut {NoStop}%
\bibitem [{\citenamefont {Olsthoorn}\ \emph {et~al.}(2020)\citenamefont
  {Olsthoorn}, \citenamefont {Hellsvik},\ and\ \citenamefont
  {Balatsky}}]{PhysRevResearch.2.043308}%
  \BibitemOpen
  \bibfield  {author} {\bibinfo {author} {\bibfnamefont {B.}~\bibnamefont
  {Olsthoorn}}, \bibinfo {author} {\bibfnamefont {J.}~\bibnamefont
  {Hellsvik}},\ and\ \bibinfo {author} {\bibfnamefont {A.~V.}\ \bibnamefont
  {Balatsky}},\ }\bibfield  {title} {\bibinfo {title} {Finding hidden order in
  spin models with persistent homology},\ }\href
  {https://doi.org/10.1103/PhysRevResearch.2.043308} {\bibfield  {journal}
  {\bibinfo  {journal} {Physical Review Res.}\ }\textbf {\bibinfo {volume}
  {2}},\ \bibinfo {pages} {043308} (\bibinfo {year} {2020})}\BibitemShut
  {NoStop}%
\bibitem [{\citenamefont {Cole}\ \emph {et~al.}(2021)\citenamefont {Cole},
  \citenamefont {Loges},\ and\ \citenamefont {Shiu}}]{PhysRevB.104.104426}%
  \BibitemOpen
  \bibfield  {author} {\bibinfo {author} {\bibfnamefont {A.}~\bibnamefont
  {Cole}}, \bibinfo {author} {\bibfnamefont {G.~J.}\ \bibnamefont {Loges}},\
  and\ \bibinfo {author} {\bibfnamefont {G.}~\bibnamefont {Shiu}},\ }\bibfield
  {title} {\bibinfo {title} {Quantitative and interpretable order parameters
  for phase transitions from persistent homology},\ }\href
  {https://doi.org/10.1103/PhysRevB.104.104426} {\bibfield  {journal} {\bibinfo
   {journal} {Physical Review B}\ }\textbf {\bibinfo {volume} {104}},\ \bibinfo
  {pages} {104426} (\bibinfo {year} {2021})}\BibitemShut {NoStop}%
\bibitem [{\citenamefont {Tran}\ \emph {et~al.}(2021)\citenamefont {Tran},
  \citenamefont {Chen},\ and\ \citenamefont {Hasegawa}}]{PhysRevE.103.052127}%
  \BibitemOpen
  \bibfield  {author} {\bibinfo {author} {\bibfnamefont {Q.~H.}\ \bibnamefont
  {Tran}}, \bibinfo {author} {\bibfnamefont {M.}~\bibnamefont {Chen}},\ and\
  \bibinfo {author} {\bibfnamefont {Y.}~\bibnamefont {Hasegawa}},\ }\bibfield
  {title} {\bibinfo {title} {Topological persistence machine of phase
  transitions},\ }\href {https://doi.org/10.1103/PhysRevE.103.052127}
  {\bibfield  {journal} {\bibinfo  {journal} {Physical Review E}\ }\textbf
  {\bibinfo {volume} {103}},\ \bibinfo {pages} {052127} (\bibinfo {year}
  {2021})}\BibitemShut {NoStop}%
\bibitem [{\citenamefont {Tirelli}\ and\ \citenamefont
  {Costa}(2021)}]{PhysRevB.104.235146}%
  \BibitemOpen
  \bibfield  {author} {\bibinfo {author} {\bibfnamefont {A.}~\bibnamefont
  {Tirelli}}\ and\ \bibinfo {author} {\bibfnamefont {N.~C.}\ \bibnamefont
  {Costa}},\ }\bibfield  {title} {\bibinfo {title} {Learning quantum phase
  transitions through topological data analysis},\ }\href
  {https://doi.org/10.1103/PhysRevB.104.235146} {\bibfield  {journal} {\bibinfo
   {journal} {Physical Review B}\ }\textbf {\bibinfo {volume} {104}},\ \bibinfo
  {pages} {235146} (\bibinfo {year} {2021})}\BibitemShut {NoStop}%
\bibitem [{\citenamefont {Sale}\ \emph {et~al.}(2022)\citenamefont {Sale},
  \citenamefont {Giansiracusa},\ and\ \citenamefont
  {Lucini}}]{PhysRevE.105.024121}%
  \BibitemOpen
  \bibfield  {author} {\bibinfo {author} {\bibfnamefont {N.}~\bibnamefont
  {Sale}}, \bibinfo {author} {\bibfnamefont {J.}~\bibnamefont {Giansiracusa}},\
  and\ \bibinfo {author} {\bibfnamefont {B.}~\bibnamefont {Lucini}},\
  }\bibfield  {title} {\bibinfo {title} {Quantitative analysis of phase
  transitions in two-dimensional $xy$ models using persistent homology},\
  }\href {https://doi.org/10.1103/PhysRevE.105.024121} {\bibfield  {journal}
  {\bibinfo  {journal} {Physical Review E}\ }\textbf {\bibinfo {volume}
  {105}},\ \bibinfo {pages} {024121} (\bibinfo {year} {2022})}\BibitemShut
  {NoStop}%
\bibitem [{\citenamefont {{Tirelli, Andrea}}\ \emph {et~al.}(2022)\citenamefont
  {{Tirelli, Andrea}}, \citenamefont {{Carvalho, Danyella O.}}, \citenamefont
  {{Oliveira, Lucas A.}}, \citenamefont {{de Lima, Jos\'e P.}}, \citenamefont
  {{Costa, Natanael C.}},\ and\ \citenamefont {{dos Santos, Raimundo
  R.}}}]{10.1140/epjb/s10051-022-00453-3}%
  \BibitemOpen
  \bibfield  {author} {\bibinfo {author} {\bibnamefont {{Tirelli, Andrea}}},
  \bibinfo {author} {\bibnamefont {{Carvalho, Danyella O.}}}, \bibinfo {author}
  {\bibnamefont {{Oliveira, Lucas A.}}}, \bibinfo {author} {\bibnamefont {{de
  Lima, Jos\'e P.}}}, \bibinfo {author} {\bibnamefont {{Costa, Natanael C.}}},\
  and\ \bibinfo {author} {\bibnamefont {{dos Santos, Raimundo R.}}},\
  }\bibfield  {title} {\bibinfo {title} {Unsupervised machine learning
  approaches to the q-state potts model},\ }\href
  {https://doi.org/10.1140/epjb/s10051-022-00453-3} {\bibfield  {journal}
  {\bibinfo  {journal} {Eur. Phys. J. B}\ }\textbf {\bibinfo {volume} {95}},\
  \bibinfo {pages} {189} (\bibinfo {year} {2022})}\BibitemShut {NoStop}%
\bibitem [{\citenamefont {Spitz}\ \emph {et~al.}(2021)\citenamefont {Spitz},
  \citenamefont {Berges}, \citenamefont {Oberthaler},\ and\ \citenamefont
  {Wienhard}}]{10.21468/SciPostPhys.11.3.060}%
  \BibitemOpen
  \bibfield  {author} {\bibinfo {author} {\bibfnamefont {D.}~\bibnamefont
  {Spitz}}, \bibinfo {author} {\bibfnamefont {J.}~\bibnamefont {Berges}},
  \bibinfo {author} {\bibfnamefont {M.}~\bibnamefont {Oberthaler}},\ and\
  \bibinfo {author} {\bibfnamefont {A.}~\bibnamefont {Wienhard}},\ }\bibfield
  {title} {\bibinfo {title} {{Finding self-similar behavior in quantum
  many-body dynamics via persistent homology}},\ }\href
  {https://doi.org/10.21468/SciPostPhys.11.3.060} {\bibfield  {journal}
  {\bibinfo  {journal} {SciPost Phys.}\ }\textbf {\bibinfo {volume} {11}},\
  \bibinfo {pages} {060} (\bibinfo {year} {2021})}\BibitemShut {NoStop}%
\bibitem [{\citenamefont {Sehayek}\ and\ \citenamefont
  {Melko}(2022)}]{PhysRevB.106.085111}%
  \BibitemOpen
  \bibfield  {author} {\bibinfo {author} {\bibfnamefont {D.}~\bibnamefont
  {Sehayek}}\ and\ \bibinfo {author} {\bibfnamefont {R.~G.}\ \bibnamefont
  {Melko}},\ }\bibfield  {title} {\bibinfo {title} {Persistent homology of
  ${\mathbb{z}}_{2}$ gauge theories},\ }\href
  {https://doi.org/10.1103/PhysRevB.106.085111} {\bibfield  {journal} {\bibinfo
   {journal} {Physical Review B}\ }\textbf {\bibinfo {volume} {106}},\ \bibinfo
  {pages} {085111} (\bibinfo {year} {2022})}\BibitemShut {NoStop}%
\bibitem [{\citenamefont {Sale}\ \emph {et~al.}(2023)\citenamefont {Sale},
  \citenamefont {Lucini},\ and\ \citenamefont
  {Giansiracusa}}]{PhysRevD.107.034501}%
  \BibitemOpen
  \bibfield  {author} {\bibinfo {author} {\bibfnamefont {N.}~\bibnamefont
  {Sale}}, \bibinfo {author} {\bibfnamefont {B.}~\bibnamefont {Lucini}},\ and\
  \bibinfo {author} {\bibfnamefont {J.}~\bibnamefont {Giansiracusa}},\
  }\bibfield  {title} {\bibinfo {title} {Probing center vortices and
  deconfinement in su(2) lattice gauge theory with persistent homology},\
  }\href {https://doi.org/10.1103/PhysRevD.107.034501} {\bibfield  {journal}
  {\bibinfo  {journal} {Physical Review D}\ }\textbf {\bibinfo {volume}
  {107}},\ \bibinfo {pages} {034501} (\bibinfo {year} {2023})}\BibitemShut
  {NoStop}%
\bibitem [{\citenamefont {Spitz}\ \emph {et~al.}(2023)\citenamefont {Spitz},
  \citenamefont {Urban},\ and\ \citenamefont
  {Pawlowski}}]{PhysRevD.107.034506}%
  \BibitemOpen
  \bibfield  {author} {\bibinfo {author} {\bibfnamefont {D.}~\bibnamefont
  {Spitz}}, \bibinfo {author} {\bibfnamefont {J.~M.}\ \bibnamefont {Urban}},\
  and\ \bibinfo {author} {\bibfnamefont {J.~M.}\ \bibnamefont {Pawlowski}},\
  }\bibfield  {title} {\bibinfo {title} {Confinement in non-abelian lattice
  gauge theory via persistent homology},\ }\href
  {https://doi.org/10.1103/PhysRevD.107.034506} {\bibfield  {journal} {\bibinfo
   {journal} {Physical Review D}\ }\textbf {\bibinfo {volume} {107}},\ \bibinfo
  {pages} {034506} (\bibinfo {year} {2023})}\BibitemShut {NoStop}%
\bibitem [{\citenamefont {Mendes-Santos}\ \emph {et~al.}(2023)\citenamefont
  {Mendes-Santos}, \citenamefont {Schmitt}, \citenamefont {Angelone},
  \citenamefont {Rodriguez}, \citenamefont {Scholl}, \citenamefont {Williams},
  \citenamefont {Barredo}, \citenamefont {Lahaye}, \citenamefont {Browaeys},
  \citenamefont {Heyl} \emph {et~al.}}]{mendes2023wave}%
  \BibitemOpen
  \bibfield  {author} {\bibinfo {author} {\bibfnamefont {T.}~\bibnamefont
  {Mendes-Santos}}, \bibinfo {author} {\bibfnamefont {M.}~\bibnamefont
  {Schmitt}}, \bibinfo {author} {\bibfnamefont {A.}~\bibnamefont {Angelone}},
  \bibinfo {author} {\bibfnamefont {A.}~\bibnamefont {Rodriguez}}, \bibinfo
  {author} {\bibfnamefont {P.}~\bibnamefont {Scholl}}, \bibinfo {author}
  {\bibfnamefont {H.}~\bibnamefont {Williams}}, \bibinfo {author}
  {\bibfnamefont {D.}~\bibnamefont {Barredo}}, \bibinfo {author} {\bibfnamefont
  {T.}~\bibnamefont {Lahaye}}, \bibinfo {author} {\bibfnamefont
  {A.}~\bibnamefont {Browaeys}}, \bibinfo {author} {\bibfnamefont
  {M.}~\bibnamefont {Heyl}}, \emph {et~al.},\ }\bibfield  {title} {\bibinfo
  {title} {Wave function network description and kolmogorov complexity of
  quantum many-body systems},\ }\href@noop {} {\bibfield  {journal} {\bibinfo
  {journal} {arXiv preprint arXiv:2301.13216}\ } (\bibinfo {year}
  {2023})}\BibitemShut {NoStop}%
\bibitem [{\citenamefont {Herrero}(2002)}]{herrero2002ising}%
  \BibitemOpen
  \bibfield  {author} {\bibinfo {author} {\bibfnamefont {C.~P.}\ \bibnamefont
  {Herrero}},\ }\bibfield  {title} {\bibinfo {title} {Ising model in
  small-world networks},\ }\href@noop {} {\bibfield  {journal} {\bibinfo
  {journal} {Physical Review E}\ }\textbf {\bibinfo {volume} {65}},\ \bibinfo
  {pages} {066110} (\bibinfo {year} {2002})}\BibitemShut {NoStop}%
\bibitem [{\citenamefont {Pekalski}(2001)}]{pekalski2001ising}%
  \BibitemOpen
  \bibfield  {author} {\bibinfo {author} {\bibfnamefont {A.}~\bibnamefont
  {Pekalski}},\ }\bibfield  {title} {\bibinfo {title} {Ising model on a small
  world network},\ }\href@noop {} {\bibfield  {journal} {\bibinfo  {journal}
  {Physical Review E}\ }\textbf {\bibinfo {volume} {64}},\ \bibinfo {pages}
  {057104} (\bibinfo {year} {2001})}\BibitemShut {NoStop}%
\bibitem [{\citenamefont {Herrero}(2004)}]{herrero2004ising}%
  \BibitemOpen
  \bibfield  {author} {\bibinfo {author} {\bibfnamefont {C.~P.}\ \bibnamefont
  {Herrero}},\ }\bibfield  {title} {\bibinfo {title} {Ising model in scale-free
  networks: A monte carlo simulation},\ }\href@noop {} {\bibfield  {journal}
  {\bibinfo  {journal} {Physical Review E}\ }\textbf {\bibinfo {volume} {69}},\
  \bibinfo {pages} {067109} (\bibinfo {year} {2004})}\BibitemShut {NoStop}%
\bibitem [{\citenamefont {Herrero}(2015)}]{herrero2015ising}%
  \BibitemOpen
  \bibfield  {author} {\bibinfo {author} {\bibfnamefont {C.~P.}\ \bibnamefont
  {Herrero}},\ }\bibfield  {title} {\bibinfo {title} {Ising model in clustered
  scale-free networks},\ }\href@noop {} {\bibfield  {journal} {\bibinfo
  {journal} {Physical Review E}\ }\textbf {\bibinfo {volume} {91}},\ \bibinfo
  {pages} {052812} (\bibinfo {year} {2015})}\BibitemShut {NoStop}%
\bibitem [{\citenamefont {Bradde}\ \emph {et~al.}(2010)\citenamefont {Bradde},
  \citenamefont {Caccioli}, \citenamefont {Dall’Asta},\ and\ \citenamefont
  {Bianconi}}]{bradde2010critical}%
  \BibitemOpen
  \bibfield  {author} {\bibinfo {author} {\bibfnamefont {S.}~\bibnamefont
  {Bradde}}, \bibinfo {author} {\bibfnamefont {F.}~\bibnamefont {Caccioli}},
  \bibinfo {author} {\bibfnamefont {L.}~\bibnamefont {Dall’Asta}},\ and\
  \bibinfo {author} {\bibfnamefont {G.}~\bibnamefont {Bianconi}},\ }\bibfield
  {title} {\bibinfo {title} {Critical fluctuations in spatial complex
  networks},\ }\href@noop {} {\bibfield  {journal} {\bibinfo  {journal}
  {Physical review letters}\ }\textbf {\bibinfo {volume} {104}},\ \bibinfo
  {pages} {218701} (\bibinfo {year} {2010})}\BibitemShut {NoStop}%
\bibitem [{\citenamefont {Onsager}(1944)}]{onsager1944crystal}%
  \BibitemOpen
  \bibfield  {author} {\bibinfo {author} {\bibfnamefont {L.}~\bibnamefont
  {Onsager}},\ }\bibfield  {title} {\bibinfo {title} {Crystal statistics. i. a
  two-dimensional model with an order-disorder transition},\ }\href@noop {}
  {\bibfield  {journal} {\bibinfo  {journal} {Physical Review}\ }\textbf
  {\bibinfo {volume} {65}},\ \bibinfo {pages} {117} (\bibinfo {year}
  {1944})}\BibitemShut {NoStop}%
\bibitem [{\citenamefont {Panda}\ \emph {et~al.}(2023)\citenamefont {Panda},
  \citenamefont {Verdel}, \citenamefont {Rodriguez}, \citenamefont {Sun},
  \citenamefont {Bianconi},\ and\ \citenamefont {Dalmonte}}]{panda23}%
  \BibitemOpen
  \bibfield  {author} {\bibinfo {author} {\bibfnamefont {R.~K.}\ \bibnamefont
  {Panda}}, \bibinfo {author} {\bibfnamefont {R.}~\bibnamefont {Verdel}},
  \bibinfo {author} {\bibfnamefont {A.}~\bibnamefont {Rodriguez}}, \bibinfo
  {author} {\bibfnamefont {H.}~\bibnamefont {Sun}}, \bibinfo {author}
  {\bibfnamefont {G.}~\bibnamefont {Bianconi}},\ and\ \bibinfo {author}
  {\bibfnamefont {M.}~\bibnamefont {Dalmonte}},\ }\bibfield  {title} {\bibinfo
  {title} {Non-parametric learning critical behavior in {I}sing partition
  functions: {PCA} entropy and intrinsic dimension},\ }\href@noop {} {\bibfield
   {journal} {\bibinfo  {journal} {arXiv preprint arXiv:2308.13636}\ }
  (\bibinfo {year} {2023})}\BibitemShut {NoStop}%
\bibitem [{\citenamefont {Wolff}(1989{\natexlab{a}})}]{wolff1989collective}%
  \BibitemOpen
  \bibfield  {author} {\bibinfo {author} {\bibfnamefont {U.}~\bibnamefont
  {Wolff}},\ }\bibfield  {title} {\bibinfo {title} {Collective {M}onte {C}arlo
  updating for spin systems},\ }\href@noop {} {\bibfield  {journal} {\bibinfo
  {journal} {Physical Review Letters}\ }\textbf {\bibinfo {volume} {62}},\
  \bibinfo {pages} {361} (\bibinfo {year} {1989}{\natexlab{a}})}\BibitemShut
  {NoStop}%
\bibitem [{\citenamefont {Wolff}(1989{\natexlab{b}})}]{wolff1989comparison}%
  \BibitemOpen
  \bibfield  {author} {\bibinfo {author} {\bibfnamefont {U.}~\bibnamefont
  {Wolff}},\ }\bibfield  {title} {\bibinfo {title} {Comparison between cluster
  {M}onte {C}arlo algorithms in the {I}sing model},\ }\href@noop {} {\bibfield
  {journal} {\bibinfo  {journal} {Physics Letters B}\ }\textbf {\bibinfo
  {volume} {228}},\ \bibinfo {pages} {379} (\bibinfo {year}
  {1989}{\natexlab{b}})}\BibitemShut {NoStop}%
\bibitem [{\citenamefont {Graham}\ and\ \citenamefont
  {Hell}(1985)}]{graham1985history}%
  \BibitemOpen
  \bibfield  {author} {\bibinfo {author} {\bibfnamefont {R.~L.}\ \bibnamefont
  {Graham}}\ and\ \bibinfo {author} {\bibfnamefont {P.}~\bibnamefont {Hell}},\
  }\bibfield  {title} {\bibinfo {title} {On the history of the minimum spanning
  tree problem},\ }\href@noop {} {\bibfield  {journal} {\bibinfo  {journal}
  {Annals of the History of Computing}\ }\textbf {\bibinfo {volume} {7}},\
  \bibinfo {pages} {43} (\bibinfo {year} {1985})}\BibitemShut {NoStop}%
\bibitem [{\citenamefont {McInnes}\ \emph {et~al.}(2018)\citenamefont
  {McInnes}, \citenamefont {Healy},\ and\ \citenamefont
  {Melville}}]{mcinnes2018umap}%
  \BibitemOpen
  \bibfield  {author} {\bibinfo {author} {\bibfnamefont {L.}~\bibnamefont
  {McInnes}}, \bibinfo {author} {\bibfnamefont {J.}~\bibnamefont {Healy}},\
  and\ \bibinfo {author} {\bibfnamefont {J.}~\bibnamefont {Melville}},\
  }\bibfield  {title} {\bibinfo {title} {Umap: Uniform manifold approximation
  and projection for dimension reduction},\ }\href@noop {} {\bibfield
  {journal} {\bibinfo  {journal} {arXiv preprint arXiv:1802.03426}\ } (\bibinfo
  {year} {2018})}\BibitemShut {NoStop}%
\bibitem [{\citenamefont {Bavelas}(1950)}]{bavelas1950communication}%
  \BibitemOpen
  \bibfield  {author} {\bibinfo {author} {\bibfnamefont {A.}~\bibnamefont
  {Bavelas}},\ }\bibfield  {title} {\bibinfo {title} {Communication patterns in
  task-oriented groups},\ }\href@noop {} {\bibfield  {journal} {\bibinfo
  {journal} {The journal of the acoustical society of America}\ }\textbf
  {\bibinfo {volume} {22}},\ \bibinfo {pages} {725} (\bibinfo {year}
  {1950})}\BibitemShut {NoStop}%
\bibitem [{\citenamefont {Dorogovtsev}\ \emph {et~al.}(2008)\citenamefont
  {Dorogovtsev}, \citenamefont {Goltsev},\ and\ \citenamefont
  {Mendes}}]{dorogovtsev2008critical}%
  \BibitemOpen
  \bibfield  {author} {\bibinfo {author} {\bibfnamefont {S.~N.}\ \bibnamefont
  {Dorogovtsev}}, \bibinfo {author} {\bibfnamefont {A.~V.}\ \bibnamefont
  {Goltsev}},\ and\ \bibinfo {author} {\bibfnamefont {J.~F.}\ \bibnamefont
  {Mendes}},\ }\bibfield  {title} {\bibinfo {title} {Critical phenomena in
  complex networks},\ }\href@noop {} {\bibfield  {journal} {\bibinfo  {journal}
  {Reviews of Modern Physics}\ }\textbf {\bibinfo {volume} {80}},\ \bibinfo
  {pages} {1275} (\bibinfo {year} {2008})}\BibitemShut {NoStop}%
\bibitem [{\citenamefont {Li}\ \emph {et~al.}(2021)\citenamefont {Li},
  \citenamefont {Liu}, \citenamefont {L{\"u}}, \citenamefont {Hu},
  \citenamefont {Xu},\ and\ \citenamefont {Zhang}}]{li2021percolation}%
  \BibitemOpen
  \bibfield  {author} {\bibinfo {author} {\bibfnamefont {M.}~\bibnamefont
  {Li}}, \bibinfo {author} {\bibfnamefont {R.-R.}\ \bibnamefont {Liu}},
  \bibinfo {author} {\bibfnamefont {L.}~\bibnamefont {L{\"u}}}, \bibinfo
  {author} {\bibfnamefont {M.-B.}\ \bibnamefont {Hu}}, \bibinfo {author}
  {\bibfnamefont {S.}~\bibnamefont {Xu}},\ and\ \bibinfo {author}
  {\bibfnamefont {Y.-C.}\ \bibnamefont {Zhang}},\ }\bibfield  {title} {\bibinfo
  {title} {Percolation on complex networks: Theory and application},\
  }\href@noop {} {\bibfield  {journal} {\bibinfo  {journal} {Physics Reports}\
  }\textbf {\bibinfo {volume} {907}},\ \bibinfo {pages} {1} (\bibinfo {year}
  {2021})}\BibitemShut {NoStop}%
\bibitem [{\citenamefont {Cohen}\ \emph {et~al.}(2000)\citenamefont {Cohen},
  \citenamefont {Erez}, \citenamefont {Ben-Avraham},\ and\ \citenamefont
  {Havlin}}]{cohen2000resilience}%
  \BibitemOpen
  \bibfield  {author} {\bibinfo {author} {\bibfnamefont {R.}~\bibnamefont
  {Cohen}}, \bibinfo {author} {\bibfnamefont {K.}~\bibnamefont {Erez}},
  \bibinfo {author} {\bibfnamefont {D.}~\bibnamefont {Ben-Avraham}},\ and\
  \bibinfo {author} {\bibfnamefont {S.}~\bibnamefont {Havlin}},\ }\bibfield
  {title} {\bibinfo {title} {Resilience of the internet to random breakdowns},\
  }\href@noop {} {\bibfield  {journal} {\bibinfo  {journal} {Physical Review
  Letters}\ }\textbf {\bibinfo {volume} {85}},\ \bibinfo {pages} {4626}
  (\bibinfo {year} {2000})}\BibitemShut {NoStop}%
\bibitem [{\citenamefont {Bobrowski}\ and\ \citenamefont
  {Skraba}(2023)}]{bobrowski2023universal}%
  \BibitemOpen
  \bibfield  {author} {\bibinfo {author} {\bibfnamefont {O.}~\bibnamefont
  {Bobrowski}}\ and\ \bibinfo {author} {\bibfnamefont {P.}~\bibnamefont
  {Skraba}},\ }\bibfield  {title} {\bibinfo {title} {A universal
  null-distribution for topological data analysis},\ }\href@noop {} {\bibfield
  {journal} {\bibinfo  {journal} {Scientific Reports}\ }\textbf {\bibinfo
  {volume} {13}},\ \bibinfo {pages} {12274} (\bibinfo {year}
  {2023})}\BibitemShut {NoStop}%
\bibitem [{\citenamefont {Baptista}\ \emph {et~al.}(2023)\citenamefont
  {Baptista}, \citenamefont {S{\'a}nchez-Garc{\'\i}a}, \citenamefont {Baudot},\
  and\ \citenamefont {Bianconi}}]{baptista2023zoo}%
  \BibitemOpen
  \bibfield  {author} {\bibinfo {author} {\bibfnamefont {A.}~\bibnamefont
  {Baptista}}, \bibinfo {author} {\bibfnamefont {R.~J.}\ \bibnamefont
  {S{\'a}nchez-Garc{\'\i}a}}, \bibinfo {author} {\bibfnamefont
  {A.}~\bibnamefont {Baudot}},\ and\ \bibinfo {author} {\bibfnamefont
  {G.}~\bibnamefont {Bianconi}},\ }\bibfield  {title} {\bibinfo {title} {Zoo
  guide to network embedding},\ }\href@noop {} {\bibfield  {journal} {\bibinfo
  {journal} {arXiv preprint arXiv:2305.03474}\ } (\bibinfo {year}
  {2023})}\BibitemShut {NoStop}%
\bibitem [{\citenamefont {Seyed-Allaei}\ \emph {et~al.}(2006)\citenamefont
  {Seyed-Allaei}, \citenamefont {Bianconi},\ and\ \citenamefont
  {Marsili}}]{seyed2006scale}%
  \BibitemOpen
  \bibfield  {author} {\bibinfo {author} {\bibfnamefont {H.}~\bibnamefont
  {Seyed-Allaei}}, \bibinfo {author} {\bibfnamefont {G.}~\bibnamefont
  {Bianconi}},\ and\ \bibinfo {author} {\bibfnamefont {M.}~\bibnamefont
  {Marsili}},\ }\bibfield  {title} {\bibinfo {title} {Scale-free networks with
  an exponent less than two},\ }\href@noop {} {\bibfield  {journal} {\bibinfo
  {journal} {Physical Review E}\ }\textbf {\bibinfo {volume} {73}},\ \bibinfo
  {pages} {046113} (\bibinfo {year} {2006})}\BibitemShut {NoStop}%
\bibitem [{\citenamefont {Courtney}\ and\ \citenamefont
  {Bianconi}(2018)}]{courtney2018dense}%
  \BibitemOpen
  \bibfield  {author} {\bibinfo {author} {\bibfnamefont {O.~T.}\ \bibnamefont
  {Courtney}}\ and\ \bibinfo {author} {\bibfnamefont {G.}~\bibnamefont
  {Bianconi}},\ }\bibfield  {title} {\bibinfo {title} {Dense power-law networks
  and simplicial complexes},\ }\href@noop {} {\bibfield  {journal} {\bibinfo
  {journal} {Physical Review E}\ }\textbf {\bibinfo {volume} {97}},\ \bibinfo
  {pages} {052303} (\bibinfo {year} {2018})}\BibitemShut {NoStop}%
\bibitem [{\citenamefont {Caron}\ and\ \citenamefont
  {Fox}(2017)}]{caron2017sparse}%
  \BibitemOpen
  \bibfield  {author} {\bibinfo {author} {\bibfnamefont {F.}~\bibnamefont
  {Caron}}\ and\ \bibinfo {author} {\bibfnamefont {E.~B.}\ \bibnamefont
  {Fox}},\ }\bibfield  {title} {\bibinfo {title} {Sparse graphs using
  exchangeable random measures},\ }\href@noop {} {\bibfield  {journal}
  {\bibinfo  {journal} {Journal of the Royal Statistical Society Series B:
  Statistical Methodology}\ }\textbf {\bibinfo {volume} {79}},\ \bibinfo
  {pages} {1295} (\bibinfo {year} {2017})}\BibitemShut {NoStop}%
\bibitem [{\citenamefont {Timar}\ \emph {et~al.}(2016)\citenamefont {Timar},
  \citenamefont {Dorogovtsev},\ and\ \citenamefont {Mendes}}]{timar2016scale}%
  \BibitemOpen
  \bibfield  {author} {\bibinfo {author} {\bibfnamefont {G.}~\bibnamefont
  {Timar}}, \bibinfo {author} {\bibfnamefont {S.~N.}\ \bibnamefont
  {Dorogovtsev}},\ and\ \bibinfo {author} {\bibfnamefont {J.~F.~F.}\
  \bibnamefont {Mendes}},\ }\bibfield  {title} {\bibinfo {title} {Scale-free
  networks with exponent one},\ }\href@noop {} {\bibfield  {journal} {\bibinfo
  {journal} {Physical Review E}\ }\textbf {\bibinfo {volume} {94}},\ \bibinfo
  {pages} {022302} (\bibinfo {year} {2016})}\BibitemShut {NoStop}%
\bibitem [{\citenamefont {Barab{\'a}si}\ and\ \citenamefont
  {Albert}(1999)}]{barabasi1999emergence}%
  \BibitemOpen
  \bibfield  {author} {\bibinfo {author} {\bibfnamefont {A.-L.}\ \bibnamefont
  {Barab{\'a}si}}\ and\ \bibinfo {author} {\bibfnamefont {R.}~\bibnamefont
  {Albert}},\ }\bibfield  {title} {\bibinfo {title} {Emergence of scaling in
  random networks},\ }\href@noop {} {\bibfield  {journal} {\bibinfo  {journal}
  {Science}\ }\textbf {\bibinfo {volume} {286}},\ \bibinfo {pages} {509}
  (\bibinfo {year} {1999})}\BibitemShut {NoStop}%
\bibitem [{\citenamefont {Albert}\ and\ \citenamefont
  {Barab{\'a}si}(2002)}]{albert2002statistical}%
  \BibitemOpen
  \bibfield  {author} {\bibinfo {author} {\bibfnamefont {R.}~\bibnamefont
  {Albert}}\ and\ \bibinfo {author} {\bibfnamefont {A.-L.}\ \bibnamefont
  {Barab{\'a}si}},\ }\bibfield  {title} {\bibinfo {title} {Statistical
  mechanics of complex networks},\ }\href@noop {} {\bibfield  {journal}
  {\bibinfo  {journal} {Reviews of Modern Physics}\ }\textbf {\bibinfo {volume}
  {74}},\ \bibinfo {pages} {47} (\bibinfo {year} {2002})}\BibitemShut {NoStop}%
\bibitem [{\citenamefont {Voitalov}\ \emph {et~al.}(2019)\citenamefont
  {Voitalov}, \citenamefont {Van Der~Hoorn}, \citenamefont {Van Der~Hofstad},\
  and\ \citenamefont {Krioukov}}]{voitalov2019scale}%
  \BibitemOpen
  \bibfield  {author} {\bibinfo {author} {\bibfnamefont {I.}~\bibnamefont
  {Voitalov}}, \bibinfo {author} {\bibfnamefont {P.}~\bibnamefont {Van
  Der~Hoorn}}, \bibinfo {author} {\bibfnamefont {R.}~\bibnamefont {Van
  Der~Hofstad}},\ and\ \bibinfo {author} {\bibfnamefont {D.}~\bibnamefont
  {Krioukov}},\ }\bibfield  {title} {\bibinfo {title} {Scale-free networks well
  done},\ }\href@noop {} {\bibfield  {journal} {\bibinfo  {journal} {Physical
  Review Research}\ }\textbf {\bibinfo {volume} {1}},\ \bibinfo {pages}
  {033034} (\bibinfo {year} {2019})}\BibitemShut {NoStop}%
\bibitem [{\citenamefont {Barth{\'e}lemy}\ \emph {et~al.}(2005)\citenamefont
  {Barth{\'e}lemy}, \citenamefont {Barrat}, \citenamefont {Pastor-Satorras},\
  and\ \citenamefont {Vespignani}}]{barthelemy2005characterization}%
  \BibitemOpen
  \bibfield  {author} {\bibinfo {author} {\bibfnamefont {M.}~\bibnamefont
  {Barth{\'e}lemy}}, \bibinfo {author} {\bibfnamefont {A.}~\bibnamefont
  {Barrat}}, \bibinfo {author} {\bibfnamefont {R.}~\bibnamefont
  {Pastor-Satorras}},\ and\ \bibinfo {author} {\bibfnamefont {A.}~\bibnamefont
  {Vespignani}},\ }\bibfield  {title} {\bibinfo {title} {Characterization and
  modeling of weighted networks},\ }\href@noop {} {\bibfield  {journal}
  {\bibinfo  {journal} {Physica a: Statistical mechanics and its applications}\
  }\textbf {\bibinfo {volume} {346}},\ \bibinfo {pages} {34} (\bibinfo {year}
  {2005})}\BibitemShut {NoStop}%
\bibitem [{\citenamefont {Barrat}\ \emph {et~al.}(2004)\citenamefont {Barrat},
  \citenamefont {Barthelemy}, \citenamefont {Pastor-Satorras},\ and\
  \citenamefont {Vespignani}}]{barrat2004architecture}%
  \BibitemOpen
  \bibfield  {author} {\bibinfo {author} {\bibfnamefont {A.}~\bibnamefont
  {Barrat}}, \bibinfo {author} {\bibfnamefont {M.}~\bibnamefont {Barthelemy}},
  \bibinfo {author} {\bibfnamefont {R.}~\bibnamefont {Pastor-Satorras}},\ and\
  \bibinfo {author} {\bibfnamefont {A.}~\bibnamefont {Vespignani}},\ }\bibfield
   {title} {\bibinfo {title} {The architecture of complex weighted networks},\
  }\href@noop {} {\bibfield  {journal} {\bibinfo  {journal} {Proceedings of the
  national academy of sciences}\ }\textbf {\bibinfo {volume} {101}},\ \bibinfo
  {pages} {3747} (\bibinfo {year} {2004})}\BibitemShut {NoStop}%
\bibitem [{\citenamefont {Watts}\ and\ \citenamefont
  {Strogatz}(1998)}]{watts1998collective}%
  \BibitemOpen
  \bibfield  {author} {\bibinfo {author} {\bibfnamefont {D.~J.}\ \bibnamefont
  {Watts}}\ and\ \bibinfo {author} {\bibfnamefont {S.~H.}\ \bibnamefont
  {Strogatz}},\ }\bibfield  {title} {\bibinfo {title} {Collective dynamics of
  ‘small-world’networks},\ }\href@noop {} {\bibfield  {journal} {\bibinfo
  {journal} {Nature}\ }\textbf {\bibinfo {volume} {393}},\ \bibinfo {pages}
  {440} (\bibinfo {year} {1998})}\BibitemShut {NoStop}%
\bibitem [{\citenamefont {Bianconi}\ and\ \citenamefont
  {Marsili}(2006)}]{bianconi2006effect}%
  \BibitemOpen
  \bibfield  {author} {\bibinfo {author} {\bibfnamefont {G.}~\bibnamefont
  {Bianconi}}\ and\ \bibinfo {author} {\bibfnamefont {M.}~\bibnamefont
  {Marsili}},\ }\bibfield  {title} {\bibinfo {title} {Effect of degree
  correlations on the loop structure of scale-free networks},\ }\href@noop {}
  {\bibfield  {journal} {\bibinfo  {journal} {Physical Review E}\ }\textbf
  {\bibinfo {volume} {73}},\ \bibinfo {pages} {066127} (\bibinfo {year}
  {2006})}\BibitemShut {NoStop}%
\bibitem [{\citenamefont {Alvarez-Hamelin}\ \emph {et~al.}(2005)\citenamefont
  {Alvarez-Hamelin}, \citenamefont {Dall'Asta}, \citenamefont {Barrat},\ and\
  \citenamefont {Vespignani}}]{alvarez2005large}%
  \BibitemOpen
  \bibfield  {author} {\bibinfo {author} {\bibfnamefont {J.}~\bibnamefont
  {Alvarez-Hamelin}}, \bibinfo {author} {\bibfnamefont {L.}~\bibnamefont
  {Dall'Asta}}, \bibinfo {author} {\bibfnamefont {A.}~\bibnamefont {Barrat}},\
  and\ \bibinfo {author} {\bibfnamefont {A.}~\bibnamefont {Vespignani}},\
  }\bibfield  {title} {\bibinfo {title} {Large scale networks fingerprinting
  and visualization using the k-core decomposition},\ }\href@noop {} {\bibfield
   {journal} {\bibinfo  {journal} {Advances in neural information processing
  systems}\ }\textbf {\bibinfo {volume} {18}} (\bibinfo {year}
  {2005})}\BibitemShut {NoStop}%
\bibitem [{\citenamefont {Carmi}\ \emph {et~al.}(2007)\citenamefont {Carmi},
  \citenamefont {Havlin}, \citenamefont {Kirkpatrick}, \citenamefont
  {Shavitt},\ and\ \citenamefont {Shir}}]{carmi2007model}%
  \BibitemOpen
  \bibfield  {author} {\bibinfo {author} {\bibfnamefont {S.}~\bibnamefont
  {Carmi}}, \bibinfo {author} {\bibfnamefont {S.}~\bibnamefont {Havlin}},
  \bibinfo {author} {\bibfnamefont {S.}~\bibnamefont {Kirkpatrick}}, \bibinfo
  {author} {\bibfnamefont {Y.}~\bibnamefont {Shavitt}},\ and\ \bibinfo {author}
  {\bibfnamefont {E.}~\bibnamefont {Shir}},\ }\bibfield  {title} {\bibinfo
  {title} {A model of internet topology using k-shell decomposition},\
  }\href@noop {} {\bibfield  {journal} {\bibinfo  {journal} {Proceedings of the
  National Academy of Sciences}\ }\textbf {\bibinfo {volume} {104}},\ \bibinfo
  {pages} {11150} (\bibinfo {year} {2007})}\BibitemShut {NoStop}%
\bibitem [{\citenamefont {Dorogovtsev}\ \emph {et~al.}(2006)\citenamefont
  {Dorogovtsev}, \citenamefont {Goltsev},\ and\ \citenamefont
  {Mendes}}]{dorogovtsev2006k}%
  \BibitemOpen
  \bibfield  {author} {\bibinfo {author} {\bibfnamefont {S.~N.}\ \bibnamefont
  {Dorogovtsev}}, \bibinfo {author} {\bibfnamefont {A.~V.}\ \bibnamefont
  {Goltsev}},\ and\ \bibinfo {author} {\bibfnamefont {J.~F.~F.}\ \bibnamefont
  {Mendes}},\ }\bibfield  {title} {\bibinfo {title} {K-core organization of
  complex networks},\ }\href@noop {} {\bibfield  {journal} {\bibinfo  {journal}
  {Physical Review Letters}\ }\textbf {\bibinfo {volume} {96}},\ \bibinfo
  {pages} {040601} (\bibinfo {year} {2006})}\BibitemShut {NoStop}%
\bibitem [{\citenamefont {Almaas}\ \emph {et~al.}(2004)\citenamefont {Almaas},
  \citenamefont {Kovacs}, \citenamefont {Vicsek}, \citenamefont {Oltvai},\ and\
  \citenamefont {Barab{\'a}si}}]{almaas2004global}%
  \BibitemOpen
  \bibfield  {author} {\bibinfo {author} {\bibfnamefont {E.}~\bibnamefont
  {Almaas}}, \bibinfo {author} {\bibfnamefont {B.}~\bibnamefont {Kovacs}},
  \bibinfo {author} {\bibfnamefont {T.}~\bibnamefont {Vicsek}}, \bibinfo
  {author} {\bibfnamefont {Z.}~\bibnamefont {Oltvai}},\ and\ \bibinfo {author}
  {\bibfnamefont {A.-L.}\ \bibnamefont {Barab{\'a}si}},\ }\bibfield  {title}
  {\bibinfo {title} {Global organization of metabolic fluxes in the bacterium
  escherichia coli},\ }\href@noop {} {\bibfield  {journal} {\bibinfo  {journal}
  {Nature}\ }\textbf {\bibinfo {volume} {427}},\ \bibinfo {pages} {839}
  (\bibinfo {year} {2004})}\BibitemShut {NoStop}%
\bibitem [{\citenamefont {Burioni}\ and\ \citenamefont
  {Cassi}(2005)}]{burioni2005random}%
  \BibitemOpen
  \bibfield  {author} {\bibinfo {author} {\bibfnamefont {R.}~\bibnamefont
  {Burioni}}\ and\ \bibinfo {author} {\bibfnamefont {D.}~\bibnamefont
  {Cassi}},\ }\bibfield  {title} {\bibinfo {title} {Random walks on graphs:
  ideas, techniques and results},\ }\href@noop {} {\bibfield  {journal}
  {\bibinfo  {journal} {Journal of Physics A: Mathematical and General}\
  }\textbf {\bibinfo {volume} {38}},\ \bibinfo {pages} {R45} (\bibinfo {year}
  {2005})}\BibitemShut {NoStop}%
\bibitem [{\citenamefont {Correia}\ and\ \citenamefont
  {Wheater}(1998)}]{correia1998spectral}%
  \BibitemOpen
  \bibfield  {author} {\bibinfo {author} {\bibfnamefont {J.~D.}\ \bibnamefont
  {Correia}}\ and\ \bibinfo {author} {\bibfnamefont {J.~F.}\ \bibnamefont
  {Wheater}},\ }\bibfield  {title} {\bibinfo {title} {The spectral dimension of
  non-generic branched polymer ensembles},\ }\href@noop {} {\bibfield
  {journal} {\bibinfo  {journal} {Physics Letters B}\ }\textbf {\bibinfo
  {volume} {422}},\ \bibinfo {pages} {76} (\bibinfo {year} {1998})}\BibitemShut
  {NoStop}%
\bibitem [{\citenamefont {Ambj{\o}rn}\ \emph {et~al.}(2005)\citenamefont
  {Ambj{\o}rn}, \citenamefont {Jurkiewicz},\ and\ \citenamefont
  {Loll}}]{ambjorn2005spectral}%
  \BibitemOpen
  \bibfield  {author} {\bibinfo {author} {\bibfnamefont {J.}~\bibnamefont
  {Ambj{\o}rn}}, \bibinfo {author} {\bibfnamefont {J.}~\bibnamefont
  {Jurkiewicz}},\ and\ \bibinfo {author} {\bibfnamefont {R.}~\bibnamefont
  {Loll}},\ }\bibfield  {title} {\bibinfo {title} {The spectral dimension of
  the universe is scale dependent},\ }\href@noop {} {\bibfield  {journal}
  {\bibinfo  {journal} {Physical Review Letters}\ }\textbf {\bibinfo {volume}
  {95}},\ \bibinfo {pages} {171301} (\bibinfo {year} {2005})}\BibitemShut
  {NoStop}%
\bibitem [{\citenamefont {Anand}\ and\ \citenamefont
  {Bianconi}(2009)}]{anand2009entropy}%
  \BibitemOpen
  \bibfield  {author} {\bibinfo {author} {\bibfnamefont {K.}~\bibnamefont
  {Anand}}\ and\ \bibinfo {author} {\bibfnamefont {G.}~\bibnamefont
  {Bianconi}},\ }\bibfield  {title} {\bibinfo {title} {Entropy measures for
  networks: Toward an information theory of complex topologies},\ }\href@noop
  {} {\bibfield  {journal} {\bibinfo  {journal} {Physical Review E}\ }\textbf
  {\bibinfo {volume} {80}},\ \bibinfo {pages} {045102} (\bibinfo {year}
  {2009})}\BibitemShut {NoStop}%
\bibitem [{\citenamefont {Anand}\ \emph {et~al.}(2011)\citenamefont {Anand},
  \citenamefont {Bianconi},\ and\ \citenamefont {Severini}}]{anand2011shannon}%
  \BibitemOpen
  \bibfield  {author} {\bibinfo {author} {\bibfnamefont {K.}~\bibnamefont
  {Anand}}, \bibinfo {author} {\bibfnamefont {G.}~\bibnamefont {Bianconi}},\
  and\ \bibinfo {author} {\bibfnamefont {S.}~\bibnamefont {Severini}},\
  }\bibfield  {title} {\bibinfo {title} {Shannon and von {N}eumann entropy of
  random networks with heterogeneous expected degree},\ }\href@noop {}
  {\bibfield  {journal} {\bibinfo  {journal} {Physical Review E}\ }\textbf
  {\bibinfo {volume} {83}},\ \bibinfo {pages} {036109} (\bibinfo {year}
  {2011})}\BibitemShut {NoStop}%
\bibitem [{\citenamefont {De~Domenico}\ \emph {et~al.}(2015)\citenamefont
  {De~Domenico}, \citenamefont {Nicosia}, \citenamefont {Arenas},\ and\
  \citenamefont {Latora}}]{de2015structural}%
  \BibitemOpen
  \bibfield  {author} {\bibinfo {author} {\bibfnamefont {M.}~\bibnamefont
  {De~Domenico}}, \bibinfo {author} {\bibfnamefont {V.}~\bibnamefont
  {Nicosia}}, \bibinfo {author} {\bibfnamefont {A.}~\bibnamefont {Arenas}},\
  and\ \bibinfo {author} {\bibfnamefont {V.}~\bibnamefont {Latora}},\
  }\bibfield  {title} {\bibinfo {title} {Structural reducibility of multilayer
  networks},\ }\href@noop {} {\bibfield  {journal} {\bibinfo  {journal} {Nature
  Communications}\ }\textbf {\bibinfo {volume} {6}},\ \bibinfo {pages} {6864}
  (\bibinfo {year} {2015})}\BibitemShut {NoStop}%
\bibitem [{\citenamefont {Mendes-Santos}\ \emph
  {et~al.}(2021{\natexlab{b}})\citenamefont {Mendes-Santos}, \citenamefont
  {Angelone}, \citenamefont {Rodriguez}, \citenamefont {Fazio},\ and\
  \citenamefont {Dalmonte}}]{mendes2021intrinsic}%
  \BibitemOpen
  \bibfield  {author} {\bibinfo {author} {\bibfnamefont {T.}~\bibnamefont
  {Mendes-Santos}}, \bibinfo {author} {\bibfnamefont {A.}~\bibnamefont
  {Angelone}}, \bibinfo {author} {\bibfnamefont {A.}~\bibnamefont {Rodriguez}},
  \bibinfo {author} {\bibfnamefont {R.}~\bibnamefont {Fazio}},\ and\ \bibinfo
  {author} {\bibfnamefont {M.}~\bibnamefont {Dalmonte}},\ }\bibfield  {title}
  {\bibinfo {title} {Intrinsic dimension of path integrals: Data-mining quantum
  criticality and emergent simplicity},\ }\href@noop {} {\bibfield  {journal}
  {\bibinfo  {journal} {PRX Quantum}\ }\textbf {\bibinfo {volume} {2}},\
  \bibinfo {pages} {030332} (\bibinfo {year} {2021}{\natexlab{b}})}\BibitemShut
  {NoStop}%
\end{thebibliography}%

\end{document}